\documentclass[twocolumn,showpacs,aps,amsmath,amssymb,superscriptaddress]{revtex4-2}  

\usepackage[latin9]{inputenc}
\usepackage{textcomp}
\usepackage{amsmath}
\usepackage{graphicx}
\PassOptionsToPackage{version=3}{mhchem}
\usepackage{mhchem}
\usepackage{xcolor}

\usepackage[ruled]{algorithm2e}

\renewcommand{\thetable}{\arabic{table}}

\makeatletter

\usepackage{subfigure}\usepackage{epstopdf}\usepackage{dcolumn}
\usepackage{bm}
\usepackage{multirow}
\usepackage{longtable}\usepackage{rotating}\usepackage{threeparttable}
\usepackage{float}

\newcommand{\addSIcontents}[2]{%
	\addcontentsline{tosi}{#1}{\protect\numberline{\csname the#1\endcsname}#2}%
}

\makeatother

\begin{document}
\title{Reduced Density Matrix Functional Theory Across Molecules and Periodic Systems: Short-Range Screening and Coupled Optimization}

\author{Shuxin Pei}
\affiliation{Center for Theoretical and Computational Chemistry, Frontiers Science Center for New Organic Matter, State Key Laboratory of Advanced Chemical Power Sources, Key Laboratory of Advanced Energy Materials Chemistry (Ministry of Education), Department of Chemistry, Nankai University, Tianjin 300071, China}

\author{Neil Qiang Su}
\email[Corresponding author. ]{nqsu@nankai.edu.cn}
\affiliation{Center for Theoretical and Computational Chemistry, Frontiers Science Center for New Organic Matter, State Key Laboratory of Advanced Chemical Power Sources, Key Laboratory of Advanced Energy Materials Chemistry (Ministry of Education), Department of Chemistry, Nankai University, Tianjin 300071, China}

\begin{abstract}

Reduced density matrix functional theory (RDMFT) provides a rigorous framework for describing many-electron systems, yet its application to periodic systems has long been hindered by two major challenges: the lack of accurate, transferable functionals and the computational inefficiency of orbital-occupation optimization. Here, we address both challenges by developing a periodic RDMFT framework that enables the simultaneous optimization of Bloch natural orbitals and their occupation numbers through coupled optimization (CO). Using this framework, we demonstrate that physically motivated short-range screening substantially improves the transferability of the Power functional across molecular and periodic systems. Beyond improving energetic accuracy, short-range screening enhances numerical robustness by smoothing energy profiles during optimization and improves computational efficiency through rapid reciprocal-space convergence of the long-range contribution. The CO algorithm consistently reduces the number of optimization iterations relative to conventional decoupled optimization, with reductions approaching an order of magnitude for representative periodic solids. Furthermore, the screened $\omega$P22 functional outperforms widely used semilocal and hybrid density functionals, including PBE and PBE0, in predicting molecular and surface reaction barriers. These findings establish an efficient computational framework for periodic RDMFT and identify short-range screening as a promising design principle for developing accurate and transferable one-body reduced-density-matrix functionals.

\end{abstract}
\maketitle

\section{\label{sec:level1}Introduction}
Over the past few decades, Kohn-Sham density functional theory (KS-DFT) \cite{HK1964,KS1965,Parr1989dft,Gross2012dft} has become the workhorse for electronic-structure calculations of molecules and solids because of its favorable balance between computational efficiency and predictive accuracy. Nevertheless, widely used density functionals in KS-DFT still struggle with systems in which strong electronic correlation effects plays a decisive role. Representative challenges include reaction energies and activation barriers \cite{johnson1994barrier,COpuzzle2001,zhao2011energy,Norskov2015,2017sbh10barrier}, molecular dissociation \cite{bally1997curve,dutoi2006curve,ruzsinszky2006curve,Yang2008curve}, and Mott insulating states \cite{terakura1984Mott,imada1998Mott,dudarev1998Mott}.
Reduced density matrix functional theory (RDMFT) \cite{gilbert1975rdmft,levy1979rdmft,valone1980rdmft} provides an  alternative framework by adopting the one-body reduced density matrix (1-RDM), or equivalently its spectral representation in terms of natural orbitals (NOs) and their occupation numbers (ONs) \cite{coleman19631rdm}, as the fundamental variable. This formulation enables an explicit description of fractional orbital occupations arising from both strong and dynamic electron correlation, thereby naturally incorporating multireference character without relying on a fictitious non-interacting Kohn-Sham reference system. Consequently, RDMFT offers an attractive framework for describing electron correlation and has demonstrated particular promise for systems dominated by strong correlation \cite{Su2021pra,gritsenko2005xc,piris2006pnof,rohr2008,lathiotakis2009,polar2020jctc,Schilling2021prl,Gibney2023prl}.

Despite these advantages, practical applications of RDMFT remain relatively limited, with most developments and benchmarks focused on finite systems, particularly molecular dissociation and other strongly correlated problems \cite{gritsenko2005xc,rohr2008,lathiotakis2009,ai2022wp22}. Applications to extended systems are considerably less developed. Pioneering studies by Gross and co-workers demonstrated that the Power functional can reproduce the insulating character of prototypical Mott insulators and predict fundamental gaps of conventional semiconductors \cite{power2008}. Subsequent investigations showed promising descriptions of spectral properties in comparison with experiments and many-body calculations \cite{sharma2013spectrum,shinohara2015spectrum}, as well as metal-insulator transitions in qualitative agreement with dynamical mean-field theory \cite{shinohara2015doping}. These advances demonstrate the potential of RDMFT for periodic systems while highlighting the need for further methodological and functional development. A central challenge is to develop approximate 1-RDM functionals that are accurate and transferable across different classes of systems. In particular, achieving reliable ground-state energetics for both molecules and extended solids within a single functional approximation remains difficult. Recently, incorporating physically motivated short-range screening into the Power functional led to the development of the screened $\omega$P22 functional \cite{ai2022wp22,ai2023wp22}, which substantially improved the description of molecular energetics, including reaction barriers \cite{ai2023wp22,Zhang2023pra}. More broadly, screening is a fundamental concept in many-electron physics, playing a central role in the description of effective electron-electron interactions and electronic correlation across a wide range of many-body theories \cite{Hedin1965pr,Sabatino2022prb,Sabatino2023prb,Mahler2022prb,Yu2025JPCL}. These developments motivate a fundamental question: Can short-range screening serve as a transferable design principle for advancing 1-RDM functionals that accurately describe both molecular and periodic systems?

Beyond the development of accurate functionals, computational efficiency presents another major challenge for RDMFT. Conventional RDMFT calculations typically employ decoupled optimization procedures, in which the total-energy functional is minimized alternately with respect to NOs and ONs. Such procedures can suffer from slow convergence because these two sets of variational variables are strongly coupled. Recently, an efficient coupled optimization (CO) framework \cite{Zhang2026} was developed by combining unitary transformations of the NOs \cite{unitary2002,unitary2009}, which preserve orbital orthonormality, with the explicit-by-implicit (EBI) method \cite{ebi2021,ebi2022}, which enforces ensemble-$N$-representability constraints on the ONs \cite{coleman19631rdm,valone1980rdmft,lowdin1955ensembleN}. By optimizing NOs and ONs simultaneously, CO substantially accelerates convergence and achieves lower variational energies in molecular benchmarks \cite{yao2024COopt,cartier2024COopt}. Extending this framework to periodic systems, however, introduces additional challenges. Periodic RDMFT requires the treatment of Bloch natural orbitals, Brillouin-zone sampling, and long-range electron interactions under periodic boundary conditions. Moreover, the global particle-number constraint couples occupation variables across different $\mathbf{k}$ points, complicating their simultaneous optimization with the Bloch orbitals. These challenges raise a second fundamental question: Can the efficiency and robustness of coupled NO-ON optimization be retained in periodic RDMFT calculations?

In this work, we address these two questions by developing a periodic implementation of RDMFT incorporating both the CO framework and conventional decoupled optimization (deCO). We systematically investigate the numerical performance of these optimization strategies and compare the original Power functional with its short-range-screened counterpart, $\omega$P22. Through benchmarks spanning molecular and periodic systems, including surface reaction barriers, we examine the transferability of short-range screening and the efficiency of simultaneous NO-ON optimization under periodic boundary conditions. Our results demonstrate that the combination of CO and short-range screening provides a promising route toward accurate and computationally efficient RDMFT calculations for extended systems.

The remainder of this paper is organized as follows. Section~\ref{sec:level2} presents the theoretical formulation of periodic RDMFT, including the derivation of the first- and second-order derivatives of the total-energy functional with respect to the optimization variables and the extension of the CO algorithm to periodic boundary conditions. Computational details are provided in Sec.~\ref{sec:level3}. Section~\ref{sec:level4} evaluates the optimization efficiency, $\mathbf{k}$-point convergence, and energetic accuracy of the proposed framework for molecular and periodic systems, including surface reaction barriers. Finally, Sec.~\ref{sec:level5} summarizes the main findings and conclusions.

\section{\label{sec:level2}Theory}
Throughout this paper, the indices $\mu, \nu, \kappa, \lambda$ refer to atomic orbitals (AOs), whereas $i, j, a, b, p, q$ represent crystalline/molecular orbitals (MOs).  

\subsection{Energy Functionals for Periodic RDMFT}
RDMFT employs the 1-RDM as its fundamental variable \cite{lowdin1955ensembleN,gilbert1975rdmft,levy1979rdmft},
\begin{align}
\label{eq:gamma_def}
\gamma(\mathbf{r},\mathbf{r}')
&=\gamma_{\alpha}(\mathbf{r},\mathbf{r}')
+\gamma_{\beta}(\mathbf{r},\mathbf{r}') \nonumber\\
&=N\int \mathrm{d}\sigma_1\,\mathrm{d}\mathbf{x}_2\cdots\mathrm{d}\mathbf{x}_N\,
\Psi(\mathbf{r}\sigma_1,\mathbf{x}_2,\ldots,\mathbf{x}_N)
\nonumber\\
&\quad\times
\Psi^*(\mathbf{r}'\sigma_1,\mathbf{x}_2,\ldots,\mathbf{x}_N).
\end{align}
Here, $\Psi$ denotes the normalized $N$-electron wavefunction, and $\mathbf{x}=(\mathbf{r},\sigma)$ represents the combined spatial and spin coordinates. The electron density is given by the diagonal of the 1-RDM,
\begin{equation}
\label{eq:density_rdmft}
\rho(\mathbf{r})
=\rho_\alpha(\mathbf{r})+\rho_\beta(\mathbf{r})
=\gamma_\alpha(\mathbf{r},\mathbf{r})
+\gamma_\beta(\mathbf{r},\mathbf{r}).
\end{equation}
The total electronic energy can be expressed as
\begin{equation}
\label{eq:energy_rdmft}
E[\gamma_\sigma]
=T[\gamma_\sigma]
+E_{\mathrm{ext}}[\gamma_\sigma]
+E_{\mathrm{H}}[\gamma_\sigma]
+E_{\mathrm{XC}}[\gamma_\sigma],
\end{equation}
where the kinetic energy is given by
\begin{equation}
\label{eq:T_rdmft}
T[\gamma_\sigma]
=-\frac{1}{2}\sum_\sigma
\int\mathrm{d}\mathbf{r}\,
\left.
\nabla_{\mathbf{r}}^2
\gamma_\sigma(\mathbf{r},\mathbf{r}')
\right|_{\mathbf{r}'=\mathbf{r}},
\end{equation}
the external-potential energy by
\begin{equation}
\label{eq:Eext_rdmft}
E_{\mathrm{ext}}[\gamma_\sigma]
=\int\mathrm{d}\mathbf{r}\,
v_{\mathrm{ext}}(\mathbf{r})\rho(\mathbf{r}),
\end{equation}
and the Hartree energy by
\begin{equation}
\label{eq:Har_rdmft}
E_{\mathrm{H}}[\gamma_\sigma]
=\frac{1}{2}
\int\mathrm{d}\mathbf{r}
\int\mathrm{d}\mathbf{r}'
\frac{\rho(\mathbf{r})\rho(\mathbf{r}')}{|\mathbf{r}-\mathbf{r}'|}.
\end{equation}
While the kinetic, external-potential, and Hartree energies are explicitly determined by the 1-RDM, the exact form of the exchange-correlation (XC) energy functional, $E_{\mathrm{XC}}$, is generally unknown and must be approximated in practical calculations.

Spectral decomposition of the 1-RDM yields the NOs and their corresponding ONs, in terms of which the RDMFT energy functional can be equivalently expressed. For periodic systems, this decomposition takes the form \cite{power2008}
\begin{equation}
\label{eq:gamma_spec}
\gamma_{\sigma}(\mathbf{r},\mathbf{r}')
=\frac{1}{N_{\mathbf{k}}}
\sum_{\mathbf{k}} \sum_i
n_i^{\mathbf{k},\sigma}
\psi_i^{\mathbf{k},\sigma}(\mathbf{r})
\psi_i^{\mathbf{k},\sigma,*}(\mathbf{r}'),
\end{equation}
where $\psi_i^{\mathbf{k},\sigma}$ and $n_i^{\mathbf{k},\sigma}$ denote the NOs and their corresponding ONs at crystal momentum $\mathbf{k}$, respectively. Owing to translational symmetry, the NOs are chosen as Bloch functions, which are orthonormal within the primitive unit cell of volume $\Omega$,
\begin{equation}
\label{eq:orb_constraint}
\left\langle \psi_i^{\mathbf{k},\sigma}
\middle|
\psi_j^{\mathbf{k},\sigma}\right\rangle
=
\int_{\Omega}\mathrm{d}\mathbf{r}\,
\psi_i^{\mathbf{k},\sigma,*}(\mathbf{r})
\psi_j^{\mathbf{k},\sigma}(\mathbf{r})
=\delta_{ij}.
\end{equation}
Furthermore, ensemble $N$-representability requires the ONs to satisfy \cite{coleman19631rdm}
\begin{equation}
\label{eq:occ_constraint}
\frac{1}{N_{\mathbf{k}}}
\sum_{\mathbf{k}} \sum_i
n_i^{\mathbf{k},\sigma}=N_e^\sigma,
\qquad
0\leq n_i^{\mathbf{k},\sigma}\leq 1,
\end{equation}
where $N_e^\sigma$ is the number of electrons with spin $\sigma$ per primitive unit cell, and $N_{\mathbf{k}}$ denotes the number of sampled $\mathbf{k}$ points in the first Brillouin zone.

In terms of the Bloch NOs and their corresponding ONs, the RDMFT energy in Eqs.~\eqref{eq:energy_rdmft}-\eqref{eq:Har_rdmft} can be expressed per primitive unit cell. In particular, the kinetic and external-potential energies are combined into the one-electron energy,
\begin{equation}
\label{eq:T+Eext_rdmft}
T[\gamma_\sigma]+E_{\mathrm{ext}}[\gamma_\sigma]
=\frac{1}{N_{\mathbf{k}}}
\sum_{\sigma}\sum_{\mathbf{k}}\sum_i
n_i^{\mathbf{k},\sigma}
h_{ii}^{\mathbf{k},\sigma},
\end{equation}
where the one-electron integrals are defined as
\begin{equation}
\label{eq:hii}
h_{ij}^{\mathbf{k},\sigma}
=
\int_{\Omega}\mathrm{d}\mathbf{r}\,
\psi_i^{\mathbf{k},\sigma,*}(\mathbf{r})
\left[-\frac{1}{2}\nabla^2
+v_{\mathrm{ext}}(\mathbf{r})\right]
\psi_j^{\mathbf{k},\sigma}(\mathbf{r}).
\end{equation}
The Hartree energy is given by
\begin{align}
\label{eq:Har_in_orb}
E_{\mathrm{H}}[\gamma_\sigma]
=&\frac{1}{2N_{\mathbf{k}}^2}
\sum_{\mathbf{k}\mathbf{k}'}
\sum_{\sigma\sigma'}\sum_{ij}
n_i^{\mathbf{k},\sigma}
n_j^{\mathbf{k}',\sigma'}\nonumber\\
&\times
(i^{\mathbf{k},\sigma}i^{\mathbf{k},\sigma}
|j^{\mathbf{k}',\sigma'}j^{\mathbf{k}',\sigma'}),
\end{align}
where the two-electron integrals are defined as
\begin{align}
\label{eq:2e_ints}
&(i^{\mathbf{k}_i,\sigma}j^{\mathbf{k}_j,\sigma}
|a^{\mathbf{k}_a,\sigma'}b^{\mathbf{k}_b,\sigma'})
\nonumber\\
&=\int_{\Omega}\mathrm{d}\mathbf{r}_1
\int\mathrm{d}\mathbf{r}_2\,
\frac{
\psi_i^{\mathbf{k}_i,\sigma,*}(\mathbf{r}_1)
\psi_j^{\mathbf{k}_j,\sigma}(\mathbf{r}_1)
}{
|\mathbf{r}_1-\mathbf{r}_2|
}
\nonumber\\
&\quad\times
\psi_a^{\mathbf{k}_a,\sigma',*}(\mathbf{r}_2)
\psi_b^{\mathbf{k}_b,\sigma'}(\mathbf{r}_2).
\end{align}
Translational symmetry imposes crystal-momentum conservation,
$-\mathbf{k}_i+\mathbf{k}_j-\mathbf{k}_a+\mathbf{k}_b=\mathbf{G}$,
where $\mathbf{G}$ is a reciprocal lattice vector.

Common approximations to the XC energy, including the Müller \cite{muller1984} and Power \cite{power2008} functionals, can be expressed as
\begin{align}
\label{eq:Exc}
E_{\mathrm{XC}}[\gamma_\sigma]
=&-\frac{1}{2N_{\mathbf{k}}^2}
\sum_{\mathbf{k}\mathbf{k}'}
\sum_{\sigma}\sum_{ij}
f(n_i^{\mathbf{k},\sigma},n_j^{\mathbf{k}',\sigma})
\nonumber\\
&\times
(i^{\mathbf{k},\sigma}j^{\mathbf{k}',\sigma}
|j^{\mathbf{k}',\sigma}i^{\mathbf{k},\sigma}),
\end{align}
where the Power functional employs
$f(n_i^{\mathbf{k},\sigma},n_j^{\mathbf{k}',\sigma})
=(n_i^{\mathbf{k},\sigma}n_j^{\mathbf{k}',\sigma})^m$.
The choices $m=1$ and $m=1/2$ recover the Hartree-Fock exchange expression and the Müller functional, respectively. These approximations belong to the class of Coulomb- and exchange-integral-based (JK-only) functionals, which are generally considered to provide an incomplete description of dynamic electron correlation \cite{Cioslowski2003jcp,Kollmar2004jcp,Pernal2015}. The range-separated $\omega$P22 functional \cite{ai2023wp22}, derived from the range separation of the electron-electron Coulomb interaction \cite{Savin1995ijqc,Thierry1997cpl}, extends the Power functional by incorporating short-range screening,
\begin{equation}
\label{eq:exc_wp22}
E_{\mathrm{XC}}^{\omega\mathrm{P}22}[\gamma_\sigma]
=
E_{\mathrm{XC}}^{\mathrm{LR}}[\gamma_\sigma]
+
E_{\mathrm{X}}^{\mathrm{KS,SR}}[\rho_\sigma]
+
U_{\mathrm{C}}^{\mathrm{KS,SR}}[\rho_\sigma].
\end{equation}
Here, $E_{\mathrm{XC}}^{\mathrm{LR}}$ has the same form as Eq.~\eqref{eq:Exc}, but with the Coulomb integrals replaced by their long-range counterparts, $(i^{\mathbf{k}_i,\sigma}j^{\mathbf{k}_j,\sigma}|a^{\mathbf{k}_a,\sigma'}b^{\mathbf{k}_b,\sigma'})_{\mathrm{LR}}$, as detailed in Sections SII of the Supplemental Material \cite{SupplementalMaterial}. The short-range exchange and correlation contributions, $E_{\mathrm{X}}^{\mathrm{KS,SR}}$ and $U_{\mathrm{C}}^{\mathrm{KS,SR}}$ make use of the short-range B88 \cite{becke1988,iikura2001B88} and LYPsr \cite{ai2021LYPsr} functionals, respectively, both of which are readily evaluated from the electron density$\rho_\sigma(\mathbf{r})$. These short-range contributions complement the long-range Power functional, improving its description of electron correlation. In $\omega$P22, the power exponent and range-separation parameter are fixed at $m=0.6$ and $\omega=0.45$, respectively \cite{ai2023wp22}. The resulting functional has demonstrated promising accuracy for a range of molecular properties.

To simplify the energy expression and its subsequent derivatives, we introduce two matrices. The first is the Coulomb matrix,
\begin{equation}
\label{eq:column_ints}
J_{pq}^{\mathbf{k},\sigma}
=
\frac{1}{N_{\mathbf{k}}}
\sum_{\mathbf{k}'}
\sum_{\sigma'}\sum_j
n_j^{\mathbf{k}',\sigma'}
(p^{\mathbf{k},\sigma}q^{\mathbf{k},\sigma}
|j^{\mathbf{k}',\sigma'}j^{\mathbf{k}',\sigma'}).
\end{equation}
The second is a generalized exchange matrix, which resembles the conventional exchange matrix but is defined through its action on the ONs,
\begin{align}
\label{eq:Kii_ni}
K_{pq}^{\mathbf{k},\sigma}n_i^{\mathbf{k},\sigma}
=&\frac{1}{N_{\mathbf{k}}}
\sum_{\mathbf{k}'}\sum_j
f(n_i^{\mathbf{k},\sigma},n_j^{\mathbf{k}',\sigma})
\nonumber\\
&\times
(p^{\mathbf{k},\sigma}j^{\mathbf{k}',\sigma}
|j^{\mathbf{k}',\sigma}q^{\mathbf{k},\sigma}).
\end{align}
With these definitions, the total electronic energy obtained by combining Eqs.~\eqref{eq:T+Eext_rdmft}, \eqref{eq:Har_in_orb}, and \eqref{eq:Exc} can be written compactly as
\begin{align}
\label{eq:e_rdmft_simple}
E[\gamma_\sigma]
=\frac{1}{N_{\mathbf{k}}}
\sum_{\mathbf{k}\sigma i}
\left(
h_{ii}^{\mathbf{k},\sigma}
+\frac{1}{2}J_{ii}^{\mathbf{k},\sigma}
-\frac{1}{2}K_{ii}^{\mathbf{k},\sigma}
\right)n_i^{\mathbf{k},\sigma}.
\end{align}
The following derivations are based on this expression, while other XC functionals can be treated analogously. In particular, the screened $\omega$P22 functional retains only the long-range electron-electron interaction in the Power functional contribution and incorporates additional short-range exchange and correlation terms. Although this range-separated construction was originally introduced to improve the description of electron correlation, we find that it also offers significant computational advantages for periodic systems. Specifically, the short-range contributions can be efficiently evaluated using the corresponding KS-based functionals, while, more importantly, the long-range contribution can be accurately evaluated in reciprocal space using a relatively small number of reciprocal lattice vectors $\mathbf{G}$ (see Supplemental Material \cite{SupplementalMaterial}, Sections SII and SIII).

\subsection{Variational Parameterization of Bloch Natural Orbitals and Occupations}

According to Gilbert's generalization \cite{gilbert1975rdmft} of the Hohenberg-Kohn theorem \cite{HK1964}, minimization of the exact energy functional with respect to the 1-RDM, or equivalently the NOs and ONs, yields the ground-state energy. For periodic systems, this requires the variational optimization of the Bloch NOs and their corresponding ONs.

For practical calculations, the Bloch NOs are expanded in a finite basis set,
\begin{equation}
\label{eq:pbc_gto}
\psi_i^{\mathbf{k},\sigma}(\mathbf{r})
=
\sum_{\mu=1}
C_{\mu i}^{\mathbf{k},\sigma}
\phi_\mu^{\mathbf{k}}(\mathbf{r}),
\end{equation}
where $C_{\mu i}^{\mathbf{k},\sigma}$ are the expansion coefficients. In this work, we employ translationally symmetry-adapted atomic orbitals (AOs), constructed as Bloch sums of atom-centered Gaussian-type orbitals (GTOs),
\begin{equation}
\label{eq:pbc_gto_symm}
\phi_\mu^{\mathbf{k}}(\mathbf{r})
=
\sum_{\mathbf{T}}
e^{i\mathbf{k}\cdot\mathbf{T}}
\phi_\mu(\mathbf{r}-\mathbf{T}),
\end{equation}
where $\phi_\mu$ denotes an atom-centered GTO and $\mathbf{T}$ runs over all lattice translation vectors.

To preserve the orthonormality constraint in Eq.~\eqref{eq:orb_constraint}, the Bloch NOs at each crystal momentum $\mathbf{k}$ are updated through unitary orbital rotations \cite{unitary2002,unitary2009}. Specifically, the orbital coefficient matrix at iteration $t$ is updated according to
\begin{equation}
\label{eq:orb_update}
\mathbf{C}^{\mathbf{k},\sigma}_{t+1}
=
\mathbf{C}^{\mathbf{k},\sigma}_{t}
\mathbf{U}^{\mathbf{k},\sigma},
\end{equation}
where the unitary rotation matrix is parameterized as
\begin{equation}
\label{eq:u_mat}
\mathbf{U}^{\mathbf{k},\sigma}
=
\exp(\mathbf{R}^{\mathbf{k},\sigma})
=
\mathbf{I}
+
\mathbf{R}^{\mathbf{k},\sigma}
+
\frac{(\mathbf{R}^{\mathbf{k},\sigma})^2}{2}
+\cdots.
\end{equation}
Here, $\mathbf{R}^{\mathbf{k},\sigma}=\mathbf{X}^{\mathbf{k},\sigma}+i\mathbf{Y}^{\mathbf{k},\sigma}$ is a skew-Hermitian matrix satisfying $(\mathbf{R}^{\mathbf{k},\sigma})^\dagger=-\mathbf{R}^{\mathbf{k},\sigma}$, where $\mathbf{X}^{\mathbf{k},\sigma}$ and $\mathbf{Y}^{\mathbf{k},\sigma}$ are real matrices obeying
\begin{equation}
\label{eq:real_imag_mat}
(\mathbf{X}^{\mathbf{k},\sigma})^{\mathrm T}
=-\mathbf{X}^{\mathbf{k},\sigma},
\qquad
(\mathbf{Y}^{\mathbf{k},\sigma})^{\mathrm T}
=\mathbf{Y}^{\mathbf{k},\sigma}.
\end{equation}
For an $M\times M$ orbital rotation matrix, $\mathbf{X}^{\mathbf{k},\sigma}$ and $\mathbf{Y}^{\mathbf{k},\sigma}$ contain $M(M-1)/2$ and $M(M+1)/2$ independent real parameters, respectively.

The ensemble $N$-representability constraints on the ONs are enforced using the EBI method \cite{ebi2021,ebi2022}, which expresses the ONs in terms of unconstrained auxiliary variables $\{x_i^{\mathbf{k},\sigma}\}$,
\begin{equation}
\label{eq:ebi_formula}
n_i^{\mathbf{k},\sigma}
=
s(x_i^{\mathbf{k},\sigma}+\mu^\sigma),
\end{equation}
where $s$ is a mapping function that restricts the ONs to the interval $[0,1]$, and $\mu^\sigma$ is an implicitly determined function of the auxiliary variables that enforces particle-number conservation. In this work, we adopt the error-function mapping,
\begin{equation}
n_i^{\mathbf{k},\sigma}
=
\frac{1}{2}
\left[
1+\operatorname{erf}(x_i^{\mathbf{k},\sigma}+\mu^\sigma)
\right].
\end{equation}
Together, the unitary orbital parameterization and the EBI mapping transform the constrained minimization of the energy functional over the admissible Bloch NOs and ONs, $\{\psi_i^{\mathbf{k},\sigma},n_i^{\mathbf{k},\sigma}\}$, into an unconstrained optimization problem in terms of the independent orbital-rotation parameters and auxiliary occupation variables, $\{R_{pq}^{\mathbf{k},\sigma},x_i^{\mathbf{k},\sigma}\}$. In the following, we derive the energy gradients with respect to these variables and develop the corresponding optimization algorithm.

\subsection{First- and Second-Order Energy Derivatives}
For conciseness, we present only the energy derivatives required for the variational optimization of Bloch NOs and ONs, leaving more derivations to the Supplemental Material \cite{SupplementalMaterial}, Section SI.

The first derivative of the total energy $E$ with respect to the orbital-rotation matrix $\mathbf{R}^{\mathbf{k},\sigma}$, evaluated at $\mathbf{R}^{\mathbf{k},\sigma}=\mathbf{0}$, is given by
\begin{equation}
\label{eq:dE_dR}
\left.
\frac{\partial E}{\partial \mathbf{R}^{\mathbf{k},\sigma}}
\right|_{\mathbf{R}^{\mathbf{k},\sigma}=\mathbf{0}}
=
(\mathbf{C}^{\mathbf{k},\sigma})^{\rm T}
\frac{\partial E}{\partial \mathbf{C}^{\mathbf{k},\sigma}}
-
\left(
\frac{\partial E}{\partial \mathbf{C}^{\mathbf{k},\sigma,*}}
\right)^{\rm T}
\mathbf{C}^{\mathbf{k},\sigma,*}.
\end{equation}
We introduce the generalized Fock matrix $\mathbf{F}^{\mathbf{k},\sigma}$, whose elements are defined as
\begin{equation}
\label{eq:Fuv}
F_{pq}^{\mathbf{k},\sigma}
=
h_{pq}^{\mathbf{k},\sigma}
+
J_{pq}^{\mathbf{k},\sigma}
-
K_{pq}^{\mathbf{k},\sigma}.
\end{equation}
Using this definition, the orbital-rotation derivative can be expressed as
\begin{equation}
\label{eq:dE_dR_fock}
\left.
\frac{\partial E}{\partial R_{pq}^{\mathbf{k},\sigma}}
\right|_{\mathbf{R}^{\mathbf{k},\sigma}=\mathbf{0}}
=
\frac{1}{N_{\mathbf{k}}}
\left[
(F_{pq}^{\mathbf{k},\sigma})^*
n_q^{\mathbf{k},\sigma}
-
F_{qp}^{\mathbf{k},\sigma}
n_p^{\mathbf{k},\sigma}
\right].
\end{equation}
Since $\mathbf{R}^{\mathbf{k},\sigma}=\mathbf{X}^{\mathbf{k},\sigma}+i\mathbf{Y}^{\mathbf{k},\sigma}$, the gradients with respect to its independent real and imaginary components are
\begin{align}
\label{eq:dE_dX}
\left(\mathbf{g}_{X}^{\mathbf{k},\sigma}\right)_{pq} &=
\left.\frac{\partial E}{\partial X_{pq}^{\mathbf{k},\sigma}}\right|_{\mathbf{R}^{\mathbf{k},\sigma}=\mathbf{0}}
=
2\operatorname{Re}
\left(\left.
\frac{\partial E}{\partial R_{pq}^{\mathbf{k},\sigma}}\right|_{\mathbf{R}^{\mathbf{k},\sigma}=\mathbf{0}}
\right) \nonumber \\
&=
-\frac{2}{N_{\mathbf{k}}}
\operatorname{Re}(F_{pq}^{\mathbf{k},\sigma})
\left(
n_p^{\mathbf{k},\sigma}
-
n_q^{\mathbf{k},\sigma}
\right),
\end{align}
and
\begin{align}
\label{eq:dE_dY}
\left(\mathbf{g}_{Y}^{\mathbf{k},\sigma}\right)_{pq} &=
\left.
\frac{\partial E}{\partial Y_{pq}^{\mathbf{k},\sigma}}\right|_{\mathbf{R}^{\mathbf{k},\sigma}=\mathbf{0}}
=
-2\operatorname{Im}
\left(\left.
\frac{\partial E}{\partial R_{pq}^{\mathbf{k},\sigma}}\right|_{\mathbf{R}^{\mathbf{k},\sigma}=\mathbf{0}}
\right) \nonumber \\
&=
-\frac{2}{N_{\mathbf{k}}}
\operatorname{Im}(F_{pq}^{\mathbf{k},\sigma})
\left(
n_p^{\mathbf{k},\sigma}
-
n_q^{\mathbf{k},\sigma}
\right).
\end{align}
The resulting gradient matrices,
$\mathbf{g}_{X}^{\mathbf{k},\sigma}$ and
$\mathbf{g}_{Y}^{\mathbf{k},\sigma}$, with respect to
$\mathbf{X}^{\mathbf{k},\sigma}$ and
$\mathbf{Y}^{\mathbf{k},\sigma}$ are antisymmetric and
symmetric, respectively. For subsequent calculations,
their independent components are vectorized and
concatenated to form the gradient vector
\begin{equation}
\label{eq:dE_dXY}
\mathbf{g}_{R}^{\mathbf{k},\sigma}
=
\left(
\operatorname{vec}(\mathbf{g}_{X}^{\mathbf{k},\sigma}),
\operatorname{vec}(\mathbf{g}_{Y}^{\mathbf{k},\sigma})
\right),
\end{equation}
where $\operatorname{vec}$ denotes the vectorization
of the independent matrix components. All vectors
defined here and hereafter are taken to be row vectors.

The energy gradient with respect to the auxiliary occupation variable $\mathbf{x}^{\mathbf{k},\sigma}$ follows from the chain rule,
\begin{equation}
\label{eq:dE_dx}
\left(\mathbf{g}_{x}^{\mathbf{k},\sigma}\right)_{p} =\frac{\partial E}{\partial x_p^{\mathbf{k},\sigma}}
=
\sum_{\mathbf{k}_q}
\sum_{q}
\frac{\partial E}{\partial n_q^{\mathbf{k}_q,\sigma}}
\frac{\partial n_q^{\mathbf{k}_q,\sigma}}
{\partial x_p^{\mathbf{k},\sigma}},
\end{equation}
where
\begin{equation}
\label{eq:dE_dn}
\frac{\partial E}{\partial n_q^{\mathbf{k}_q,\sigma}}
=
\frac{1}{N_{\mathbf{k}}}
\left(
h_{qq}^{\mathbf{k}_q,\sigma}
+
J_{qq}^{\mathbf{k}_q,\sigma}
\right)
+
\frac{\partial E_{\mathrm{XC}}}
{\partial n_q^{\mathbf{k}_q,\sigma}},
\end{equation}
and
\begin{align}
\label{eq:dExc_dn}
\frac{\partial E_{\mathrm{XC}}}
{\partial n_q^{\mathbf{k}_q,\sigma}}
=-\frac{1}{N_{\mathbf{k}}^2}
\sum_{\mathbf{k}'}&
\sum_{j=1}
(q^{\mathbf{k}_q,\sigma}j^{\mathbf{k}',\sigma}
|
j^{\mathbf{k}',\sigma}q^{\mathbf{k}_q,\sigma})
\nonumber \\
&\times
\frac{\partial f(n_q^{\mathbf{k}_q,\sigma},
n_j^{\mathbf{k}',\sigma})}
{\partial n_q^{\mathbf{k}_q,\sigma}}.
\end{align}
Under the EBI parameterization, the derivative of an ON with respect to an auxiliary variable is
\begin{equation}
\label{eq:dn_dx}
\begin{split}
\frac{\partial n_q^{\mathbf{k}_q,\sigma}}
{\partial x_p^{\mathbf{k},\sigma}}
=&
\delta_{\mathbf{k}\mathbf{k}_q}
\delta_{pq}
s_x(x_q^{\mathbf{k}_q,\sigma},\mu^\sigma)
\\
&+
s_\mu(x_q^{\mathbf{k}_q,\sigma},\mu^\sigma)
\frac{\partial\mu^\sigma}
{\partial x_p^{\mathbf{k},\sigma}},
\end{split}
\end{equation}
where the implicit dependence of $\mu^\sigma$ on the auxiliary variables is determined by particle-number conservation:
\begin{equation}
\label{eq:du_dx}
\frac{\partial\mu^\sigma}
{\partial x_p^{\mathbf{k},\sigma}}
=
-\frac{
s_x(x_p^{\mathbf{k},\sigma},\mu^\sigma)
}{V^\sigma}.
\end{equation}
Here, $s_x(x,\mu)=\frac{\partial s(x,\mu)}{\partial x}$, $s_\mu(x,\mu)=\frac{\partial s(x,\mu)}{\partial\mu}$, and $V^\sigma=
\sum_{\mathbf{k}_i}
\sum_{i=1}
s_\mu(x_i^{\mathbf{k}_i,\sigma},\mu^\sigma)$.
These expressions provide the first-order energy derivatives required for the joint optimization of Bloch NOs and ONs.

The second-order energy derivatives with respect to the orbital-rotation parameters $\mathbf{X}^{\mathbf{k},\sigma}$ and $\mathbf{Y}^{\mathbf{k},\sigma}$ define the corresponding orbital Hessian blocks. For a given crystal momentum $\mathbf{k}$ and spin $\sigma$, these blocks are expressed as
\begin{align}
	\label{eq:d2E_dX2}
		(\mathbf{H}_{XX}^{\mathbf{k},\sigma})_{ab,pq}
&=
\left.
\frac{\partial^2 E}
{\partial X_{ab}^{\mathbf{k},\sigma}
 \partial X_{pq}^{\mathbf{k},\sigma}}
\right|_{\mathbf{R}^{\mathbf{k},\sigma}=\mathbf{0}} \nonumber \\
		= & \frac{1}{N_{\mathbf{k}}} \text{Re} \Big\{\delta_{bp} F_{aq}^{\mathbf{k},\sigma} (n_a^{\mathbf{k},\sigma} + n_q^{\mathbf{k},\sigma} - 2n_b^{\mathbf{k},\sigma}) \nonumber \\
		& - \delta_{bq} F_{ap}^{\mathbf{k},\sigma} (n_a^{\mathbf{k},\sigma} + n_p^{\mathbf{k},\sigma} - 2n_b^{\mathbf{k},\sigma}) \nonumber \\
		& - \delta_{ap} F_{bq}^{\mathbf{k},\sigma} (n_b^{\mathbf{k},\sigma} + n_q^{\mathbf{k},\sigma} - 2n_a^{\mathbf{k},\sigma}) \nonumber \\
		& + \delta_{aq} F_{bp}^{\mathbf{k},\sigma} (n_b^{\mathbf{k},\sigma} + n_p^{\mathbf{k},\sigma} - 2n_a^{\mathbf{k},\sigma}) \nonumber \\ 
		& + 2\frac{1}{N_{\mathbf{k}}} (t_{abpq}^{\mathbf{k},\sigma} + t_{abqp}^{\mathbf{k},\sigma} )\Delta n_{abpq}^{\mathbf{k},\sigma}\nonumber \\
		& - 2\frac{1}{N_{\mathbf{k}}} (t_{aqpb}^{\mathbf{k},\sigma} + t_{apqb}^{\mathbf{k},\sigma} ) \Delta f_{abpq}^{\mathbf{k},\sigma} \Big\},
\end{align}
\begin{align}
\label{eq:d2E_dY2}
(\mathbf{H}_{YY}^{\mathbf{k},\sigma})_{ab,pq}
&=
\left.
\frac{\partial^2 E}
{\partial Y_{ab}^{\mathbf{k},\sigma}
 \partial Y_{pq}^{\mathbf{k},\sigma}}
\right|_{\mathbf{R}^{\mathbf{k},\sigma}=\mathbf{0}}
\nonumber \\
		= &  \frac{1}{N_{\mathbf{k}}} \text{Re} \Big\{ -\delta_{bp} F_{aq}^{\mathbf{k},\sigma} (n_a^{\mathbf{k},\sigma} + n_q^{\mathbf{k},\sigma} - 2n_b^{\mathbf{k},\sigma}) \nonumber \\
		& - \delta_{bq} F_{ap}^{\mathbf{k},\sigma} (n_a^{\mathbf{k},\sigma} + n_p^{\mathbf{k},\sigma} - 2n_b^{\mathbf{k},\sigma}) \nonumber \\
		& - \delta_{ap} F_{bq}^{\mathbf{k},\sigma} (n_b^{\mathbf{k},\sigma} + n_q^{\mathbf{k},\sigma} - 2n_a^{\mathbf{k},\sigma}) \nonumber \\
		& - \delta_{aq} F_{bp}^{\mathbf{k},\sigma} (n_b^{\mathbf{k},\sigma} + n_p^{\mathbf{k},\sigma} - 2n_a^{\mathbf{k},\sigma}) \nonumber \\
		& - 2 \frac{1}{N_{\mathbf{k}}} (t_{abpq}^{\mathbf{k},\sigma} - t_{abqp}^{\mathbf{k},\sigma}) \Delta n_{abpq}^{\mathbf{k},\sigma} \nonumber \\
		& + 2 \frac{1}{N_{\mathbf{k}}} (t_{aqpb}^{\mathbf{k},\sigma} - t_{apqb}^{\mathbf{k},\sigma}) \Delta f_{abpq}^{\mathbf{k},\sigma} \Big\},
\end{align}
\begin{align}
\label{eq:d2E_dXdY}
(\mathbf{H}_{XY}^{\mathbf{k},\sigma})_{ab,pq}
&=
\left.
\frac{\partial^2 E}
{\partial X_{ab}^{\mathbf{k},\sigma}
 \partial Y_{pq}^{\mathbf{k},\sigma}}
\right|_{\mathbf{R}^{\mathbf{k},\sigma}=\mathbf{0}}
\nonumber \\
		= & \frac{1}{N_{\mathbf{k}}} \text{Im} \Big\{ 
		\delta_{bp} F_{aq}^{\mathbf{k},\sigma} (n_a^{\mathbf{k},\sigma} + n_q^{\mathbf{k},\sigma} - 2n_b^{\mathbf{k},\sigma}) \nonumber \\
		& + \delta_{bq} F_{ap}^{\mathbf{k},\sigma} (n_a^{\mathbf{k},\sigma} + n_p^{\mathbf{k},\sigma} - 2n_b^{\mathbf{k},\sigma}) \nonumber \\
		& - \delta_{ap} F_{bq}^{\mathbf{k},\sigma} (n_b^{\mathbf{k},\sigma} + n_q^{\mathbf{k},\sigma} - 2n_a^{\mathbf{k},\sigma}) \nonumber \\
		& - \delta_{aq} F_{bp}^{\mathbf{k},\sigma} (n_b^{\mathbf{k},\sigma} + n_p^{\mathbf{k},\sigma} - 2n_a^{\mathbf{k},\sigma}) \nonumber \\
		& + 2 \frac{1}{N_{\mathbf{k}}} (t_{abpq}^{\mathbf{k},\sigma} - t_{abqp}^{\mathbf{k},\sigma}) \Delta n_{abpq}^{\mathbf{k},\sigma} \nonumber \\
		& - 2 \frac{1}{N_{\mathbf{k}}} (t_{aqpb}^{\mathbf{k},\sigma} - t_{apqb}^{\mathbf{k},\sigma}) \Delta f_{abpq}^{\mathbf{k},\sigma} \Big\},
\end{align}
where $t_{abpq}^{\mathbf{k},\sigma} = (a^{\mathbf{k},\sigma} b^{\mathbf{k},\sigma} | p^{\mathbf{k},\sigma} q^{\mathbf{k},\sigma})$, $\Delta n_{abpq}^{\mathbf{k},\sigma} = (n_a^{\mathbf{k},\sigma} - n_b^{\mathbf{k},\sigma}) (n_p^{\mathbf{k},\sigma} - n_q^{\mathbf{k},\sigma})$, and $\Delta f_{abpq}^{\mathbf{k},\sigma}= f(n_p^{\mathbf{k},\sigma}, n_a^{\mathbf{k},\sigma}) - f(n_q^{\mathbf{k},\sigma}, n_a^{\mathbf{k},\sigma}) - f(n_p^{\mathbf{k},\sigma}, n_b^{\mathbf{k},\sigma}) + f(n_q^{\mathbf{k},\sigma}, n_b^{\mathbf{k},\sigma})$.

The second-order energy derivatives with respect to the auxiliary occupation variables $\{x_p^{\mathbf{k},\sigma}\}$ define the occupation Hessian for each spin channel $\sigma$,
\begin{align}
\label{eq:d2E_dx2}
(\mathbf{H}_{xx}^{\mathbf{k}\mathbf{k}', \sigma})_{pq}
&=
\frac{\partial^2 E}
{\partial x_p^{\mathbf{k},\sigma}
\partial x_q^{\mathbf{k}',\sigma}}
\nonumber \\
	&= \sum_{\mathbf{k}_i} \sum_{i} \frac{\partial E}{\partial n_{i}^{\mathbf{k}_i,\sigma}} \frac{\partial^2 n_{i}^{\mathbf{k}_i,\sigma}}{\partial x_p^{\mathbf{k},\sigma} \partial x_q^{\mathbf{k}',\sigma}} \nonumber \\
	& + \sum_{\mathbf{k}_i \mathbf{k}_j} \sum_{ij} \frac{\partial^2 E}{\partial n_{i}^{\mathbf{k}_i,\sigma} \partial n_{j}^{\mathbf{k}_j,\sigma}} \frac{\partial n_{i}^{\mathbf{k}_i,\sigma}}{\partial x_p^{\mathbf{k},\sigma}} \frac{\partial n_{j}^{\mathbf{k}_j,\sigma}}{\partial x_q^{\mathbf{k}',\sigma}}.
\end{align}
The second derivatives of the ONs are given by
\begin{align}
	\label{eq:d2n_dx2}
		\frac{\partial^2 n_i^{\mathbf{k}_i,\sigma}}{\partial x_p^{\mathbf{k},\sigma} \partial x_q^{\mathbf{k}',\sigma}} = & \delta_{\mathbf{k}_i \mathbf{k}} \delta_{\mathbf{k}_i \mathbf{k}'} \delta_{ip} \delta_{iq} s_{xx} (x_i^{\mathbf{k}_i,\sigma}, \mu^\sigma) \nonumber \\ 
		& + \delta_{\mathbf{k}_i \mathbf{k}} \delta_{ip} s_{\mu x} (x_i^{\mathbf{k}_i,\sigma}, \mu^\sigma) \frac{\partial \mu^\sigma}{\partial x_q^{\mathbf{k}',\sigma}}\nonumber \\
		& + \delta_{\mathbf{k}_i \mathbf{k}'} \delta_{iq} s_{\mu x} (x_i^{\mathbf{k}_i,\sigma}, \mu^\sigma) \frac{\partial \mu^\sigma}{\partial x_p^{\mathbf{k},\sigma}} \nonumber \\
		& + s_{\mu\mu} (x_i^{\mathbf{k}_i,\sigma}, \mu^\sigma) \frac{\partial \mu^\sigma}{\partial x_p^{\mathbf{k},\sigma}} \frac{\partial \mu^\sigma}{\partial x_q^{\mathbf{k}',\sigma}}\nonumber \\
		& + s_{\mu} (x_i^{\mathbf{k}_i,\sigma}, \mu^\sigma) \frac{\partial^2 \mu^\sigma}{\partial x_p^{\mathbf{k},\sigma} \partial x_q^{\mathbf{k}',\sigma}}.
\end{align}
The second derivative of the implicit function $\mu^\sigma$ follows from the particle-number conservation constraint,
\begin{align}
	\label{eq:d2mu_dx2}
	\frac{\partial^2 \mu^\sigma}{\partial x_p^{\mathbf{k},\sigma} \partial x_q^{\mathbf{k}',\sigma}} = & -\frac{1}{V^\sigma} \Big[ \delta_{\mathbf{k} \mathbf{k}'} \delta_{pq} s_{xx}(x_p^{\mathbf{k},\sigma}, \mu^\sigma) \nonumber \\
	& + W^\sigma \frac{\partial \mu^\sigma}{\partial x_p^{\mathbf{k},\sigma} } \frac{\partial \mu^\sigma}{\partial x_q^{\mathbf{k}',\sigma}}\nonumber \\
	& + s_{x\mu}(x_q^{\mathbf{k}',\sigma}, \mu^\sigma) \frac{\partial \mu^\sigma}{\partial x_p^{\mathbf{k},\sigma}}\nonumber \\
	& + s_{x\mu}(x_p^{\mathbf{k},\sigma}, \mu^\sigma) \frac{\partial \mu^\sigma}{\partial x_q^{\mathbf{k}',\sigma}} \Big], 
\end{align}
where $s_{xx}$, $s_{x\mu}$, and $s_{\mu\mu}$ denote the corresponding second-order partial derivatives of the mapping function $s(x,\mu)$, and
$W^\sigma
=
\sum_{\mathbf{k}_i}\sum_{i}
s_{\mu\mu}(x_i^{\mathbf{k}_i,\sigma},\mu^\sigma)$.
Here, $V^\sigma$ is defined as in Eq.~\eqref{eq:du_dx}. These expressions account for the implicit dependence of $\mu^\sigma$ on the auxiliary occupation variables, thereby ensuring particle-number conservation throughout the optimization.

The mixed orbital-occupation Hessian blocks are obtained by applying the chain rule to the EBI parameterization,
\begin{align}
	\label{eq:d2E_dXdx}
	(&\mathbf{H}_{Xx}^{\mathbf{k}\mathbf{k}', \sigma})_{pq,b}
    =\frac{\partial^2 E}{\partial X_{pq}^{\mathbf{k},\sigma} \partial x_b^{\mathbf{k}',\sigma}} \Big|_{\mathbf{R}^{\mathbf{k},\sigma}=\mathbf{0}} \nonumber \\
    &= \sum_{\mathbf{k}_a} \sum_a \left(\frac{\partial^2 E}{\partial X_{pq}^{\mathbf{k},\sigma} \partial n_a^{\mathbf{k}_a,\sigma}}\Big|_{\mathbf{R}^{\mathbf{k},\sigma}=\mathbf{0}}\right) \frac{\partial n_a^{\mathbf{k}_a,\sigma}}{\partial x_b^{\mathbf{k}',\sigma}} ,
\end{align}
and
\begin{align}
	\label{eq:d2E_dYdx}
	(&\mathbf{H}_{Yx}^{\mathbf{k}\mathbf{k}', \sigma})_{pq,b}
    =\frac{\partial^2 E}{\partial Y_{pq}^{\mathbf{k},\sigma} \partial x_b^{\mathbf{k}',\sigma}} \Big|_{\mathbf{R}^{\mathbf{k},\sigma}=\mathbf{0}} \nonumber \\
    &= \sum_{\mathbf{k}_a} \sum_a \left(\frac{\partial^2 E}{\partial Y_{pq}^{\mathbf{k},\sigma} \partial n_a^{\mathbf{k}_a,\sigma}}\Big|_{\mathbf{R}^{\mathbf{k},\sigma}=\mathbf{0}}\right) \frac{\partial n_a^{\mathbf{k}_a,\sigma}}{\partial x_b^{\mathbf{k}',\sigma}} ,
\end{align}
Here
\begin{align}
	\label{eq:d2E_dXdn}
	&\frac{\partial^2 E}{\partial X_{pq}^{\mathbf{k},\sigma} \partial n_a^{\mathbf{k}_a,\sigma}}\Big|_{\mathbf{R}^{\mathbf{k},\sigma}=\mathbf{0}} =  \delta_{\mathbf{k} \mathbf{k}_a}  \frac{2}{N_{\mathbf{k}}}  \text{Re}(F_{pq}^{\mathbf{k},\sigma})(\delta_{aq} - \delta_{ap}) \nonumber \\ 
		& + \frac{2}{N_{\mathbf{k}}^2} \text{Re}(p^{\mathbf{k}} q^{\mathbf{k}}|a^{\mathbf{k}_a} a^{\mathbf{k}_a})  (n_q^{\mathbf{k},\sigma} - n_p^{\mathbf{k},\sigma}) \nonumber \\
		 &-\frac{2}{N_{\mathbf{k}}^2} \text{Re}(p^{\mathbf{k}} a^{\mathbf{k}_a}|a^{\mathbf{k}_a}q^{\mathbf{k}}) \nonumber \\
		&\ \ \ \ \ \ \ \ \ \times \left[  \frac{\partial f(n_q^{\mathbf{k},\sigma}, n_a^{\mathbf{k}_a,\sigma})}{\partial n_a^{\mathbf{k}_a,\sigma}} - \frac{\partial f(n_p^{\mathbf{k},\sigma}, n_a^{\mathbf{k}_a,\sigma})}{\partial n_a^{\mathbf{k}_a,\sigma}} \right],
\end{align} 
and
\begin{align}
	\label{eq:d2E_dYdn}
		&\frac{\partial^2 E}{\partial Y_{pq}^{\mathbf{k}, \sigma} \partial n_{a}^{\mathbf{k}_a, \sigma}} \Big|_{\mathbf{R}^{\mathbf{k},\sigma}=\mathbf{0}} =  \delta_{\mathbf{k} \mathbf{k}_a}  2 \frac{1}{N_{\mathbf{k}}} \text{Im}(F_{pq}^{\mathbf{k}, \sigma}) (\delta_{aq} + \delta_{ap}) \nonumber \\
		 & + \frac{2}{N_{\mathbf{k}}^2} \text{Im}(p^{\mathbf{k}} q^{\mathbf{k}}|a^{\mathbf{k}_a} a^{\mathbf{k}_a})  (n_q^{\mathbf{k}, \sigma} + n_p^{\mathbf{k}, \sigma}) \nonumber \\
		&- \frac{2}{N_{\mathbf{k}}^2} \text{Im}(p^{\mathbf{k}} a^{\mathbf{k}_a} |a^{\mathbf{k}_a} q^{\mathbf{k}} )\nonumber \\
		&\ \ \ \ \ \ \ \ \ \times \left[ \frac{\partial f(n_q^{\mathbf{k}, \sigma}, n_a^{\mathbf{k}_a, \sigma})}{\partial n_a^{\mathbf{k}_a, \sigma}} + \frac{\partial f(n_p^{\mathbf{k}, \sigma}, n_a^{\mathbf{k}_a, \sigma})}{\partial n_a^{\mathbf{k}_a, \sigma}} \right] .
\end{align}
In addition to its action on the ONs defined in Eq.~\eqref{eq:Kii_ni}, the action of the generalized exchange operator \(K_{pq}^{\mathbf{k},\sigma}\) on the Kronecker delta is further defined as
\begin{equation}
\label{eq:Kpq_delta_aq}
\begin{split}
K_{pq}^{\mathbf{k},\sigma}\delta_{aq}
=&\frac{1}{N_{\mathbf{k}}}
\sum_{\mathbf{k}'}\sum_j
(p^{\mathbf{k},\sigma}j^{\mathbf{k}',\sigma}
\mid j^{\mathbf{k}',\sigma}q^{\mathbf{k},\sigma})
\\
&\times\delta_{aq}
\frac{\partial f(n_q^{\mathbf{k},\sigma},
n_j^{\mathbf{k}',\sigma})}
{\partial n_q^{\mathbf{k},\sigma}}.
\end{split}
\end{equation}

Combining the Hessian expressions in Eqs.~\eqref{eq:d2E_dX2}-\eqref{eq:d2E_dXdY}, \eqref{eq:d2E_dx2}, \eqref{eq:d2E_dXdx}, and \eqref{eq:d2E_dYdx}, the Hessian with respect to the orbital-rotation and occupation variables can be written in block form as
\begin{equation}
\label{eq:hess}
\mathbf{H}^{\mathbf{k},\sigma}
=
\begin{bmatrix}
\mathbf{H}_{XX}^{\mathbf{k},\sigma}
&
\mathbf{H}_{XY}^{\mathbf{k},\sigma}
&
\mathbf{H}_{Xx}^{\mathbf{k},\sigma}
\\
\mathbf{H}_{YX}^{\mathbf{k},\sigma}
&
\mathbf{H}_{YY}^{\mathbf{k},\sigma}
&
\mathbf{H}_{Yx}^{\mathbf{k},\sigma}
\\
\mathbf{H}_{xX}^{\mathbf{k},\sigma}
&
\mathbf{H}_{xY}^{\mathbf{k},\sigma}
&
\mathbf{H}_{xx}^{\mathbf{k},\sigma}
\end{bmatrix}.
\end{equation}
Since only the diagonal Hessian elements are used to construct the preconditioner in the present work, we do not explicitly present the full Hessian, including couplings between different $\mathbf{k}$ points, which can be derived analogously. Nevertheless, these Hessian expressions provide a general foundation for developing efficient optimization algorithms. In the following subsection, we introduce a preconditioned algorithm for the coupled optimization of Bloch NOs and ONs, using the energy gradients in Eqs.~\eqref{eq:dE_dX}-\eqref{eq:dE_dx} and the Hessian information in Eq.~\eqref{eq:hess}. The preconditioner is designed to capture the essential curvature information while avoiding the computationally expensive evaluation of complicated Hessian terms.

\subsection{Coupled Optimization of Bloch Natural Orbitals and Occupations}
The preconditioned nonlinear conjugate gradient (PCG) method, combined with a line search, is employed for the coupled optimization of Bloch NOs and ONs \cite{nocedal2006numerical}. At iteration $t$, the search direction is updated as
\begin{equation}
\label{eq:pcg_pk}
\mathbf{p}_t
=
-\mathbf{g}_t\mathbf{P}_t^{-1}
+
\beta_t\mathbf{p}_{t-1},
\end{equation}
where $\mathbf{P}_t$ is the preconditioner, $\mathbf{g}_t$ is the energy gradient, and $\beta_t$ is the conjugate-gradient coefficient. The latter is determined using the preconditioned Polak-Ribière formula,
\begin{equation}
\label{eq:PR_beta}
\beta_t^{\mathrm{PR}}
=
\frac{
\left\langle
\mathbf{g}_t-\mathbf{g}_{t-1},
\mathbf{g}_t\mathbf{P}_t^{-1}
\right\rangle
}{
\left\langle
\mathbf{g}_{t-1},
\mathbf{g}_{t-1}\mathbf{P}_{t-1}^{-1}
\right\rangle
},
\end{equation}
with $\beta_t=\max(\beta_t^{\mathrm{PR}},0)$. When $\beta_t^{\mathrm{PR}}\leq 0$, the search direction is reset to the negative preconditioned gradient.

To account for the distinct physical roles of Bloch NOs and ONs, separate search directions, step sizes, and conjugate-gradient coefficients are employed for the orbital-rotation parameters $(\mathbf{X}^{\mathbf{k},\sigma},\mathbf{Y}^{\mathbf{k},\sigma})$ and the auxiliary occupation variables $\mathbf{x}^{\mathbf{k},\sigma}$. For periodic systems with multiple $\mathbf{k}$ points, we consider two strategies for evaluating the conjugate-gradient coefficients. In the first strategy, termed $k$-resolved PCG, the coefficients are determined independently at each $\mathbf{k}$ point but shared between the two spin channels. The corresponding orbital and occupation gradient vectors, $\mathbf{g}_{R}^{\mathbf{k},\sigma}$ and $\mathbf{g}_{x}^{\mathbf{k},\sigma}$, are obtained from Eqs.~\eqref{eq:dE_dXY} and \eqref{eq:dE_dx}, respectively. Combining the gradients from both spin channels as $\mathbf{g}_{R}^{\mathbf{k}}=\left(\mathbf{g}_{R}^{\mathbf{k},\alpha},\mathbf{g}_{R}^{\mathbf{k},\beta}\right)$ and $\mathbf{g}_{x}^{\mathbf{k}}=\left(\mathbf{g}_{x}^{\mathbf{k},\alpha},\mathbf{g}_{x}^{\mathbf{k},\beta}\right)$, and applying Eq.~\eqref{eq:PR_beta} separately to these two vectors yields the $k$-dependent coefficients $\beta_R^{\mathbf{k}}$ and $\beta_x^{\mathbf{k}}$. In the second strategy, termed $k$-global PCG, common conjugate-gradient coefficients are employed across all $\mathbf{k}$ points and spin channels for the orbital and occupation variables, respectively. The corresponding gradients are assembled into the global vectors
\begin{align}
\label{eq:grad_k_global}
\mathbf{g}_{R}
&=
\left(
\mathbf{g}_{R}^{\mathbf{k}_1},
\mathbf{g}_{R}^{\mathbf{k}_2},
\ldots,
\mathbf{g}_{R}^{\mathbf{k}_{N_{\mathbf{k}}}}
\right)^T,\\
\mathbf{g}_{x}
&=
\left(
\mathbf{g}_{x}^{\mathbf{k}_1},
\mathbf{g}_{x}^{\mathbf{k}_2},
\ldots,
\mathbf{g}_{x}^{\mathbf{k}_{N_{\mathbf{k}}}}
\right).
\end{align}
The coefficients $\beta_R$ and $\beta_x$ are then evaluated using Eq.~\eqref{eq:PR_beta} with the respective global gradient vectors. Thus, the two strategies differ in whether the conjugate-gradient coefficients are determined independently at each $\mathbf{k}$ point or collectively across all $\mathbf{k}$ points, while both employ spin-shared coefficients. Their numerical performance and robustness are evaluated in the following section.

Although the exact Hessian of Eq. \eqref{eq:hess} captures the local curvature of the energy landscape, its evaluation and storage become prohibitively expensive for large systems. An effective preconditioner is therefore essential for accelerating convergence while maintaining computational efficiency. Following the preconditioning strategy of Ref.~\cite{yao2024COopt}, we construct a diagonal preconditioner. For orbital optimization, its diagonal elements are approximated from those of the orbital Hessian by neglecting computationally expensive contributions:
\begin{align}
\label{eq:preconditioner}
(\mathbf{P}_{X}^{\mathbf{k},\sigma})_{pq,pq}
&=
(\mathbf{P}_{Y}^{\mathbf{k},\sigma})_{pq,pq}
\nonumber\\
&=
\frac{2}{N_{\mathbf{k}}}
\operatorname{Re}
\left(F_{pp}^{\mathbf{k},\sigma}
-F_{qq}^{\mathbf{k},\sigma}\right)
\left(n_q^{\mathbf{k},\sigma}
-n_p^{\mathbf{k},\sigma}\right).
\end{align}
The orbital preconditioner is then assembled as
\begin{equation}
\label{eq:P_R}
\mathbf{P}_{R}^{\mathbf{k},\sigma}
=
\begin{bmatrix}
\mathbf{P}_{X}^{\mathbf{k},\sigma} & \mathbf{0}\\
\mathbf{0} & \mathbf{P}_{Y}^{\mathbf{k},\sigma}
\end{bmatrix}.
\end{equation}
To ensure positive definiteness, the diagonal elements are shifted by the magnitude of their minimum value if it is negative. The resulting elements are then bounded below by $10^{-5}$ to prevent numerical instabilities.

For the auxiliary occupation variables, the diagonal elements of the Hessian are given by
\begin{align}
\label{eq:d2e_dx2_diag}
\frac{\partial^2 E}
{\partial x_p^{\mathbf{k},\sigma} \partial x_p^{\mathbf{k},\sigma}}
={}&
\sum_{\mathbf{k}_a,\mathbf{k}_b}
\sum_{a,b}
\frac{\partial^2 E}
{\partial n_a^{\mathbf{k}_a,\sigma}
\partial n_b^{\mathbf{k}_b,\sigma}}
\frac{\partial n_a^{\mathbf{k}_a,\sigma}}
{\partial x_p^{\mathbf{k},\sigma}}
\frac{\partial n_b^{\mathbf{k}_b,\sigma}}
{\partial x_p^{\mathbf{k},\sigma}}
\nonumber\\
&+
\sum_{\mathbf{k}_a}\sum_a
\frac{\partial E}
{\partial n_a^{\mathbf{k}_a,\sigma}}
\frac{\partial^2 n_a^{\mathbf{k}_a,\sigma}}
{\partial x_p^{\mathbf{k},\sigma}\partial x_p^{\mathbf{k},\sigma}}.
\end{align}
Neglecting the first term yields a computationally inexpensive diagonal approximation,
\begin{equation}
\label{eq:occ_prec_1}
(\mathbf{P'}_x^{\mathbf{k},\sigma})_{pp}
=
\sum_{\mathbf{k}_a}\sum_a
\frac{\partial E}
{\partial n_a^{\mathbf{k}_a,\sigma}}
\frac{\partial^2 n_a^{\mathbf{k}_a,\sigma}}
{\partial x_p^{\mathbf{k},\sigma}\partial x_p^{\mathbf{k},\sigma}}.
\end{equation}
To further improve convergence, this approximation is combined with an inverse-Hessian estimate obtained from the Broyden-Fletcher-Goldfarb-Shanno (BFGS) method \cite{yao2024COopt}:
\begin{equation}
\label{eq:occ_prec_tot}
\left(\mathbf{P}_x^{\mathbf{k},\sigma}\right)^{-1}
=
d\left(\mathbf{P}_x^{\mathrm{BFGS},\mathbf{k},\sigma}\right)^{-1}
+
(1-d)\left(\mathbf{P'}_x^{\mathbf{k},\sigma}\right)^{-1},
\end{equation}
where the inverse BFGS preconditioner\cite{bfgs1970b,bfgs1970f,bfgs1970g,bfgs1970s} at iteration $t$ is constructed from the diagonal elements of the approximate inverse Hessian,
\begin{equation}
\label{eq:occ_prec_bfgs_inv}
\left(\mathbf{P}_x^{\mathrm{BFGS},\mathbf{k},\sigma}\right)^{-1}
=
\operatorname{diag}
\left(\mathbf{B}_t^{\mathbf{k},\sigma,-1}\right).
\end{equation}
The approximate inverse Hessian is updated according to
\begin{equation}
\label{eq:bfgs_update}
\begin{aligned}
\mathbf{B}_t^{\mathbf{k},\sigma,-1}
={}&
\left(
\mathbf{I}-
\frac{(\mathbf{s}_t^{\mathbf{k},\sigma})^T
\mathbf{y}_t^{\mathbf{k},\sigma}}
{\mathbf{y}_t^{\mathbf{k},\sigma}
(\mathbf{s}_t^{\mathbf{k},\sigma})^T}
\right)
\mathbf{B}_{t-1}^{\mathbf{k},\sigma,-1}
\\
&\times
\left(
\mathbf{I}-
\frac{(\mathbf{y}_t^{\mathbf{k},\sigma})^T
\mathbf{s}_t^{\mathbf{k},\sigma}}
{\mathbf{y}_t^{\mathbf{k},\sigma}
(\mathbf{s}_t^{\mathbf{k},\sigma})^T}
\right)
\\
&+
\frac{(\mathbf{s}_t^{\mathbf{k},\sigma})^T
\mathbf{s}_t^{\mathbf{k},\sigma}}
{\mathbf{y}_t^{\mathbf{k},\sigma}
(\mathbf{s}_t^{\mathbf{k},\sigma})^T},
\end{aligned}
\end{equation}
with
\begin{align}
\mathbf{s}_t^{\mathbf{k},\sigma}
&=
\mathbf{x}_t^{\mathbf{k},\sigma}
-
\mathbf{x}_{t-1}^{\mathbf{k},\sigma},\\
\mathbf{y}_t^{\mathbf{k},\sigma}
&=
\mathbf{g}_{x,t}^{\mathbf{k},\sigma}
-
\mathbf{g}_{x,t-1}^{\mathbf{k},\sigma}.
\end{align}
Here, $\mathbf{x}_t^{\mathbf{k},\sigma}$ and $\mathbf{g}_{x,t}^{\mathbf{k},\sigma}$ denote the auxiliary occupation variables and their corresponding energy gradients at iteration $t$ for a given $\mathbf{k}$ point and spin $\sigma$. The mixing parameter $d$ controls the relative contributions of the two inverse-Hessian approximations. We use $d=0.6$, which provides a favorable balance between convergence speed and numerical robustness.

We employ a quadratic line-search procedure to determine the optimization step sizes. Specifically, the energy along a given search direction is approximated by
\begin{equation}
\label{eq:e_alpha}
E(\alpha) \approx a\alpha^2 + b\alpha + c.
\end{equation}
The minimum of the convex quadratic approximation yields the optimal step size,
\begin{equation}
\label{eq:optimal_alpha}
\alpha = \frac{E'(0)\tilde{\alpha}}{E'(0)-E'(\tilde{\alpha})},
\end{equation}
where $\tilde{\alpha}$ denotes the trial step size. As discussed above, the orbital-rotation parameters $\mathbf{R}^{\mathbf{k},\sigma}$ and auxiliary occupation variables $\mathbf{x}^{\mathbf{k},\sigma}$ represent distinct variational degrees of freedom, and separate step sizes are therefore employed for their optimization. The initial trial step size is set to 1.0. If the quadratic approximation is not convex, the trial step size is reduced by a factor of 0.05. This procedure is repeated until a convex quadratic approximation is obtained or the sufficient-decrease condition is satisfied. For periodic systems with multiple $\mathbf{k}$ points, we consider two line-search strategies analogous to those used for evaluating the conjugate-gradient coefficients. In the first strategy, termed $k$-resolved line search, separate step sizes $\alpha_R^{\mathbf{k}}$ and $\alpha_x^{\mathbf{k}}$ are determined at each $\mathbf{k}$ point, with both spin channels sharing the same step size within each variable group. In the second strategy, termed $k$-global line search, common step sizes $\alpha_R$ and $\alpha_x$ are employed across all $\mathbf{k}$ points and spin channels, determined from the total energy along the corresponding combined search directions. Thus, both strategies retain separate step sizes for orbital and occupation optimization while differing in whether the step sizes are determined independently at each $\mathbf{k}$ point or collectively across all $\mathbf{k}$ points.

To efficiently determine the step sizes for both $k$-resolved and $k$-global line searches, we extend the approach of Ref.~\cite{yao2024COopt} to multiple $\mathbf{k}$ points. Rather than performing separate trial evaluations for each step size, we simultaneously displace the orbital and occupation variables along their respective search directions at all $\mathbf{k}$ points using trial step sizes $\tilde{\alpha}^{(\mathbf{k})}$. A single additional total-energy and gradient evaluation at the trial point provides the energy derivatives with respect to all step-size variables. Together with the corresponding derivatives at the initial point, these quantities allow Eq.~\eqref{eq:optimal_alpha} to be applied to each step-size variable, yielding all the required step sizes simultaneously. This procedure can be understood by expanding the multidimensional total energy $E(\{\alpha_R^{(\mathbf{k})},\alpha_x^{(\mathbf{k})}\})$ to second order in all step sizes and neglecting explicit mixed quadratic terms. The resulting separable quadratic model allows each step size to be determined using Eq.~\eqref{eq:optimal_alpha}, with its effective curvature estimated from the gradient change at the simultaneous trial point. This approach substantially reduces the number of trial evaluations while accounting for the coupling among orbital and occupation variables across all $\mathbf{k}$ points through the total-energy and gradient evaluations. Its effectiveness is demonstrated in the following numerical tests.

In practice, the search directions $\mathbf{p}_R^{\mathbf{k},\sigma}$ and $\mathbf{p}_x^{\mathbf{k},\sigma}$ obtained from Eq.~\eqref{eq:pcg_pk} are not necessarily descent directions, as their inner products with the corresponding gradients may become nonnegative. Consequently, the quadratic line-search procedure described above may fail to yield an acceptable step size. To address this issue, a restart procedure is employed whenever $\langle \mathbf{p}_R^{\mathbf{k},\sigma},\mathbf{g}_R^{\mathbf{k},\sigma}\rangle \geq 0$ or $\langle \mathbf{p}_x^{\mathbf{k},\sigma},\mathbf{g}_x^{\mathbf{k},\sigma}\rangle \geq 0$. In either case, the corresponding search direction is reset to the negative preconditioned gradient, thereby restoring the descent condition. The complete CO procedure is summarized in Algorithm~\ref{alg:pcg}.

\begin{algorithm}
\caption{\raggedright Coupled optimization of Bloch
NOs and ONs with multiple $\mathbf{k}$ points.}
\label{alg:pcg}
\SetInd{0.5em}{1em}
\SetAlgoNoLine

Initialize $\{\mathbf C_0^{\mathbf k,\sigma},
\mathbf x_0^{\mathbf k,\sigma}\}$ and set $t=0$\;
Set $\beta_{R,0}^{(\mathbf k)}=\beta_{x,0}^{(\mathbf k)}=0$\;
Select $k$-resolved or $k$-global strategies
for PCG and line search\;

Compute $E_0$ and gradients
$\{\mathbf g_{R,0}^{\mathbf k,\sigma},
\mathbf g_{x,0}^{\mathbf k,\sigma}\}$\;

\While{\rm not converged}{

\ForEach{$(\mathbf k,\sigma)$}{
Construct preconditioners
$\mathbf P_{R,t}^{\mathbf k,\sigma}$
and $\mathbf P_{x,t}^{\mathbf k,\sigma}$\;

$\mathbf z_{R,t}^{\mathbf k,\sigma}
=-\mathbf g_{R,t}^{\mathbf k,\sigma}(\mathbf P_{R,t}^{\mathbf k,\sigma})^{-1}
$\;

$\mathbf z_{x,t}^{\mathbf k,\sigma}
=-\mathbf g_{x,t}^{\mathbf k,\sigma}(\mathbf P_{x,t}^{\mathbf k,\sigma})^{-1}
$\;
}

Compute $\beta_{R,t}^{(\mathbf k)}$ and $\beta_{x,t}^{(\mathbf k)}$
using Eq.~\eqref{eq:PR_beta} with
$k$-resolved or $k$-global gradients\;

\ForEach{$(\mathbf k,\sigma)$}{

$\mathbf p_{R,t}^{\mathbf k,\sigma}
=\mathbf z_{R,t}^{\mathbf k,\sigma}
+\beta_{R,t}^{(\mathbf k)}
\mathbf p_{R,t-1}^{\mathbf k,\sigma}$\;

$\mathbf p_{x,t}^{\mathbf k,\sigma}
=\mathbf z_{x,t}^{\mathbf k,\sigma}
+\beta_{x,t}^{(\mathbf k)}
\mathbf p_{x,t-1}^{\mathbf k,\sigma}$\;

\If{$\langle\mathbf p_{R,t}^{\mathbf k,\sigma},
\mathbf g_{R,t}^{\mathbf k,\sigma}\rangle\geq0$}{
$\mathbf p_{R,t}^{\mathbf k,\sigma}
=\mathbf z_{R,t}^{\mathbf k,\sigma}$\;
}

\If{$\langle\mathbf p_{x,t}^{\mathbf k,\sigma},
\mathbf g_{x,t}^{\mathbf k,\sigma}\rangle\geq0$}{
$\mathbf p_{x,t}^{\mathbf k,\sigma}
=\mathbf z_{x,t}^{\mathbf k,\sigma}$\;
}
}

Determine $\alpha_{R,t}^{(\mathbf k)}$
and $\alpha_{x,t}^{(\mathbf k)}$
using the quadratic line search
[Eq.~\eqref{eq:optimal_alpha}]
with the selected $k$-resolved or
$k$-global strategy\;

\ForEach{$(\mathbf k,\sigma)$}{

Construct the skew-Hermitian matrix
$\mathbf R_t^{\mathbf k,\sigma}$
from the search direction
$\mathbf p_{R,t}^{\mathbf k,\sigma}$\;

$\mathbf C_{t+1}^{\mathbf k,\sigma}
=\mathbf C_t^{\mathbf k,\sigma}
\exp\left(
\alpha_{R,t}^{(\mathbf k)}
\mathbf R_t^{\mathbf k,\sigma}
\right)$\;

$\mathbf x_{t+1}^{\mathbf k,\sigma}
=\mathbf x_t^{\mathbf k,\sigma}
+\alpha_{x,t}^{(\mathbf k)}
(\mathbf p_{x,t}^{\mathbf k,\sigma})^T$\;

}

Update ONs using the EBI parameterization\;
Update BFGS inverse-Hessian approximations
for occupation preconditioning\;

Compute $E_{t+1}$ and gradients
$\{\mathbf g_{R,t+1}^{\mathbf k,\sigma},
\mathbf g_{x,t+1}^{\mathbf k,\sigma}\}$\;

$t\leftarrow t+1$\;

}
\end{algorithm}

\section{\label{sec:level3}Computational details}
Except where otherwise noted, all calculations presented below were performed using a local software package, which used the Libcint \cite{sun2015libcint} library for Gaussian integral evaluation and the LibXC \cite{2018libxc} library for the KS-DFT XC functional calculations. Unless specified otherwise, the GTH-PBE pseudopotential \cite{1996gthpseudo,1998gthpseudo} and GTH-SZV-MOLOPT-SR basis set \cite{2007gthbasis} are used. The range-separated Gaussian density fitting \cite{ye2021rsgdf} in conjunction with the AutoAux \cite{2017autoaux} auxiliary basis set is utilized to calculate two-center and three-center integrals (see Supplemental Material \cite{SupplementalMaterial}, Sections SII and SIII). The integrable Coulomb singularity in the exchange integrals is treated using a finite-size correction based on the electrostatic energy difference between isolated and periodically repeated probe charges \cite{paier2005perdew}. Convergence is considered achieved when the energy change between successive cycles drops below $10^{-8}$ a.u. and the gradients of NOs and ONs are smaller than $10^{-4}$ a.u. An exception is made for reaction barrier calculations, for which the gradient convergence threshold is set to $5 \times 10^{-4}$ a.u. The calculations were carried out with the same initial setup described below. The superposition of atomic densities (SAD) method \cite{sad1982,sad2006} is used to generate a density matrix. This density matrix is then used to construct the Fock matrix, whose diagonalization yield the initial orbital coefficients and orbital energies. The initial guesses for the ONs are assigned as follows: $\{x_p^{\mathbf{k},\sigma}\}$ is set to 2 for orbitals below the Fermi energy level, while $\{x_p^{\mathbf{k},\sigma}\}$ for the remaining unoccupied orbitals is set to $-2$. The corresponding ONs $\{n_p^{\mathbf{k},\sigma}\}$ which satisfy the ensemble $N$-representability constraints are subsequently obtained using the EBI method. The nuclear repulsion energy is evaluated via Ewald summation method \cite{ewald1987allen, martin2020electronic, mcclain2017gaussian}. For the calculations of 2D materials, the Wigner-Seitz truncation \cite{2Dtruncation2013} scheme has been used to handle the Coulomb interaction in the present study. 

In the deCO scheme, the initial NOs and ONs may be far from their optimal values. Fully optimizing the NOs (or ONs) while keeping the ONs (or NOs) fixed can therefore result in an inefficient optimization path toward the coupled minimum. Moreover, even after one variable has been fully converged, subsequent updates of the other variable generally shift the location of the optimum, requiring repeated re-optimization and thus substantially increasing the computational cost. To alleviate this inefficiency, a staged iteration protocol is adopted throughout this work. Specifically, the maximum numbers of iterations for both NO and ON optimization are initially set to 2. After each complete optimization cycle, these limits are increased by 2 until they reach 100 and 200 for NO and ON optimization, respectively, allowing progressively more extensive optimization as the calculation approaches self-consistency.

\section{\label{sec:level4}Results and discussion}

\subsection{Validation of Periodic RDMFT at the \(\Gamma\) Point}

To evaluate the performance of the CO algorithm and the effects of short-range screening in the periodic implementation, we performed comparative molecular and $\Gamma$-point periodic calculations for 10 representative molecules: $\mathrm{H}_2$, $\mathrm{H}_2\mathrm{O}$, $\mathrm{O}_2$, $\mathrm{CO}$, $\mathrm{CO}_2$, $\mathrm{CH}_4$, $\mathrm{CH}_2\mathrm{O}$, $\mathrm{C}_2\mathrm{H}_4$, $\mathrm{NH}_3$, and $\mathrm{N}_2$. All molecular geometries were taken from experimental data in the Computational Chemistry Comparison and Benchmark Database (CCCBDB) \cite{cccbdb}, and both implementations employed the cc-pVTZ basis set \cite{ccpvtz1992,ccpvtz1993}. For the periodic calculations, each molecule was placed at the center of a cubic cell with a lattice constant of 20 $\mathrm{\AA}$ to minimize interactions between periodic images, with Brillouin-zone sampling restricted to the $\Gamma$ point. The workflow of the PCG algorithm with CO and deCO schemes is illustrated in Figure~S1 \cite{SupplementalMaterial}.

\begin{figure}
\includegraphics[width=0.45\textwidth]{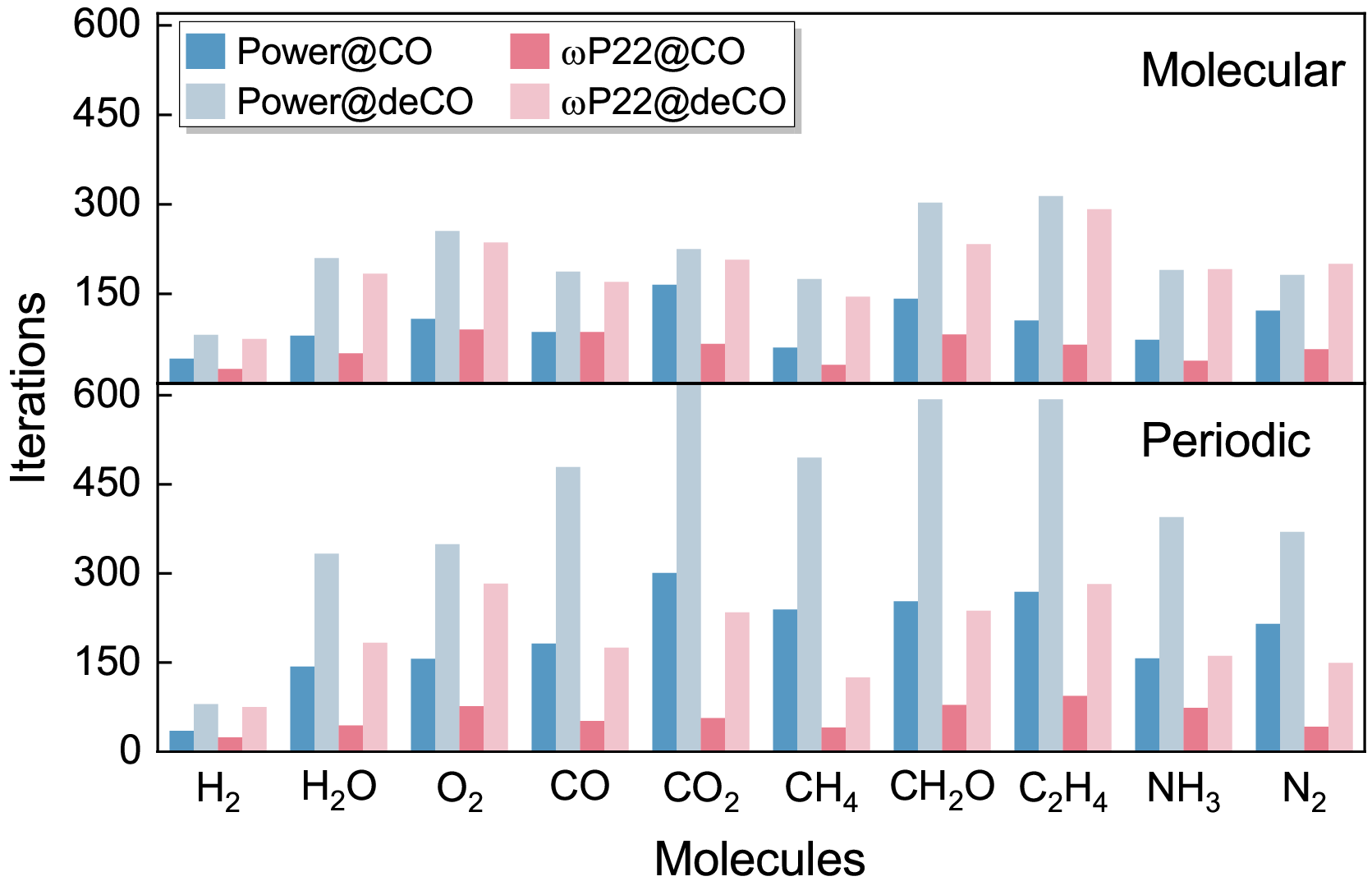}
\caption{\label{fig:20_mol}
Comparison of NO and ON optimization iteration counts between the molecular and $\Gamma$-point periodic implementations for 10 representative molecules using the Power(0.6) and $\omega$P22 functionals with the CO and deCO schemes.}
\end{figure}

Figure~\ref{fig:20_mol} compares the iteration counts of the molecular and periodic implementations using the Power \cite{power2008} and $\omega$P22 \cite{ai2023wp22} functionals with both CO and deCO schemes. To isolate the effect of short-range screening, the same Power exponent ($m=0.6$) was adopted for both functionals, with the corresponding unscreened functional denoted as Power(0.6). The periodic implementation exhibits iteration counts comparable to those of the molecular implementation, demonstrating consistent convergence behavior in the $\Gamma$-point limit. Moreover, CO substantially reduces the iteration counts relative to deCO in most cases, regardless of the functional employed. Following the recommendation of Gross and co-workers \cite{power2008} for periodic systems, we also examined the Power functional with $m=0.7$, denoted as Power(0.7). The corresponding results, together with those for Power(0.6) and $\omega$P22, are summarized in Table~S1 \cite{SupplementalMaterial}.

Notably, within the CO framework, the screened $\omega$P22 functional consistently exhibits faster convergence than the unscreened Power functional. The origin of this improvement is further investigated in the subsequent numerical tests.

\subsection{Coupled Optimization and \(\mathbf{k}\)-Point Convergence}

Unlike molecular calculations, calculations of periodic systems generally require sampling multiple $\mathbf{k}$ points in the Brillouin zone, introducing additional complexity into the optimization of NOs and ONs. Within the CO framework, we consider four optimization schemes that differ in how the conjugate-gradient coefficients $\beta$ and line-search step sizes $\alpha$ are assigned across $\mathbf{k}$ points: (i) Uniform, employing $k$-global PCG and $k$-global line searches for both NOs and ONs; (ii) Distinct, employing $k$-resolved PCG and $k$-resolved line searches for both NOs and ONs; (iii) Partially Distinct, combining $k$-resolved PCG with $k$-global line searches for both NOs and ONs; and (iv) Hybrid, employing $k$-resolved PCG and $k$-resolved line searches for NOs, but $k$-global PCG and $k$-global line searches for ONs.

\begin{figure}
\includegraphics[width=0.45\textwidth]{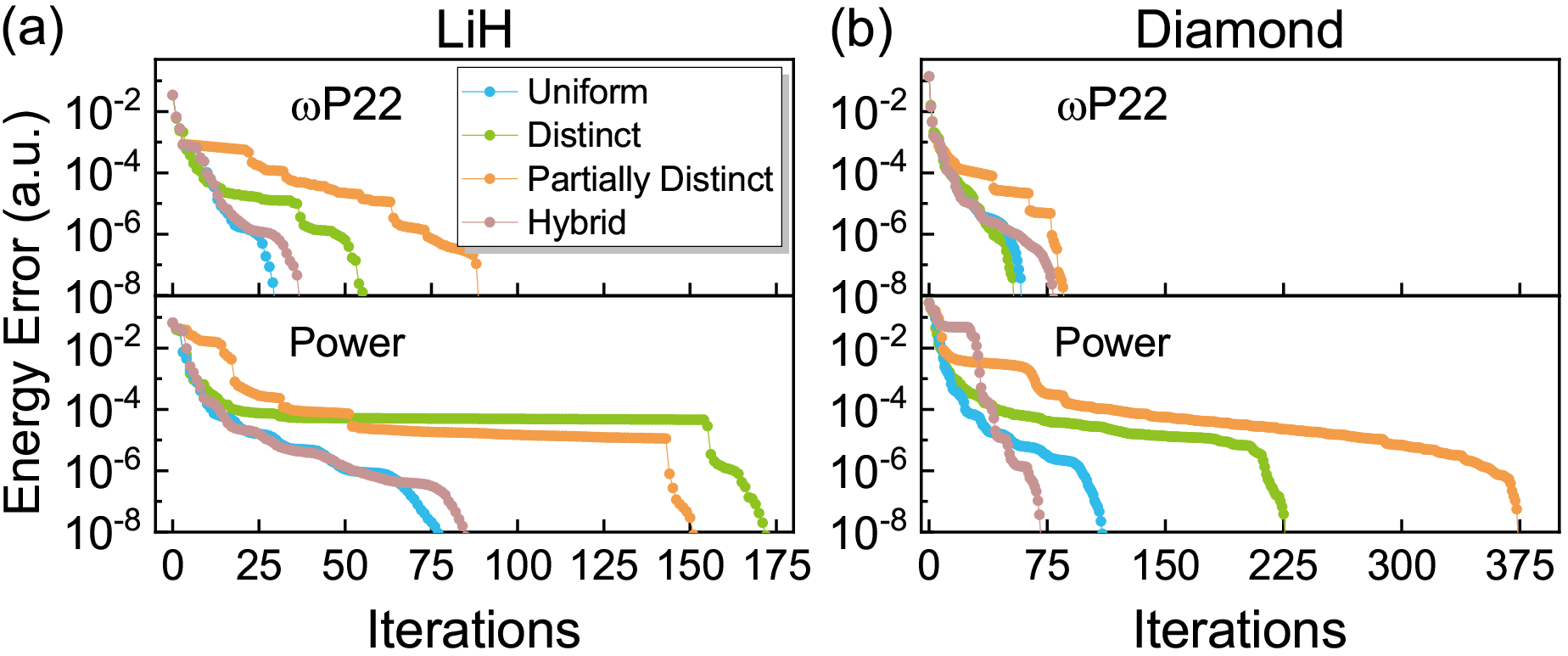}
\caption{\label{fig:21_schemes_2pic}
Comparison of four optimization schemes for implementing the CO method in periodic systems with multiple $\mathbf{k}$ points: Uniform, Distinct, Partially Distinct, and Hybrid. These schemes differ in how the conjugate-gradient coefficients $\beta$ and line-search step sizes $\alpha$ are assigned across $\mathbf{k}$ points. Energy errors as functions of optimization iteration are shown for (a) LiH with a $3\times3\times3$ $\mathbf{k}$-point mesh and (b) diamond with a $2\times2\times2$ mesh, using the Power(0.6) and $\omega$P22 functionals. For each scheme, the fully converged energy serves as the reference.}
\end{figure}
To assess the convergence behaviors of the above four schemes, we selected three-dimensional LiH and diamond as representative periodic systems. Figure~\ref{fig:21_schemes_2pic} compares the energy convergence of the four schemes using the Power(0.6) and $\omega$P22 functionals. As shown in Figures.~\ref{fig:21_schemes_2pic}(a) and \ref{fig:21_schemes_2pic}(b), the Uniform and Hybrid schemes exhibit rapid energy convergence for both functionals. The corresponding iteration counts, including results obtained with Power(0.7), are summarized in Table~S2 \cite{SupplementalMaterial}. To further assess convergence accuracy, Table~S3 \cite{SupplementalMaterial} reports the energy differences relative to the lowest converged energy among the four schemes for each system and functional. Both Uniform and Hybrid achieve small energy errors, demonstrating favorable convergence behavior and numerical robustness. Considering their overall performance, we adopt the Uniform scheme for all subsequent calculations.

We next compare the convergence efficiency of CO and deCO for four representative periodic systems: graphene, hexagonal boron nitride (h-BN), LiH, and diamond. Graphene is a two-dimensional semimetal with a vanishing band gap \cite{graphene2012,graphene2019}, whereas h-BN is a wide-band-gap insulator with a similar honeycomb lattice structure \cite{hBN2016hexagonal}. LiH and diamond are representative three-dimensional ionic and covalent insulators, respectively, with direct and indirect fundamental band gaps \cite{LiHbandgap1997,diamond2004bandgap}. Experimental lattice constants obtained from the Crystallography Open Database (COD) \cite{COD2012} were used: 2.467 $\mathrm{\AA}$ for graphene, 2.498 $\mathrm{\AA}$ for h-BN, 4.085 $\mathrm{\AA}$ for LiH, and 3.567 $\mathrm{\AA}$ for diamond.

\begin{figure}
\includegraphics[width=0.45\textwidth]{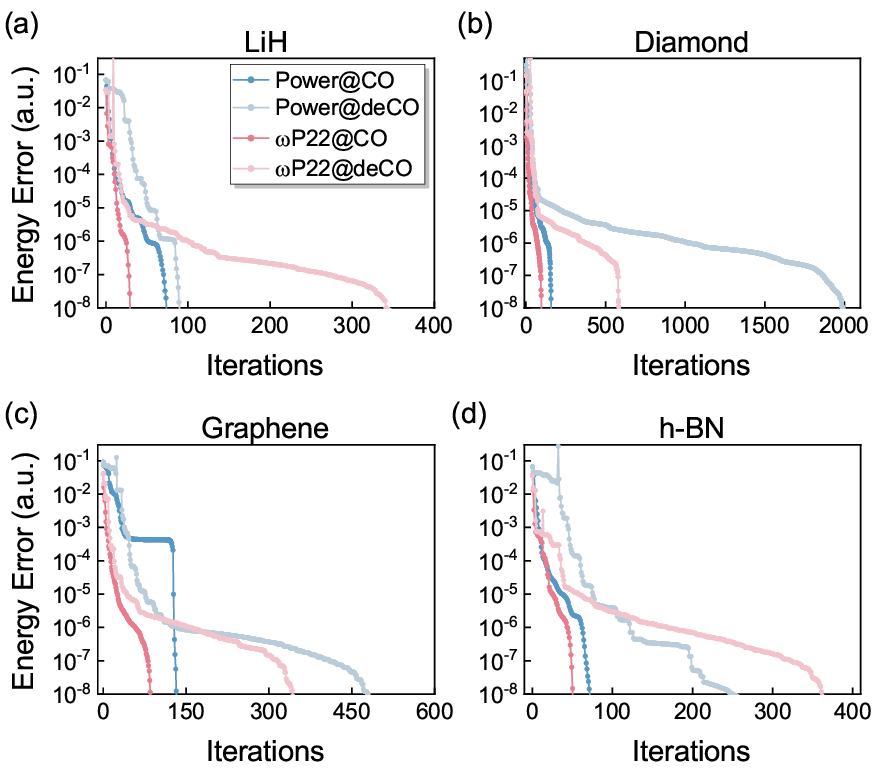}
\caption{\label{fig:22_deCOvsCO_pcg06}
Comparison of energy convergence between the CO and deCO algorithms for four representative periodic systems. Energy errors relative to the fully converged energies are plotted as functions of optimization iteration. The $\mathbf{k}$-point meshes are $3\times3\times3$ for three-dimensional LiH and diamond and $6\times6\times1$ for two-dimensional graphene and h-BN.}
\end{figure}
Figure~\ref{fig:22_deCOvsCO_pcg06} compares the energy convergence profiles of CO and deCO for these systems. For both functionals, CO exhibits substantially faster convergence than deCO across all four systems. In particular, deCO displays pronounced convergence plateaus and requires up to 12 times as many iterations as CO to reach convergence. The iteration counts obtained with Power(0.6), Power(0.7), and $\omega$P22 are summarized in Table~S4 \cite{SupplementalMaterial}. The improved convergence of CO can be attributed to the simultaneous optimization of NOs and ONs. In contrast, the alternating optimization in deCO may lead to redundant iterations, as updating one variable group can partially invalidate the optimization achieved for the other. By accounting for their mutual dependence during optimization, CO more efficiently approaches the variational minimum.

Finally, we investigate the convergence of RDMFT total energies toward the thermodynamic limit (TDL) with increasing $\mathbf{k}$-point sampling under Born-von Karman (BvK) periodic boundary conditions. For comparison, calculations using the PBE \cite{pbe1996} and PBE0 \cite{pbe01999} functionals within KS-DFT were also performed. For each system, the convergence of the total energy per unit cell, $E_{N_{\mathbf{k}}}$, is examined as a function of the number of sampled $\mathbf{k}$ points, $N_{\mathbf{k}}$.

\begin{figure}
\includegraphics[width=0.45\textwidth]{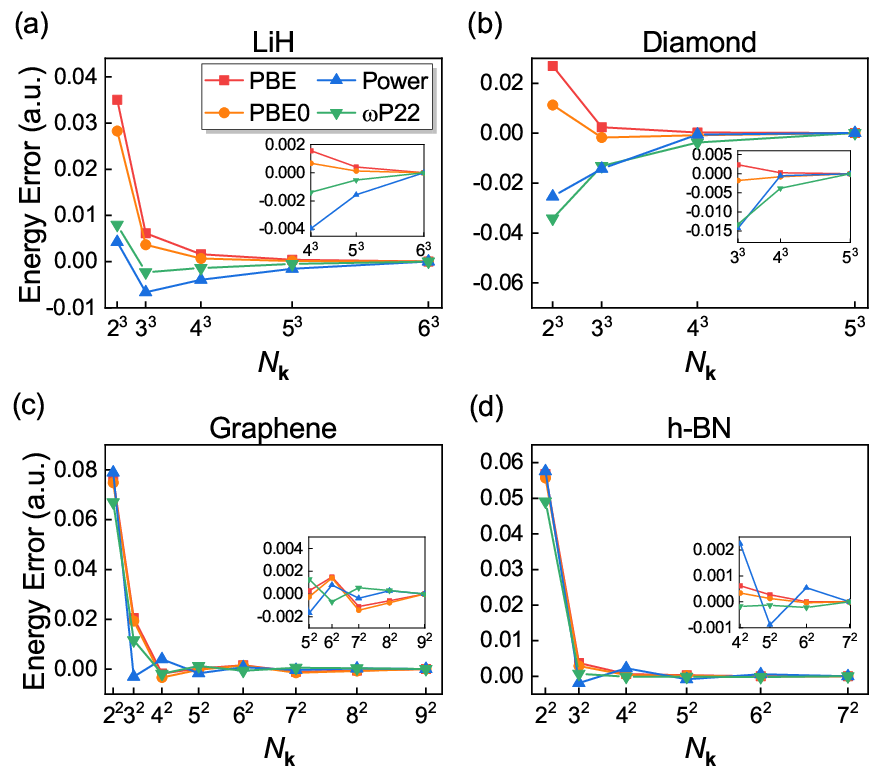}
\caption{\label{fig:24_merge_Nk_dzvp}
Convergence of the total energy per unit cell with respect to $\mathbf{k}$-point sampling for four representative periodic systems using different functionals. The energy difference, $E_{N_{\mathbf{k}}}-E_{\mathrm{ref}}$, is plotted against the number of $\mathbf{k}$ points, $N_{\mathbf{k}}$, where $E_{\mathrm{ref}}$ is the energy obtained using the densest $\mathbf{k}$-point mesh considered for each system and functional. Insets show enlarged views of the high-$N_{\mathbf{k}}$ region. Results are shown for PBE@KS-DFT (red squares), PBE0@KS-DFT (orange circles), Power@RDMFT (blue triangles), and $\omega$P22@RDMFT (green triangles), using the GTH-DZVP-MOLOPT-SR basis set.}
\end{figure}

Figure~\ref{fig:24_merge_Nk_dzvp} compares the $\mathbf{k}$-point convergence of PBE, PBE0, Power(0.6), and $\omega$P22. Overall, the RDMFT functionals exhibit convergence behavior comparable to that of the KS-DFT functionals, indicating that the explicit dependence on the nonlocal 1-RDM does not substantially deteriorate convergence with respect to $\mathbf{k}$-point sampling for the systems examined. Furthermore, the screened $\omega$P22 functional exhibits overall faster $\mathbf{k}$-point convergence than the unscreened Power(0.6) functional, suggesting that short-range screening can facilitate convergence toward the TDL.

\subsection{Short-Range Screening: Reaction Barriers and Numerical Robustness}

The ability of RDMFT to capture electron correlation motivates its application to surface reactions, for which accurate prediction of activation barriers remains challenging for approximate KS-DFT functionals \cite{diaz2009science,2017sbh10barrier,wijzenbroek2014jcp,oudot2024jcp}. To assess the performance of periodic RDMFT, we investigate $\mathrm{H}_2$ dissociative adsorption and its reverse associative desorption on Cu(111), as well as $\mathrm{H}_2$ dissociative adsorption on Al(110). These reactions involve H--H bond breaking and formation, during which stretched molecular configurations can exhibit significant static correlation arising from near-degenerate electronic states. Accurately describing such correlation presents a challenge for conventional KS-DFT approximations and provides an important benchmark for RDMFT functionals.

\begin{figure}
\centering
\includegraphics[width=0.45\textwidth]{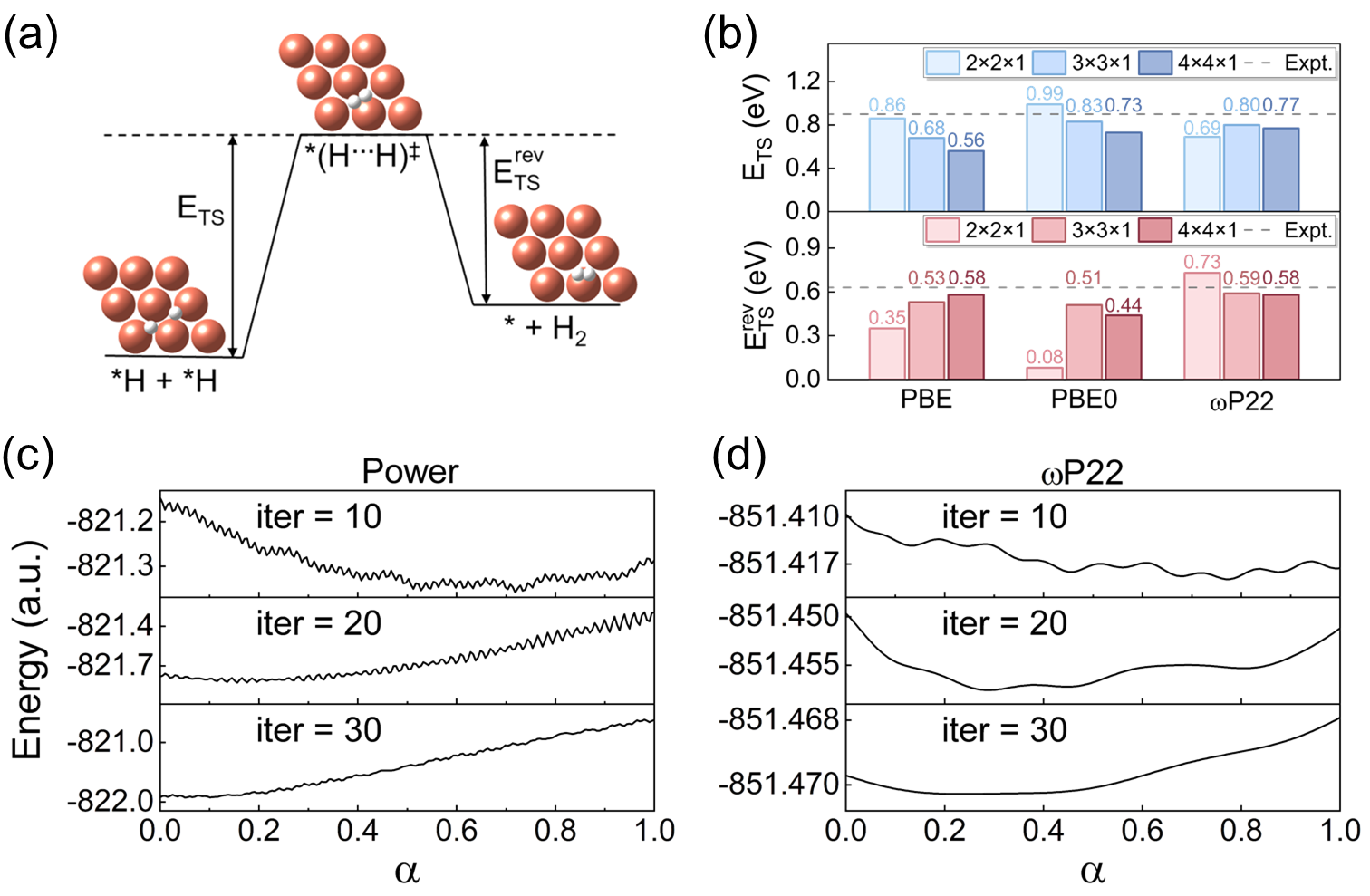}
\caption{\label{fig:26_H2Cu111_MEP_v2}
Comparison of different functionals for the $\mathrm{H}_2$/Cu(111) surface reaction. (a) Schematic illustration of $\mathrm{H}_2$ associative desorption from Cu(111) and the reverse dissociative adsorption process. (b) Forward and reverse reaction barriers calculated using PBE and PBE0 within KS-DFT and $\omega$P22 within RDMFT with different $\Gamma$-centered $\mathbf{k}$-point meshes. (c,d) Energy profiles obtained by varying the common step size $\alpha_R=\alpha_x=\alpha$ for the NO and ON updates with the unscreened Power(0.6) and screened $\omega$P22 functionals, respectively, at optimization iterations 10, 20, and 30, using a $2\times2\times1$ $\mathbf{k}$-point mesh. The step size $\alpha$ was sampled at 200 uniformly spaced points over the interval $[0,1]$.}
\end{figure}

Figure~\ref{fig:26_H2Cu111_MEP_v2}(a) illustrates the reaction pathway for $\mathrm{H}_2$ associative desorption from Cu(111) and its reverse dissociative adsorption process. The reaction geometries were taken from previous work \cite{zhou2020nature}, in which the pathway was determined using PBE \cite{pbe1996} with D3 dispersion corrections \cite{grimme2010d3,grimme2011d3} and Becke--Johnson damping \cite{beckeJohnson2005}, together with the climbing-image nudged elastic band (CI-NEB) method \cite{cineb2000}. In the initial state of associative desorption, two chemisorbed H atoms occupy adjacent hexagonal-close-packed (hcp) and face-centered-cubic (fcc) hollow sites, separated by 1.98 $\mathrm{\AA}$. As the reaction proceeds, the H atoms detach from the Cu surface and approach each other to form an H--H bond, ultimately producing a desorbed $\mathrm{H}_2$ molecule. As a proof of concept, Cu(111) was modeled using a $3\times3$ periodic slab consisting of two atomic layers and 18 Cu atoms, with a vacuum region of approximately 16 $\mathrm{\AA}$ along the surface-normal direction. The slab geometry is illustrated in Figure~S2 \cite{SupplementalMaterial}, with further structural and optimization details provided in Ref.~\cite{zhou2020nature}. To reduce computational cost, a mixed-basis strategy was adopted: the GTH-DZVP-MOLOPT-SR basis set was used for H atoms and Cu atoms near the reaction sites, while the GTH-SZV-MOLOPT-SR basis set was employed for the remaining Cu atoms.

Figure~\ref{fig:26_H2Cu111_MEP_v2}(b) compares the forward and reverse reaction barriers predicted by PBE, PBE0, and $\omega$P22 using different $\Gamma$-centered $\mathbf{k}$-point meshes. The PBE and PBE0 calculations were performed using PySCF \cite{pyscf2018,pyscf2020recent}, with Fermi-Dirac occupation smearing characterized by $\sigma=0.001$ Ha to facilitate convergence. For comparison, temperature-programmed desorption experiments report an $\mathrm{H}_2$ associative desorption barrier of approximately 0.9 eV at low coverage \cite{roberts1988,anger1989}, while a dissociative adsorption barrier of 0.63 eV was obtained from a semiempirical model fitted to experimental observables \cite{diaz2009science}. At the densest $\mathbf{k}$-point mesh considered, $4\times4\times1$, PBE predicts a forward desorption barrier of 0.56 eV, substantially underestimating the experimental reference. Incorporating exact exchange through PBE0 increases the barrier to 0.73 eV, partially reducing this discrepancy. The $\omega$P22 functional yields a barrier of 0.77 eV, providing the closest agreement among the three functionals. For the reverse dissociative adsorption process, $\omega$P22 predicts a barrier of 0.58 eV, also in close agreement with the reference value of 0.63 eV. Interestingly, PBE gives the same reverse barrier of 0.58 eV, potentially reflecting error cancellation, whereas PBE0 predicts a lower value of 0.44 eV. These results demonstrate the potential of periodic RDMFT, particularly the screened $\omega$P22 functional, for improving the description of surface reaction energetics.

Calculations using the unscreened Power functional were also attempted. However, its optimization exhibited extremely slow energy descent, accompanied by very small NO update step sizes. This behavior is consistent with the molecular and periodic benchmarks discussed above, where Power(0.6) generally showed slower optimization convergence than $\omega$P22. To investigate the origin of this difference, we selected the initial dissociated configuration on Cu(111) and examined the energy profiles along the common NO and ON update direction, $\alpha_R=\alpha_x=\alpha$, at representative optimization iterations. Figures~\ref{fig:26_H2Cu111_MEP_v2}(c) and \ref{fig:26_H2Cu111_MEP_v2}(d) show the energy variation with $\alpha$ for Power(0.6) and $\omega$P22, respectively. For Power(0.6), the energy profiles exhibit pronounced oscillations over the interval $\alpha\in[0,1]$, indicating a highly irregular energy landscape along the search direction. Such behavior complicates the quadratic line search and can lead to extremely small accepted step sizes, resulting in slow energy descent and poor convergence. In contrast, the corresponding profiles for $\omega$P22 are substantially smoother, allowing the line-search algorithm to identify more effective update steps and achieve faster, more robust convergence. Similar behavior is observed for the transition and final states along the $\mathrm{H}_2$/Cu(111) reaction pathway (Figure~S3 \cite{SupplementalMaterial}). Owing to the prohibitively slow convergence of Power(0.6), all subsequent periodic RDMFT surface-reaction-barrier calculations were performed exclusively using $\omega$P22.

To further assess the transferability of the screened $\omega$P22 functional, we investigated the reaction barriers for $\mathrm{H}_2$ dissociative adsorption on Al(110), a challenging benchmark for electronic-structure methods. The initial-state and four transition-state structures are shown in Figure~S4 \cite{SupplementalMaterial}, with geometrical parameters taken from Ref.~\cite{Al110DMC2020} and summarized in Tables~S5 and S6 \cite{SupplementalMaterial}. A three-layer Al(110) slab was employed for these calculations. The convergence of the reaction barriers with respect to $\mathbf{k}$-point sampling was first examined using PBE. As shown in Table~S7 \cite{SupplementalMaterial}, the barrier for TS-A is converged to within 30 meV with a $7\times7\times1$ $\mathbf{k}$-point mesh, which was adopted for all subsequent calculations. Table~S8 \cite{SupplementalMaterial} summarizes the reaction barriers predicted by different functionals, together with the diffusion Monte Carlo (DMC) benchmarks reported in Ref.~\cite{Al110DMC2020}. Since experimental data are available only for TS-A, the DMC results serve as the primary references for the remaining transition states. As expected, the semilocal PBE functional substantially underestimates the reaction barrier of TS-A, predicting a value of only 0.54 eV compared with the experimentally motivated value of 1.15 eV. The hybrid PBE0 functional partially alleviates this error through the inclusion of exact exchange, but the predicted barriers remain noticeably lower than the benchmark values. In contrast, the screened $\omega$P22 functional consistently yields significantly improved reaction barriers. For TS-A, the predicted barrier increases to 0.80 eV, representing a substantial improvement over both PBE and PBE0. Similar improvements are observed for the remaining transition states when compared with the DMC benchmarks. 

These benchmarks highlight two important advantages of the screened $\omega$P22 functional: improved accuracy in predicting surface reaction barriers and substantially smoother energy profiles along the optimization directions. The latter helps explain its superior convergence behavior compared with the unscreened Power functional. These findings reveal the dual role of short-range screening in RDMFT, improving both the description of electron correlation and the numerical efficiency of functional optimization. Furthermore, the improved reaction-barrier predictions across different metal surfaces demonstrate the potential transferability of $\omega$P22, highlighting the promise of screened 1-RDM functionals for predictive heterogeneous catalysis.

\subsection{Optimization Efficiency and Functional Transferability}

\begin{figure}
\centering
\includegraphics[width=0.45\textwidth]{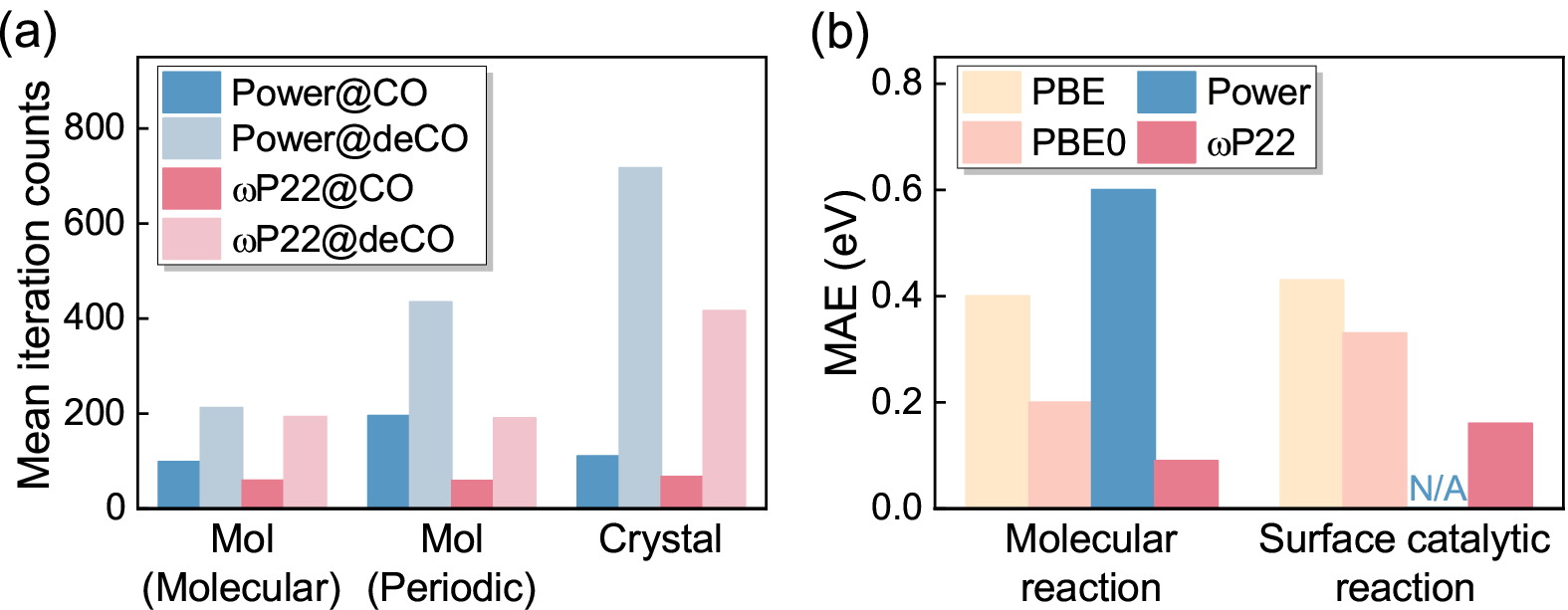}
\caption{\label{fig:27_sum_fig}
Overall performance of the periodic RDMFT framework. (a) Mean optimization iteration counts for ten molecules ($\mathrm{H}_2$, $\mathrm{H}_2\mathrm{O}$, $\mathrm{O}_2$, CO, $\mathrm{CO}_2$, $\mathrm{CH}_4$, $\mathrm{CH}_2\mathrm{O}$, $\mathrm{C}_2\mathrm{H}_4$, $\mathrm{NH}_3$, and $\mathrm{N}_2$), calculated using both molecular and periodic implementations, and four periodic solids (LiH, diamond, graphene, and h-BN), with the Power(0.6) and $\omega$P22 functionals under CO and deCO optimization. (b) Mean absolute errors (MAEs) of molecular and surface reaction barriers calculated using PBE, PBE0, Power(0.6), and $\omega$P22. The molecular benchmark comprises six reactions and their reverse processes: H + HCl $\rightarrow$ $\mathrm{H}_2$ + Cl, H + $\mathrm{H}_2$ $\rightarrow$ $\mathrm{H}_2$ + H, F + $\mathrm{H}_2$ $\rightarrow$ HF + H, H + $\mathrm{H}_2\mathrm{S}$ $\rightarrow$ $\mathrm{H}_2$ + HS, H + $\mathrm{F}_2$ $\rightarrow$ HF + F, and H + $\mathrm{N}_2$ $\rightarrow$ $\mathrm{HN}_2$, calculated using the cc-pVTZ basis set. The surface-reaction benchmark includes $\mathrm{H}_2$ dissociative adsorption on Cu(111), its reverse associative desorption, and $\mathrm{H}_2$ dissociative adsorption on Al(110). For the Al(110) transition states TS-B, TS-C, and TS-D, DMC results were used as reference values because experimental data are unavailable.}
\end{figure}

Finally, we assess the overall performance of the periodic RDMFT framework in terms of computational efficiency and energetic accuracy. Figure~\ref{fig:27_sum_fig}(a) compares the mean numbers of optimization iterations for ten molecules, calculated using both molecular and periodic implementations, and four representative periodic solids with the Power(0.6) and $\omega$P22 functionals under the CO and conventional deCO schemes. Across all benchmark systems, CO consistently requires substantially fewer iterations than deCO. The improvement is particularly pronounced for periodic solids with the Power(0.6) functional, where the mean iteration count is reduced by nearly an order of magnitude. These results demonstrate that the efficiency advantage of CO extends from molecular to periodic RDMFT calculations, providing a practical optimization strategy for extended systems.

Figure~\ref{fig:27_sum_fig}(b) summarizes the mean absolute errors (MAEs) of reaction barriers predicted by PBE, PBE0, Power(0.6), and $\omega$P22. The molecular benchmark comprises twelve forward and reverse reaction barriers from the HTBH and NHTBH test sets \cite{HTBH2009,HTBH2014}, while the periodic benchmark includes six surface reaction barriers investigated in this work. Among the functionals considered, the screened $\omega$P22 functional yields the lowest MAEs for both molecular and periodic systems. In contrast, the unscreened Power(0.6) functional exhibits the largest MAE for molecular reaction barriers, even exceeding that of PBE. The substantial improvement achieved by $\omega$P22 highlights the importance of short-range screening for accurately describing reaction energetics across molecular and periodic systems.

Taken together, these results demonstrate that short-range screening plays a central role in the practical success of RDMFT, improving not only energetic accuracy but also the numerical robustness and convergence of functional optimization. Combined with the CO algorithm, the screened $\omega$P22 functional provides an accurate, transferable, and computationally efficient RDMFT approach for both molecular and periodic systems. These findings highlight short-range screening as a promising strategy for developing transferable 1-RDM functionals, with the CO algorithm providing an efficient and broadly applicable optimization framework for their practical implementation.

\section{\label{sec:level5}Conclusions}

In this work, we have developed a periodic implementation of reduced density matrix functional theory (RDMFT) that combines coupled optimization (CO) of Bloch natural orbitals (NOs) and their occupation numbers (ONs) with short-range-screened 1-RDM functionals. This framework enables a systematic investigation of two central challenges in practical RDMFT: efficient orbital-occupation optimization and the development of accurate, transferable functionals across molecular and periodic systems.

Our results demonstrate that CO substantially improves optimization efficiency compared with conventional decoupled optimization (deCO), reducing the number of iterations by nearly an order of magnitude for representative periodic solids. Furthermore, the $k$-point convergence of RDMFT is comparable to that of conventional KS-DFT for the systems examined, supporting its practical applicability to extended systems.

Short-range screening improves both the accuracy and computational performance of RDMFT. Short-range screening improves both the accuracy and computational efficiency of RDMFT. Compared with the unscreened Power functional, the screened $\omega$P22 functional exhibits smoother energy profiles during optimization and generally faster $k$-point convergence. Furthermore, its range-separated construction enables efficient evaluation of the short-range contributions using semilocal density functionals and rapid convergence of the long-range contribution in reciprocal space with relatively few reciprocal lattice vectors. Importantly, $\omega$P22 yields the lowest mean absolute errors in reaction barriers among the functionals examined for both molecular and surface reaction benchmarks, highlighting the potential of short-range screening to improve functional transferability across finite and extended systems.

Taken together, these findings establish the complementary roles of coupled optimization and short-range screening in advancing practical RDMFT. While CO provides an efficient and broadly applicable optimization framework, short-range screening offers a promising design principle for developing accurate and transferable 1-RDM functionals. Their combination provides a foundation for extending RDMFT to more complex periodic systems and opens new opportunities for investigating electronic correlation and chemical reactivity in solids and at surfaces. More broadly, the methodological insights gained in this work may also guide the development and application of orbital functionals and computational methods within the framework of hierarchically correlated orbital functional theory (HCOFT) \cite{Su2024prl,Zhang2025pra}.

%
%
\begin{acknowledgments}
Support from Fundamental and Interdisciplinary Disciplines Breakthrough Plan of the Ministry of Education of China (JYB2025XDXM410), National Natural Science Foundation of China (Grants No. 22473061, No. 22122303, and No. 22073049), and Fundamental Research Funds for the Central Universities (Nankai University: No. 63206008) is appreciated.
\end{acknowledgments}

\bibliographystyle{aip}
\bibliography{ref}

\onecolumngrid




%
\newcommand{\ti}{\Tilde}
\newcommand{\Sch}{Schr\"{o}dinger\ }
\newcommand{\Sec}[1]{Sec.\;\ref{#1}}
\newcommand{\App}[1]{Appendix\;\ref{#1}}
\newcommand{\be}{\begin{equation}}
	\newcommand{\ee}{\end{equation}}
\newcommand{\bea}{\begin{eqnarray}}
	\newcommand{\eea}{\end{eqnarray}}
\newcommand{\bsube}{\begin{subequations}}
	\newcommand{\esube}{\end{subequations}}
\newcommand{\Eq}[1]{Eq.\,(\ref{#1})}
\newcommand{\Eqs}[1]{Eqs.\,(\ref{#1})}
\newcommand{\Fig}[1]{Fig.\,\ref{#1}}
\newcommand{\Tab}[1]{Table\,\ref{#1}}
\newcommand{\sub}[1]{$_{#1}$}
\newcommand{\upp}[1]{$^{#1}$}
\newcommand{\dg}{\dagger}
\newcommand{\la}{\langle}
\newcommand{\ra}{\rangle}
\newcommand{\kb}{\rangle\langle}
\newcommand{\w}{\omega}
\newcommand{\ep}{\epsilon}
\newcommand{\B}{\mbox{\tiny B}}
\newcommand{\bfGam}{\mbox{\boldmath $\Gamma$}}

\newpage
\vspace{20cm}

\draft

Supporting Information of
%
\vspace{0.1cm}

\begin{center}
{\large \textbf{Reduced Density Matrix Functional Theory Across Molecules and Periodic Systems: Short-Range Screening and Coupled Optimization}}
\vspace{0.2cm}

Shuxin Pei and Neil Qiang Su$^*$

{\small
\emph{Center for Theoretical and Computational Chemistry, Frontiers Science Center for New Organic Matter, State Key Laboratory of Advanced Chemical Power Sources, Key Laboratory of Advanced Energy Materials Chemistry (Ministry of Education), Department of Chemistry, Nankai University, Tianjin 300071, China\\}
\emph{\rm $^*$ Email: nqsu@nankai.edu.cn}
}

	\vspace{0.1cm}
	%
\end{center}

\linespread{1.0}



\setcounter{section}{0}
\setcounter{page}{1}
\setcounter{figure}{0}
\setcounter{table}{0}

\renewcommand{\thesection}{S\Roman{section}}
\renewcommand{\thepage}{S\arabic{page}}
\renewcommand{\thefigure}{S\arabic{figure}}
\renewcommand{\thetable}{S\arabic{table}}
\renewcommand{\theequation}{S\arabic{equation}}

\makeatletter
\renewcommand{\theenumiv}{S\arabic{enumiv}}
\renewcommand{\@biblabel}[1]{[S#1]} %
\renewcommand{\citenumfont}[1]{S#1}
\renewcommand{\bibnumfmt}[1]{[S#1]}
\makeatother

\begin{center}
	\large\textbf{Contents}
\end{center}

\makeatletter
\@starttoc{tosi} 
\makeatother

\newpage

\section{Energy Expressions and Orbital Derivatives in RDMFT for Periodic and Molecular Systems}
\addSIcontents{section}{Energy Expressions and Orbital Derivatives in RDMFT for Periodic and Molecular Systems}

For periodic systems, the total electronic energy within RDMFT is given by

\begin{equation}
	\label{eq:E_cell}
    \begin{aligned}
    	E[\gamma_{\sigma}^{\mathbf{k}}] = & \frac{1}{N_{\mathbf{k}}} \sum_{\mathbf{k}} \sum_{\sigma} \sum_{\mu\nu} h_{\mu\nu}^{\mathbf{k}} \gamma_{\nu\mu,\sigma}^{\mathbf{k}} + \frac{1}{2N_{\mathbf{k}}^2} \sum_{\mathbf{k}\mathbf{k}'} \sum_{\sigma\sigma'} \sum_{\mu\nu\kappa\lambda} \gamma_{\nu\mu,\sigma}^{\mathbf{k}} (\mu^{\mathbf{k}} \nu^{\mathbf{k}} \mid \kappa^{\mathbf{k}'} \lambda^{\mathbf{k}'}) \gamma_{\lambda\kappa,\sigma'}^{\mathbf{k}'} \\
    	& - \frac{1}{2N_{\mathbf{k}}^2} \sum_{\mathbf{k}\mathbf{k}'} \sum_{\sigma} \sum_{ij} \sum_{\mu\nu\kappa\lambda} C_{\mu i}^{\mathbf{k},\sigma,*} C_{\lambda j}^{\mathbf{k}',\sigma} (\mu^{\mathbf{k}} \lambda^{\mathbf{k}'} \mid \kappa^{\mathbf{k}'} \nu^{\mathbf{k}}) C_{\kappa j}^{\mathbf{k}',\sigma,*} C_{\nu i}^{\mathbf{k},\sigma} f(n_i^{\mathbf{k},\sigma}, n_j^{\mathbf{k}',\sigma})
    \end{aligned}
\end{equation}
where $\gamma_{\mu\nu,\sigma}^{\mathbf{k}} = \sum_{i} C_{\mu i}^{\mathbf{k},\sigma} n_i^{\mathbf{k},\sigma} C_{\nu i}^{\mathbf{k},\sigma,*}$ is the one-particle reduced density matrix at the $\mathbf{k}$-point. To maintain the orthonormality of the Bloch natural orbitals (NOs) during optimization, a unitary transformation $\mathbf{U}^{\mathbf{k},\sigma} = \exp(\mathbf{R}^{\mathbf{k},\sigma})$ is employed, where $\mathbf{R}^{\mathbf{k},\sigma}$ is a skew-Hermitian matrix satisfying $(\mathbf{R}^{\mathbf{k},\sigma})^\dagger = -\mathbf{R}^{\mathbf{k},\sigma}$. Up to second order, the Taylor expansion of $\mathbf{U}^{\mathbf{k},\sigma}$ is expressed as 
\begin{equation}
	\label{eq:U_taylor}
	\begin{aligned}
		U_{ij}^{\mathbf{k},\sigma} = 1 + R_{ij}^{\mathbf{k},\sigma} + \frac{1}{2} \sum_{k} R_{ik}^{\mathbf{k},\sigma} R_{kj}^{\mathbf{k},\sigma} + O((R^{\mathbf{k},\sigma})^3)
    \end{aligned}
\end{equation}
The first derivative of the energy with respect to $\mathbf{R}^{\mathbf{k},\sigma}$ is given by
\begin{equation}
	\label{eq:dE_dR}
		\left. \frac{\partial E}{\partial R_{ab}^{\mathbf{k},\sigma}} \right|_{\mathbf{R}^{\mathbf{k},\sigma}=\mathbf{0}}  = \sum_{\mu=1} \Big(C_{\mu a}^{\mathbf{k},\sigma} \frac{\partial E}{\partial C_{\mu b}^{\mathbf{k},\sigma}} - \frac{\partial E}{\partial C_{\mu a}^{\mathbf{k},\sigma,*}} C_{\mu b}^{\mathbf{k},\sigma,*} \Big)
\end{equation}
For simplicity, we define the generalized Fock matrix $\mathbf{F}^{\mathbf{k},\sigma}$ with elements given by
\begin{equation}
	\label{eq:Fpq}
	F_{pq}^{\mathbf{k},\sigma} = h_{pq}^{\mathbf{k},\sigma} + J_{pq}^{\mathbf{k},\sigma} - K_{pq}^{\mathbf{k},\sigma}
\end{equation}
where the three terms on the right-hand side correspond to the one-electron Hamiltonian matrix, the Coulomb matrix, and the generalized exchange matrix, respectively. Their explicit expressions are given by
\begin{equation}
	\label{eq:hpq}
	h_{pq}^{\mathbf{k},\sigma} = \sum_{\mu\nu} C_{\mu p}^{\mathbf{k},\sigma,*} h_{\mu\nu}^{\mathbf{k}} C_{\nu q}^{\mathbf{k},\sigma}
\end{equation}

\begin{equation}
	\label{eq:Jpq}
	J_{pq}^{\mathbf{k},\sigma} = \frac{1}{N_{\mathbf{k}}} \sum_{\mathbf{k}'} \sum_{\sigma'}  \sum_{\mu\nu\kappa\lambda} C_{\mu p}^{\mathbf{k},\sigma,*} C_{\nu q}^{\mathbf{k},\sigma} (\mu^{\mathbf{k}} \nu^{\mathbf{k}} \mid \kappa^{\mathbf{k}'} \lambda^{\mathbf{k}'}) \gamma_{\lambda\kappa,\sigma'}^{\mathbf{k}'}
\end{equation}

\begin{equation}
	\label{eq:Kpq}
	K_{pq}^{\mathbf{k},\sigma} n_i^{\mathbf{k},\sigma} = \frac{1}{N_{\mathbf{k}}} \sum_{\mathbf{k}'} \sum_{j} \sum_{\mu\nu\kappa\lambda} C_{\mu p}^{\mathbf{k},\sigma,*} C_{\nu j}^{\mathbf{k}',\sigma} (\mu^{\mathbf{k}} \nu^{\mathbf{k}'} \mid \kappa^{\mathbf{k}'} \lambda^{\mathbf{k}}) C_{\kappa j}^{\mathbf{k}',\sigma,*} C_{\lambda q}^{\mathbf{k},\sigma} f(n_i^{\mathbf{k},\sigma}, n_j^{\mathbf{k}',\sigma})
\end{equation}
The one-electron Hamiltonian matrix in the basis of translationally symmetry-adapted atomic orbitals (AOs) is defined as 
\begin{equation}
	\label{eq:huv}
	h_{\mu\nu}^{\mathbf{k}} = T_{\mu\nu}^{\mathbf{k}} + V_{\mu\nu}^{\mathbf{k}}
\end{equation}
where the kinetic energy component
\begin{equation}
	\label{eq:Tuv}
	T_{\mu\nu}^{\mathbf{k}} = -\frac{1}{2} \int_{\Omega} d\mathbf{r} \phi_{\mu}^{\mathbf{k},*}(\mathbf{r})\nabla^{2}_{\mathbf{r}} \phi_{\nu}^{\mathbf{k}}(\mathbf{r})\, 
\end{equation}
is evaluated analytically, while the external potential integral 
\begin{equation}
	\label{eq:Vuv}
	V_{\mu\nu}^{\mathbf{k}} = \int_{\Omega} d\mathbf{r} \phi_{\mu}^{\mathbf{k},*}(\mathbf{r}) v_{\text{ext}}(\mathbf{r})  \phi_{\nu}^{\mathbf{k}}(\mathbf{r})\, 
\end{equation}
is evaluated using a three-center integration scheme analogous to that described in ~\ref{sec: rsgdf_alg}. $(\mu^{\mathbf{k}_\mu} \nu^{\mathbf{k}_\nu} \mid \kappa^{\mathbf{k}_\kappa} \lambda^{\mathbf{k}_\lambda})$ is the two-electron repulsion integrals (ERIs) which is evaluated by the density fitting approximation described in ~\ref{sec: rsgdf_alg}. Thus, the first derivative of $E$ with respect to $\mathbf{R}^{\mathbf{k},\sigma}$ can be written as 
\begin{equation}
	\label{eq:dE_dRpq}
	\left. \frac{\partial E}{\partial R_{pq}^{\mathbf{k},\sigma}} \right|_{\mathbf{R}^{\mathbf{k},\sigma}=\mathbf{0}} = \frac{1}{N_{\mathbf{k}}} \left[ (F_{pq}^{\mathbf{k},\sigma})^* n_q^{\mathbf{k},\sigma} - F_{qp}^{\mathbf{k},\sigma} n_p^{\mathbf{k},\sigma} \right]
\end{equation}
The first derivatives of $E$ with respect to the real and the imaginary parts of $\mathbf{R}^{\mathbf{k},\sigma}$, $\mathbf{X}^{\mathbf{k},\sigma}$ and $\mathbf{Y}^{\mathbf{k},\sigma}$, respectively, are given by

\begin{equation}
	\label{eq:dE_dX}
	\left. \frac{\partial E}{\partial X_{pq}^{\mathbf{k},\sigma}} \right|_{\mathbf{R}^{\mathbf{k},\sigma}=\mathbf{0}} = 2\text{Re}\Bigl(\frac{\partial E}{\partial R_{pq}^{\mathbf{k},\sigma}}\Bigr) = -2\frac{1}{N_{\mathbf{k}}} \text{Re}(F_{pq}^{\mathbf{k},\sigma})(n_p^{\mathbf{k},\sigma} - n_q^{\mathbf{k},\sigma})
\end{equation}

\begin{equation}
	\label{eq:dE_dY}
	\left. \frac{\partial E}{\partial Y_{pq}^{\mathbf{k},\sigma}} \right|_{\mathbf{R}^{\mathbf{k},\sigma}=\mathbf{0}} = -2\text{Im}\Bigl(\frac{\partial E}{\partial R_{pq}^{\mathbf{k},\sigma}}\Bigr) = -2\frac{1}{N_{\mathbf{k}}} \text{Im}(F_{pq}^{\mathbf{k},\sigma})(n_p^{\mathbf{k},\sigma} - n_q^{\mathbf{k},\sigma})
\end{equation}
The second derivative of $E$ with respect to $\mathbf{R}^{\mathbf{k},\sigma}$ and its complex conjugate $\mathbf{R}^{\mathbf{k},\sigma,*}$ are
\begin{equation}
	\label{eq:HRR}
	\begin{aligned}
		(H_{RR}^{\mathbf{k},\sigma})_{ab,pq} = \left. \frac{\partial^2 E}{\partial R_{ab}^{\mathbf{k},\sigma} \partial R_{pq}^{\mathbf{k},\sigma}} \right|_{\mathbf{R}^{\mathbf{k},\sigma}=\mathbf{0}} 
		= & \frac{1}{2} \sum_{\mu=1} \Bigl[ \delta_{bp}  \Bigl( C_{\mu a}^{\mathbf{k},\sigma} \frac{\partial E}{\partial C_{\mu q}^{\mathbf{k},\sigma}} + \frac{\partial E}{\partial C_{\mu a}^{\mathbf{k},\sigma,*}} C_{\mu q}^{\mathbf{k},\sigma,*} \Bigr) + \delta_{aq} \Bigl( \frac{\partial E}{\partial C_{\mu b}^{\mathbf{k},\sigma}} C_{\mu p}^{\mathbf{k},\sigma} + C_{\mu b}^{\mathbf{k},\sigma,*} \frac{\partial E}{\partial C_{\mu p}^{\mathbf{k},\sigma,*}} \Bigr) \Bigr] \\
		& + \sum_{\mu, \nu=1} \Bigl( \frac{\partial^2 E}{\partial C_{\mu b}^{\mathbf{k},\sigma} \partial C_{\nu q}^{\mathbf{k},\sigma}} C_{\mu a}^{\mathbf{k},\sigma} C_{\nu p}^{\mathbf{k},\sigma} + \frac{\partial^2 E}{\partial C_{\mu a}^{\mathbf{k},\sigma,*} \partial C_{\nu p}^{\mathbf{k},\sigma,*}} C_{\mu b}^{\mathbf{k},\sigma,*} C_{\nu q}^{\mathbf{k},\sigma,*} \\
		& - \frac{\partial^2 E}{\partial C_{\mu b}^{\mathbf{k},\sigma} \partial C_{\nu p}^{\mathbf{k},\sigma,*}} C_{\mu a}^{\mathbf{k},\sigma} C_{\nu q}^{\mathbf{k},\sigma,*} - \frac{\partial^2 E}{\partial C_{\mu q}^{\mathbf{k},\sigma} \partial C_{\nu a}^{\mathbf{k},\sigma,*}} C_{\mu p}^{\mathbf{k},\sigma} C_{\nu b}^{\mathbf{k},\sigma,*} \Bigr)
		\end{aligned}
\end{equation}

\begin{equation}
	\label{eq:HRR*}
	\begin{aligned}
		(H_{R R^*}^{\mathbf{k},\sigma})_{ab,pq} = \left. \frac{\partial^2 E}{\partial R_{ab}^{\mathbf{k},\sigma} \partial R_{pq}^{\mathbf{k},\sigma,*}} \right|_{\mathbf{R}^{\mathbf{k},\sigma}=\mathbf{0}} 
		= & -\frac{1}{2} \sum_{\mu=1} \Bigl[ \delta_{bq} \Bigl( C_{\mu a}^{\mathbf{k},\sigma} \frac{\partial E}{\partial C_{\mu p}^{\mathbf{k},\sigma}} + \frac{\partial E}{\partial C_{\mu a}^{\mathbf{k},\sigma,*}} C_{\mu p}^{\mathbf{k},\sigma,*} \Bigr) + \delta_{ap} \Bigl( \frac{\partial E}{\partial C_{\mu b}^{\mathbf{k},\sigma}} C_{\mu q}^{\mathbf{k},\sigma} + C_{\mu b}^{\mathbf{k},\sigma,*} \frac{\partial E}{\partial C_{\mu q}^{\mathbf{k},\sigma,*}} \Bigr) \Bigr] \\
		& - \sum_{\mu, \nu=1} \Bigl( \frac{\partial^2 E}{\partial C_{\mu b}^{\mathbf{k},\sigma} \partial C_{\nu p}^{\mathbf{k},\sigma}} C_{\mu a}^{\mathbf{k},\sigma} C_{\nu q}^{\mathbf{k},\sigma} + \frac{\partial^2 E}{\partial C_{\mu a}^{\mathbf{k},\sigma,*} \partial C_{\nu q}^{\mathbf{k},\sigma,*}} C_{\mu b}^{\mathbf{k},\sigma,*} C_{\nu p}^{\mathbf{k},\sigma,*} \\
		& - \frac{\partial^2 E}{\partial C_{\mu b}^{\mathbf{k},\sigma} \partial C_{\nu q}^{\mathbf{k},\sigma,*}} C_{\mu a}^{\mathbf{k},\sigma} C_{\nu p}^{\mathbf{k},\sigma,*} - \frac{\partial^2 E}{\partial C_{\mu p}^{\mathbf{k},\sigma} \partial C_{\nu a}^{\mathbf{k},\sigma,*}} C_{\mu q}^{\mathbf{k},\sigma} C_{\nu b}^{\mathbf{k},\sigma,*} \Bigr)
		\end{aligned}
\end{equation}
with $H_{R^*R}^{\mathbf{k},\sigma} = (H_{RR^*}^{\mathbf{k},\sigma})^*$ and $H_{R^* R^*}^{\mathbf{k},\sigma} = (H_{RR}^{\mathbf{k},\sigma})^*$.
The second-order derivatives of $E$ with respect to  $\mathbf{X}^{\mathbf{k},\sigma}$ and $\mathbf{Y}^{\mathbf{k},\sigma}$ are related to those with respect to $\mathbf{R}^{\mathbf{k},\sigma}$ and $\mathbf{R}^{\mathbf{k},\sigma,*}$ as follows

\begin{equation}
	\label{eq:HXX}
	H_{XX}^{\mathbf{k},\sigma} = H_{RR}^{\mathbf{k},\sigma} + H_{RR^*}^{\mathbf{k},\sigma} + H_{R^* R}^{\mathbf{k},\sigma} + H_{R^* R^*}^{\mathbf{k},\sigma} = 2\text{Re}\left(H_{RR}^{\mathbf{k},\sigma} + H_{R R^*}^{\mathbf{k},\sigma}\right)
\end{equation}

\begin{equation}
	\label{eq:HYY}
	H_{YY}^{\mathbf{k},\sigma} = -H_{RR}^{\mathbf{k},\sigma} + H_{RR^*}^{\mathbf{k},\sigma} + H_{R^* R}^{\mathbf{k},\sigma} - H_{R^* R^*}^{\mathbf{k},\sigma} = 2\text{Re}\left(-H_{RR}^{\mathbf{k},\sigma} + H_{R R^*}^{\mathbf{k},\sigma}\right)
\end{equation}

\begin{equation}
	\label{eq:HXY}
	H_{XY}^{\mathbf{k},\sigma} = i\left(H_{RR}^{\mathbf{k},\sigma} - H_{RR^*}^{\mathbf{k},\sigma} + H_{R^* R}^{\mathbf{k},\sigma} - H_{R^* R^*}^{\mathbf{k},\sigma}\right) = 2\text{Im}\left(-H_{RR}^{\mathbf{k},\sigma} + H_{R R^*}^{\mathbf{k},\sigma}\right)
\end{equation}
with $H_{YX}^{\mathbf{k},\sigma} = (H_{XY}^{\mathbf{k},\sigma})^T$. 
Substituting Eq.~(\ref{eq:HRR}) and ~(\ref{eq:HRR*}) into the above equation yields

\begin{equation}
	\label{eq:HXX_C}
	\begin{aligned}
		\left. (H_{XX}^{\mathbf{k},\sigma})_{ab,pq} = \frac{\partial^2 E}{\partial X_{ab}^{\mathbf{k},\sigma} \partial X_{pq}^{\mathbf{k},\sigma}} \right|_{\mathbf{R}^{\mathbf{k},\sigma}=\mathbf{0}} =& 2 \operatorname{Re} \biggl\{ \frac{1}{2} \sum_{\mu=1} \Bigl[ \delta_{bp} \Bigl( C_{\mu a}^{\mathbf{k},\sigma} \frac{\partial E}{\partial C_{\mu q}^{\mathbf{k},\sigma}} + \frac{\partial E}{\partial C_{\mu a}^{\mathbf{k},\sigma,*}} C_{\mu q}^{\mathbf{k},\sigma,*} \Bigl) + \delta_{aq} \Bigl( \frac{\partial E}{\partial C_{\mu b}^{\mathbf{k},\sigma}} C_{\mu p}^{\mathbf{k},\sigma} + C_{\mu b}^{\mathbf{k},\sigma,*} \frac{\partial E}{\partial C_{\mu p}^{\mathbf{k},\sigma,*}} \Bigl) \\
		& - \delta_{bq} \Bigl( C_{\mu a}^{\mathbf{k},\sigma} \frac{\partial E}{\partial C_{\mu p}^{\mathbf{k},\sigma}} + \frac{\partial E}{\partial C_{\mu a}^{\mathbf{k},\sigma,*}} C_{\mu p}^{\mathbf{k},\sigma,*} \Bigl) - \delta_{ap} \Bigl( \frac{\partial E}{\partial C_{\mu b}^{\mathbf{k},\sigma}} C_{\mu q}^{\mathbf{k},\sigma} + C_{\mu b}^{\mathbf{k},\sigma,*} \frac{\partial E}{\partial C_{\mu q}^{\mathbf{k},\sigma,*}} \Bigl) \Bigr] \\
		& + \sum_{\mu,\nu=1} \Big( \frac{\partial^2 E}{\partial C_{\mu b}^{\mathbf{k},\sigma} \partial C_{\nu q}^{\mathbf{k},\sigma}} C_{\mu a}^{\mathbf{k},\sigma} C_{\nu p}^{\mathbf{k},\sigma} + \frac{\partial^2 E}{\partial C_{\mu a}^{\mathbf{k},\sigma,*} \partial C_{\nu p}^{\mathbf{k},\sigma,*}} C_{\mu b}^{\mathbf{k},\sigma,*} C_{\nu q}^{\mathbf{k},\sigma,*} \\
		& - \frac{\partial^2 E}{\partial C_{\mu b}^{\mathbf{k},\sigma} \partial C_{\nu p}^{\mathbf{k},\sigma}} C_{\mu a}^{\mathbf{k},\sigma} C_{\nu q}^{\mathbf{k},\sigma} - \frac{\partial^2 E}{\partial C_{\mu a}^{\mathbf{k},\sigma,*} \partial C_{\nu q}^{\mathbf{k},\sigma,*}} C_{\mu b}^{\mathbf{k},\sigma,*} C_{\nu p}^{\mathbf{k},\sigma,*} \\
		& - \frac{\partial^2 E}{\partial C_{\mu b}^{\mathbf{k},\sigma} \partial C_{\nu p}^{\mathbf{k},\sigma,*}} C_{\mu a}^{\mathbf{k},\sigma} C_{\nu q}^{\mathbf{k},\sigma,*} - \frac{\partial^2 E}{\partial C_{\mu q}^{\mathbf{k},\sigma} \partial C_{\nu a}^{\mathbf{k},\sigma,*}} C_{\mu p}^{\mathbf{k},\sigma} C_{\nu b}^{\mathbf{k},\sigma,*} \\
		& + \frac{\partial^2 E}{\partial C_{\mu b}^{\mathbf{k},\sigma} \partial C_{\nu q}^{\mathbf{k},\sigma,*}} C_{\mu a}^{\mathbf{k},\sigma} C_{\nu p}^{\mathbf{k},\sigma,*} + \frac{\partial^2 E}{\partial C_{\mu p}^{\mathbf{k},\sigma} \partial C_{\nu a}^{\mathbf{k},\sigma,*}} C_{\mu q}^{\mathbf{k},\sigma} C_{\nu b}^{\mathbf{k},\sigma,*} \Big) \biggr\} 
	\end{aligned}
\end{equation}

\begin{equation}
	\label{eq:HYY_C}
	\begin{aligned}
		\left. (H_{YY}^{\mathbf{k},\sigma} )_{ab,pq} = \frac{\partial^2 E}{\partial Y_{ab}^{\mathbf{k},\sigma} \partial Y_{pq}^{\mathbf{k},\sigma}} \right|_{\mathbf{R}^{\mathbf{k},\sigma}=\mathbf{0}} =& 2 \operatorname{Re} \biggl\{ \frac{1}{2} \sum_{\mu=1} \Bigl[ -\delta_{bp} \Bigl( C_{\mu a}^{\mathbf{k},\sigma} \frac{\partial E}{\partial C_{\mu q}^{\mathbf{k},\sigma}} + \frac{\partial E}{\partial C_{\mu a}^{\mathbf{k},\sigma,*}} C_{\mu q}^{\mathbf{k},\sigma,*} \Bigr) - \delta_{aq} \Bigl( \frac{\partial E}{\partial C_{\mu b}^{\mathbf{k},\sigma}} C_{\mu p}^{\mathbf{k},\sigma} + C_{\mu b}^{\mathbf{k},\sigma,*} \frac{\partial E}{\partial C_{\mu p}^{\mathbf{k},\sigma,*}} \Bigr) \\
		& - \delta_{bq} \Bigl( C_{\mu a}^{\mathbf{k},\sigma} \frac{\partial E}{\partial C_{\mu p}^{\mathbf{k},\sigma}} + \frac{\partial E}{\partial C_{\mu a}^{\mathbf{k},\sigma,*}} C_{\mu p}^{\mathbf{k},\sigma,*} \Bigr) - \delta_{ap} \Bigl( \frac{\partial E}{\partial C_{\mu b}^{\mathbf{k},\sigma}} C_{\mu q}^{\mathbf{k},\sigma} + C_{\mu b}^{\mathbf{k},\sigma,*} \frac{\partial E}{\partial C_{\mu q}^{\mathbf{k},\sigma,*}} \Bigr) \Bigr] \\
		& - \sum_{\mu,\nu=1} \Big( \frac{\partial^2 E}{\partial C_{\mu b}^{\mathbf{k},\sigma} \partial C_{\nu q}^{\mathbf{k},\sigma}} C_{\mu a}^{\mathbf{k},\sigma} C_{\nu p}^{\mathbf{k},\sigma} + \frac{\partial^2 E}{\partial C_{\mu a}^{\mathbf{k},\sigma,*} \partial C_{\nu p}^{\mathbf{k},\sigma,*}} C_{\mu b}^{\mathbf{k},\sigma,*} C_{\nu q}^{\mathbf{k},\sigma,*} \\
		& + \frac{\partial^2 E}{\partial C_{\mu b}^{\mathbf{k},\sigma} \partial C_{\nu p}^{\mathbf{k},\sigma}} C_{\mu a}^{\mathbf{k},\sigma} C_{\nu q}^{\mathbf{k},\sigma} + \frac{\partial^2 E}{\partial C_{\mu a}^{\mathbf{k},\sigma,*} \partial C_{\nu q}^{\mathbf{k},\sigma,*}} C_{\mu b}^{\mathbf{k},\sigma,*} C_{\nu p}^{\mathbf{k},\sigma,*} \\
		& - \frac{\partial^2 E}{\partial C_{\mu b}^{\mathbf{k},\sigma} \partial C_{\nu p}^{\mathbf{k},\sigma,*}} C_{\mu a}^{\mathbf{k},\sigma} C_{\nu q}^{\mathbf{k},\sigma,*} - \frac{\partial^2 E}{\partial C_{\mu q}^{\mathbf{k},\sigma} \partial C_{\nu a}^{\mathbf{k},\sigma,*}} C_{\mu p}^{\mathbf{k},\sigma} C_{\nu b}^{\mathbf{k},\sigma,*} \\
		& - \frac{\partial^2 E}{\partial C_{\mu b}^{\mathbf{k},\sigma} \partial C_{\nu q}^{\mathbf{k},\sigma,*}} C_{\mu a}^{\mathbf{k},\sigma} C_{\nu p}^{\mathbf{k},\sigma,*} - \frac{\partial^2 E}{\partial C_{\mu p}^{\mathbf{k},\sigma} \partial C_{\nu a}^{\mathbf{k},\sigma,*}} C_{\mu q}^{\mathbf{k},\sigma} C_{\nu b}^{\mathbf{k},\sigma,*} \Big) \biggr\}
	\end{aligned}
\end{equation}

\begin{equation}
	\label{eq:HXY_C}
	\begin{aligned}
		\left. (H_{XY}^{\mathbf{k},\sigma} )_{ab,pq} = \frac{\partial^2 E}{\partial X_{ab}^{\mathbf{k},\sigma} \partial Y_{pq}^{\mathbf{k},\sigma}}  \right|_{\mathbf{R}^{\mathbf{k},\sigma}=\mathbf{0}} =& 2 \operatorname{Im} \biggl\{ \frac{1}{2} \sum_{\mu=1} \Bigl[ -\delta_{bp} \Bigl( C_{\mu a}^{\mathbf{k},\sigma} \frac{\partial E}{\partial C_{\mu q}^{\mathbf{k},\sigma}} + \frac{\partial E}{\partial C_{\mu a}^{\mathbf{k},\sigma,*}} C_{\mu q}^{\mathbf{k},\sigma,*} \Bigr) - \delta_{aq} \Bigl( \frac{\partial E}{\partial C_{\mu b}^{\mathbf{k},\sigma}} C_{\mu p}^{\mathbf{k},\sigma} + C_{\mu b}^{\mathbf{k},\sigma,*} \frac{\partial E}{\partial C_{\mu p}^{\mathbf{k},\sigma,*}} \Bigr) \\
		& - \delta_{bq} \Bigl( C_{\mu a}^{\mathbf{k},\sigma} \frac{\partial E}{\partial C_{\mu p}^{\mathbf{k},\sigma}} + \frac{\partial E}{\partial C_{\mu a}^{\mathbf{k},\sigma,*}} C_{\mu p}^{\mathbf{k},\sigma,*} \Bigr) - \delta_{ap} \Bigl( \frac{\partial E}{\partial C_{\mu b}^{\mathbf{k},\sigma}} C_{\mu q}^{\mathbf{k},\sigma} + C_{\mu b}^{\mathbf{k},\sigma,*} \frac{\partial E}{\partial C_{\mu q}^{\mathbf{k},\sigma,*}} \Bigr) \Bigr] \\
		& - \sum_{\mu,\nu=1} \Big( \frac{\partial^2 E}{\partial C_{\mu b}^{\mathbf{k},\sigma} \partial C_{\nu q}^{\mathbf{k},\sigma}} C_{\mu a}^{\mathbf{k},\sigma} C_{\nu p}^{\mathbf{k},\sigma} + \frac{\partial^2 E}{\partial C_{\mu a}^{\mathbf{k},\sigma,*} \partial C_{\nu p}^{\mathbf{k},\sigma,*}} C_{\mu b}^{\mathbf{k},\sigma,*} C_{\nu q}^{\mathbf{k},\sigma,*} \\
		& + \frac{\partial^2 E}{\partial C_{\mu b}^{\mathbf{k},\sigma} \partial C_{\nu p}^{\mathbf{k},\sigma}} C_{\mu a}^{\mathbf{k},\sigma} C_{\nu q}^{\mathbf{k},\sigma} + \frac{\partial^2 E}{\partial C_{\mu a}^{\mathbf{k},\sigma,*} \partial C_{\nu q}^{\mathbf{k},\sigma,*}} C_{\mu b}^{\mathbf{k},\sigma,*} C_{\nu p}^{\mathbf{k},\sigma,*} \\
		& - \frac{\partial^2 E}{\partial C_{\mu b}^{\mathbf{k},\sigma} \partial C_{\nu p}^{\mathbf{k},\sigma,*}} C_{\mu a}^{\mathbf{k},\sigma} C_{\nu q}^{\mathbf{k},\sigma,*} - \frac{\partial^2 E}{\partial C_{\mu q}^{\mathbf{k},\sigma} \partial C_{\nu a}^{\mathbf{k},\sigma,*}} C_{\mu p}^{\mathbf{k},\sigma} C_{\nu b}^{\mathbf{k},\sigma,*} \\
		& - \frac{\partial^2 E}{\partial C_{\mu b}^{\mathbf{k},\sigma} \partial C_{\nu q}^{\mathbf{k},\sigma,*}} C_{\mu a}^{\mathbf{k},\sigma} C_{\nu p}^{\mathbf{k},\sigma,*} - \frac{\partial^2 E}{\partial C_{\mu p}^{\mathbf{k},\sigma} \partial C_{\nu a}^{\mathbf{k},\sigma,*}} C_{\mu q}^{\mathbf{k},\sigma} C_{\nu b}^{\mathbf{k},\sigma,*} \Big) \biggr\}
	\end{aligned}
\end{equation}

Expanding the energy derivatives with respect to the Bloch natural orbital coefficients directly yields the explicit forms for the Hessian blocks $H_{XX}^{\mathbf{k},\sigma}$, $H_{YY}^{\mathbf{k},\sigma}$, and $H_{XY}^{\mathbf{k},\sigma}$:

\begin{equation}
	\label{eq:HXX_F}
	\begin{aligned}
		(H_{XX}^{\mathbf{k},\sigma} )_{ab,pq} ={} & \frac{1}{N_{\mathbf{k}}} \text{Re} \Big\{ 
		\delta_{bp} F_{aq}^{\mathbf{k},\sigma} (n_a^{\mathbf{k},\sigma} + n_q^{\mathbf{k},\sigma} - 2n_b^{\mathbf{k},\sigma}) 
		- \delta_{bq} F_{ap}^{\mathbf{k},\sigma} (n_a^{\mathbf{k},\sigma} + n_p^{\mathbf{k},\sigma} - 2n_b^{\mathbf{k},\sigma}) \\
		& - \delta_{ap} F_{bq}^{\mathbf{k},\sigma} (n_b^{\mathbf{k},\sigma} + n_q^{\mathbf{k},\sigma} - 2n_a^{\mathbf{k},\sigma}) 
		+ \delta_{aq} F_{bp}^{\mathbf{k},\sigma} (n_b^{\mathbf{k},\sigma} + n_p^{\mathbf{k},\sigma} - 2n_a^{\mathbf{k},\sigma}) \\
		& + 2\delta_{\mathbf{k}\mathbf{k}'}\frac{1}{N_{\mathbf{k}'}} \big[(a^{\mathbf{k},\sigma} b^{\mathbf{k},\sigma} \mid p^{\mathbf{k},\sigma} q^{\mathbf{k},\sigma}) + (a^{\mathbf{k},\sigma} b^{\mathbf{k},\sigma} \mid q^{\mathbf{k},\sigma} p^{\mathbf{k},\sigma})\big] (n_a^{\mathbf{k},\sigma} - n_b^{\mathbf{k},\sigma}) (n_p^{\mathbf{k},\sigma} - n_q^{\mathbf{k},\sigma}) \\
		& - 2\delta_{\mathbf{k}\mathbf{k}'}\frac{1}{N_{\mathbf{k}'}} \big[(a^{\mathbf{k},\sigma} q^{\mathbf{k},\sigma} \mid p^{\mathbf{k},\sigma} b^{\mathbf{k},\sigma}) + (a^{\mathbf{k},\sigma} p^{\mathbf{k},\sigma} \mid q^{\mathbf{k},\sigma} b^{\mathbf{k},\sigma})\big] \\
		& \times \big[f(n_p^{\mathbf{k},\sigma}, n_a^{\mathbf{k},\sigma}) - f(n_q^{\mathbf{k},\sigma}, n_a^{\mathbf{k},\sigma}) - f(n_p^{\mathbf{k},\sigma}, n_b^{\mathbf{k},\sigma}) + f(n_q^{\mathbf{k},\sigma}, n_b^{\mathbf{k},\sigma})\big]
		\Big\}
	\end{aligned}
\end{equation}

\begin{equation}
	\label{eq:HYY_F}
	\begin{aligned}
		(H_{YY}^{\mathbf{k},\sigma} )_{ab,pq} ={} & \frac{1}{N_{\mathbf{k}}} \text{Re} \Big\{ 
		-\delta_{bp} F_{aq}^{\mathbf{k},\sigma} (n_a^{\mathbf{k},\sigma} + n_q^{\mathbf{k},\sigma} - 2n_b^{\mathbf{k},\sigma}) 
		- \delta_{bq} F_{ap}^{\mathbf{k},\sigma} (n_a^{\mathbf{k},\sigma} + n_p^{\mathbf{k},\sigma} - 2n_b^{\mathbf{k},\sigma}) \\
		& - \delta_{ap} F_{bq}^{\mathbf{k},\sigma} (n_b^{\mathbf{k},\sigma} + n_q^{\mathbf{k},\sigma} - 2n_a^{\mathbf{k},\sigma}) 
		- \delta_{aq} F_{bp}^{\mathbf{k},\sigma} (n_b^{\mathbf{k},\sigma} + n_p^{\mathbf{k},\sigma} - 2n_a^{\mathbf{k},\sigma}) \\
		& - 2\delta_{\mathbf{k}\mathbf{k}'} \frac{1}{N_{\mathbf{k}'}} \big[(a^{\mathbf{k},\sigma} b^{\mathbf{k},\sigma} \mid p^{\mathbf{k},\sigma} q^{\mathbf{k},\sigma}) - (a^{\mathbf{k},\sigma} b^{\mathbf{k},\sigma} \mid q^{\mathbf{k},\sigma} p^{\mathbf{k},\sigma})\big] (n_a^{\mathbf{k},\sigma} - n_b^{\mathbf{k},\sigma}) (n_p^{\mathbf{k},\sigma} - n_q^{\mathbf{k},\sigma}) \\
		& + 2\delta_{\mathbf{k}\mathbf{k}'} \frac{1}{N_{\mathbf{k}'}} \big[(a^{\mathbf{k},\sigma} q^{\mathbf{k},\sigma} \mid p^{\mathbf{k},\sigma} b^{\mathbf{k},\sigma}) - (a^{\mathbf{k},\sigma} p^{\mathbf{k},\sigma} \mid q^{\mathbf{k},\sigma} b^{\mathbf{k},\sigma})\big] \\
		& \times \big[f(n_p^{\mathbf{k},\sigma}, n_a^{\mathbf{k},\sigma}) - f(n_q^{\mathbf{k},\sigma}, n_a^{\mathbf{k},\sigma}) - f(n_p^{\mathbf{k},\sigma}, n_b^{\mathbf{k},\sigma}) + f(n_q^{\mathbf{k},\sigma}, n_b^{\mathbf{k},\sigma})\big]
		\Big\}
	\end{aligned}
\end{equation}

\begin{equation}
	\label{eq:HXY_F}
	\begin{aligned}
		(H_{XY}^{\mathbf{k},\sigma} )_{ab,pq}  ={} & \frac{1}{N_{\mathbf{k}}} \text{Im} \Big\{ 
		\delta_{bp} F_{aq}^{\mathbf{k},\sigma} (n_a^{\mathbf{k},\sigma} + n_q^{\mathbf{k},\sigma} - 2n_b^{\mathbf{k},\sigma}) 
		+ \delta_{bq} F_{ap}^{\mathbf{k},\sigma} (n_a^{\mathbf{k},\sigma} + n_p^{\mathbf{k},\sigma} - 2n_b^{\mathbf{k},\sigma}) \\
		& - \delta_{ap} F_{bq}^{\mathbf{k},\sigma} (n_b^{\mathbf{k},\sigma} + n_q^{\mathbf{k},\sigma} - 2n_a^{\mathbf{k},\sigma}) 
		- \delta_{aq} F_{bp}^{\mathbf{k},\sigma} (n_b^{\mathbf{k},\sigma} + n_p^{\mathbf{k},\sigma} - 2n_a^{\mathbf{k},\sigma}) \\
		& + 2\delta_{\mathbf{k}\mathbf{k}'} \frac{1}{N_{\mathbf{k}'}} \big[(a^{\mathbf{k},\sigma} b^{\mathbf{k},\sigma} \mid p^{\mathbf{k},\sigma} q^{\mathbf{k},\sigma}) - (a^{\mathbf{k},\sigma} b^{\mathbf{k},\sigma} \mid q^{\mathbf{k},\sigma} p^{\mathbf{k},\sigma})\big] (n_a^{\mathbf{k},\sigma} - n_b^{\mathbf{k},\sigma}) (n_p^{\mathbf{k},\sigma} - n_q^{\mathbf{k},\sigma}) \\
		& - 2\delta_{\mathbf{k}\mathbf{k}'} \frac{1}{N_{\mathbf{k}'}} \big[(a^{\mathbf{k},\sigma} q^{\mathbf{k},\sigma} \mid p^{\mathbf{k},\sigma} b^{\mathbf{k},\sigma}) - (a^{\mathbf{k},\sigma} p^{\mathbf{k},\sigma} \mid q^{\mathbf{k},\sigma} b^{\mathbf{k},\sigma})\big] \\
		& \times \big[f(n_p^{\mathbf{k},\sigma}, n_a^{\mathbf{k},\sigma}) - f(n_q^{\mathbf{k},\sigma}, n_a^{\mathbf{k},\sigma}) - f(n_p^{\mathbf{k},\sigma}, n_b^{\mathbf{k},\sigma}) + f(n_q^{\mathbf{k},\sigma}, n_b^{\mathbf{k},\sigma})\big]
		\Big\}
	\end{aligned}
\end{equation}

Without the $\mathbf{k}$-point sampling, the energy expression and its first- and second-order derivatives with respect to the orbital rotation matrix $\mathbf{R}^{\sigma}$ for molecular systems can be obtained as a special case of the corresponding expressions for periodic systems. The RDMFT energy functional for a molecular system is given by

\begin{equation}
	\label{eq:E_mol}
	\begin{aligned}
		E ={} & \sum_{\sigma} \sum_{\mu \nu} h_{\mu\nu} \gamma_{\nu\mu,\sigma} 
		+ \frac{1}{2} \sum_{\sigma,\sigma'}  \sum_{\mu\nu\kappa\lambda} \gamma_{\nu\mu,\sigma} (\mu\nu \mid \kappa\lambda) \gamma_{\lambda\kappa,\sigma'} 
		- \frac{1}{2} \sum_{\sigma} \sum_{ij} \sum_{\mu\nu\kappa\lambda} C_{\mu i}^{\sigma,*} C_{\nu j}^{\sigma} (\mu\nu \mid \kappa\lambda) C_{\kappa j}^{\sigma,*} C_{\lambda i}^{\sigma} f(n_i^{\sigma}, n_j^{\sigma})
	\end{aligned}
\end{equation}
where $\gamma_{\mu\nu,\sigma} = \sum_{i} C_{\mu i}^{\sigma} n_i^{\sigma} C_{\nu i}^{\sigma,*}$. 
For molecular systems, $\mathbf{R}^{\sigma}$ is an antisymmetric matrix satisfying $(\mathbf{R}^{\sigma})^T = -\mathbf{R}^{\sigma}$. The first-order derivative of $E$ with respect to $\mathbf{R}^{\sigma}$ is given by
\begin{equation}
	\label{eq:dE_dR_mol}
	\left. \frac{\partial E}{\partial R_{ab}^\sigma}\right|_{\mathbf{R}^\sigma=\mathbf{0}} = \sum_{\mu=1} \Big( C_{\mu a}^\sigma \frac{\partial E}{\partial C_{\mu b}^\sigma} - \frac{\partial E}{\partial C_{\mu a}^\sigma} C_{\mu b}^\sigma  \Big)
\end{equation}
which can be expressed in terms of general Fock matrix elements as
\begin{equation}
	\label{eq:dE_dR_F_mol}
	\left. \frac{\partial E}{\partial R^\sigma_{pq}} \right|_{\mathbf{R}^\sigma=\mathbf{0}} = -2F_{pq}^\sigma (n_p^\sigma - n_q^\sigma)
\end{equation}
The second-order derivative of $E$ with respect to $\mathbf{R}^{\sigma}$ is given by
\begin{equation}
	\label{eq:HRR_mol}
	\begin{aligned}
		(H_{RR}^\sigma)_{ab,pq} ={} & \delta_{bp} F_{aq}^\sigma (n_a^\sigma + n_q^\sigma - 2n_b^\sigma) - \delta_{bq} F_{ap}^\sigma (n_a^\sigma + n_p^\sigma - 2n_b^\sigma) \\
		& - \delta_{ap} F_{bq}^\sigma (n_b^\sigma + n_q^\sigma - 2n_a^\sigma) + \delta_{aq} F_{bp}^\sigma (n_b^\sigma + n_p^\sigma - 2n_a^\sigma) \\
		& + 4(a^\sigma b^\sigma \mid p^\sigma q^\sigma) (n_a^\sigma - n_b^\sigma)(n_p^\sigma - n_q^\sigma) \\
		& - 4(a^\sigma q^\sigma \mid p^\sigma b^\sigma) \big[f(n_p^\sigma, n_a^{\sigma}) - f(n_q^\sigma, n_a^\sigma) - f(n_p^\sigma, n_b^\sigma) + f(n_q^\sigma, n_b^\sigma)\big]
	\end{aligned}
\end{equation}

where

\begin{equation}
	\label{eq:Fpq_mol}
	F_{pq}^\sigma = h_{pq}^\sigma + J_{pq}^\sigma - K_{pq}^\sigma
\end{equation}

\begin{equation}
	\label{eq:hpq_mol}
	h_{pq}^\sigma = \sum_{\mu\nu}^{K} C_{\mu p}^{\sigma,*} h_{\mu\nu} C_{\nu q}^\sigma
\end{equation}

\begin{equation}
	\label{eq:Jpq_mol}
	J_{pq}^\sigma = \sum_{\sigma'} \sum_{\mu\nu\kappa\lambda} C_{\mu p}^{\sigma,*} C_{\nu q}^\sigma (\mu\nu \mid \kappa\lambda) \gamma_{\lambda \kappa, \sigma'}
\end{equation}

\begin{equation}
	\label{eq:Kpq_mol}
	K_{pq}^\sigma n_i^\sigma = \sum_{j} \sum_{\mu\nu\kappa\lambda} C_{\mu p}^{\sigma,*} C_{\nu j}^\sigma (\mu\nu \mid \kappa\lambda) C_{\kappa j}^{\sigma,*} C_{\lambda q}^\sigma f(n_i^\sigma, n_j^\sigma)
\end{equation}

\clearpage

\section{Range-separated Gaussian density fitting}
\label{sec: rsgdf_alg}
\addSIcontents{section}{Range-separated Gaussian Density Fitting}

Evaluating four-center two-electron repulsion integrals (ERIs) for Coulomb and exchange operators in periodic systems using Gaussian basis sets is notoriously challenging due to prohibitive computational costs and significant storage challenge. To mitigate this bottleneck, various approximation schemes have been developed, typically employing two- and three-center integrals to represent the original four-center ERIs. In this work, the range-separated Gaussian density fitting (RSGDF) \cite{ye2021rsgdf} approach is employed to calculate the ERIs.  

The ERIs for periodic systems are intrinsically complicated due to the introduction of $\mathbf{k}$-point dependence. These integrals are defined as
\begin{equation}
	\label{eq:eri}
	(\mu^{\mathbf{k}_\mu} \nu^{\mathbf{k}_\nu} | \kappa^{\mathbf{k}_\kappa} \lambda^{\mathbf{k}_\lambda}) = \int_{\Omega} d\mathbf{r}_1 \int d\mathbf{r}_2 \frac{\phi_{\mu}^{\mathbf{k}_\mu,*} (\mathbf{r}_1) \phi_{\nu}^{\mathbf{k}_\nu} (\mathbf{r}_1) \phi_{\kappa}^{\mathbf{k}_\kappa,*} (\mathbf{r}_2) \phi_{\lambda}^{\mathbf{k}_\lambda} (\mathbf{r}_2)}{|\mathbf{r}_1 - \mathbf{r}_2|} 
\end{equation}
where the crystal momentum conservation requires $-\mathbf{k}_{\mu}+\mathbf{k}_{\nu}-\mathbf{k}_{\kappa}+\mathbf{k}_{\lambda}=\mathbf{G}$ , with $\mathbf{G}$ representing a reciprocal lattice vector. 
In RSGDF, the pair density $\rho_{\mu\nu}^{\mathbf{k}_1\mathbf{k}_2} (\mathbf{r}) = \phi_{\mu}^{\mathbf{k}_\mu,*} (\mathbf{r}) \phi_{\nu}^{\mathbf{k}_\nu} (\mathbf{r})$ is expanded using a set of auxiliary Gaussian basis functions $\{\chi_{P}\}$
\begin{equation}
	\label{eq:pair_rho}
	\rho_{\mu \nu}^{\mathbf{k}_{\mu} \mathbf{k}_{\nu}}(\mathbf{r}) \approx \sum_{P}^{n_{\text{aux}}} d_{P \mu \nu}^{\mathbf{k}_{\mu} \mathbf{k}_{\nu}} \chi_{P}^{\mathbf{k}_{\mu \nu}}(\mathbf{r})
\end{equation}
with $\mathbf{k}_{\mu \nu}=-\mathbf{k}_\mu + \mathbf{k}_\nu$. $d_{P \mu \nu}^{\mathbf{k}_{\mu} \mathbf{k}_{\nu}}$ are obtained by solving the following linear equation
\begin{equation}
	\label{eq:line_eq}
	\sum_{Q} \mathcal{J}_{PQ}^{\mathbf{k}_{\mu \nu}} d_{Q\mu\nu}^{\mathbf{k}_\mu \mathbf{k}_\nu} = V_{P\mu\nu}^{\mathbf{k}_\mu\mathbf{k}_\nu}
\end{equation}
where $\mathcal{J}_{P Q}^{\mathbf{k}} = (\chi_{P}^{-\mathbf{k}} | \chi_{Q}^{\mathbf{k}})$ is the two-center Coulomb integral and $V_{P \mu \nu}^{\mathbf{k}_{\mu} \mathbf{k}_{\nu}} = (\chi_{P}^{-\mathbf{k}_{\mu \nu}} | \phi_{\mu}^{\mathbf{k}_{\mu}} \phi_{\nu}^{\mathbf{k}_{\nu}})$ is the three-center Coulomb integral. This allows the ERIs to be expressed as
\begin{equation}
	\label{eq:eri_expand}
	(\mu^{\mathbf{k}_\mu} \nu^{\mathbf{k}_\nu} | \kappa^{\mathbf{k}_\kappa} \lambda^{\mathbf{k}_\lambda}) = (\rho_{\mu \nu}^{\mathbf{k}_{\mu} \mathbf{k}_{\nu}}| \rho_{\kappa \lambda}^{\mathbf{k}_{\kappa} \mathbf{k}_{\lambda}}) \approx \sum_{PQ} V_{P\mu\nu}^{\mathbf{k}_\mu \mathbf{k}_\nu} (\mathcal{J}_{PQ}^{\mathbf{k}_{\kappa \lambda}})^{-1} V_{Q\kappa\lambda}^{\mathbf{k}_\kappa \mathbf{k}_\lambda} 
\end{equation}
In this way, the memory footprint is significantly reduced compared with that of the original four-center ERIs. 

To address the slow convergence of $V_{P \mu \nu}^{\mathbf{k}_\mu \mathbf{k}_\nu}$, the RSGDF scheme splits the Coulomb operator into short-range (SR) and long-range (LR) components with $\omega$ controlling their relative weights 
\begin{equation}
	\label{eq:coulomb_operator}
	\frac{1}{\mathbf{r}} = \underbrace{\frac{\operatorname{erfc}(\omega \mathbf{r})}{\mathbf{r}}}_{\text{SR}} + \underbrace{\frac{\operatorname{erf}(\omega \mathbf{r})}{\mathbf{r}}}_{\text{LR}}
\end{equation}

The SR part can be efficiently evaluated via lattice summation in real space
\begin{equation}
	\label{eq:3c_eri_sr}
	\left(V_{P \mu \nu}^{\mathbf{k}_\mu \mathbf{k}_\nu}\right)_{\mathrm{SR}}=\sum_{\mathbf{M N}} \mathrm{e}^{-i \mathbf{k}_\mu \cdot \mathbf{M}} \mathrm{e}^{i \mathbf{k}_\nu \cdot \mathbf{N}}\left(V_{P \mu \nu}^{\mathbf{0}, \mathbf{M N}}\right)_{\mathrm{SR}}-\delta_{\mathbf{k}_\mu \mathbf{k}_\nu} \frac{\pi}{\Omega \omega^2} \int_{\Omega} d \mathbf{r}_1 \int d \mathbf{r}_2 \chi_P^{\mathbf{k}_{\mu \nu}}\left(\mathbf{r}_1\right) \rho_{\mu \nu}^{\mathbf{k}_\mu \mathbf{k}_\nu}\left(\mathbf{r}_2\right) 
\end{equation}
where
\begin{equation}
	(V_{P \mu \nu}^{\mathbf{0, MN}})_{\mathrm{SR}} = \int_{\Omega} d \mathbf{r}_1 \int d \mathbf{r}_2 \chi_P(\mathbf{r}_1) \frac{\operatorname{erfc}(\omega |\mathbf{r}_1 - \mathbf{r}_2|)}{|\mathbf{r}_1 - \mathbf{r}_2|} \phi_\mu^*(\mathbf{r}_2-\mathbf{M}) \phi_\nu(\mathbf{r}_2-\mathbf{N}) 
\end{equation}
On the other hand, the LR integrals are treated in reciprocal space using Fourier transforms
\begin{equation}
	\left(V_{P \mu \nu}^{\mathbf{k}_\mu \mathbf{k}_\nu}\right)_{\mathrm{LR}} = 4\pi \sum_{\mathbf{G}}^{\prime} \frac{e^{-|\mathbf{G} + \mathbf{k}_{\mu \nu}|^2 / 4\omega^2}}{|\mathbf{G} + \mathbf{k}_{\mu \nu}|^2} \chi_P^{\mathbf{k}_{\mu \nu}}(-\mathbf{G}) \rho_{\mu \nu}^{\mathbf{k}_\mu \mathbf{k}_\nu}(\mathbf{G})
\end{equation}
where the primed summation indicates the exclusion of $\mathbf{G} = \mathbf{0}$ term for $\mathbf{k}_\mu = \mathbf{k}_\nu$ and $\rho_{\mu \nu}^{\mathbf{k}_\mu \mathbf{k}_\nu}(\mathbf{G})$ and $\chi_P^{\mathbf{k}}(\mathbf{G})$ are the analytical Fourier transform of $\rho$ and $\chi$
\begin{equation} 
	\label{eq:rho_uv_G}
	\rho_{\mu \nu}^{\mathbf{k}_\mu \mathbf{k}_\nu}(\mathbf{G}) = \sum_{\mathbf{N}} e^{i\mathbf{k}_\nu \cdot \mathbf{N}} \int_{\Omega} d\mathbf{r} \phi_\mu^*(\mathbf{r}) \phi_\nu(\mathbf{r}-\mathbf{N}) e^{-i(\mathbf{G}+\mathbf{k}_{\mu \nu}) \cdot \mathbf{r}} 
\end{equation}
\begin{equation}
	\chi_P^{\mathbf{k}}(\mathbf{G}) = \int_\Omega d\mathbf{r} \chi_P(\mathbf{r}) e^{-i(\mathbf{k}+\mathbf{G}) \cdot \mathbf{r}} 
\end{equation}

For the further improvement of convergence rate, both the auxiliary and atomic orbital (AO) basis sets are partitioned into compact and diffuse subsets according to the exponents of the primitive Gaussian functions. This partitioning results in six types of three-center integrals. Four types involving diffuse bra or ket are evaluated in reciprocal space, whereas the remaining two types consisting compact functions are computed using the range-separation scheme defined above. Implementation details can be found in Ref.~\cite{ye2021rsgdf}.

For the $\omega$P22 functional, the evaluation of long-range exchange-correlation terms relies only on the LR ERIs. Herein, Eq.~(\ref{eq:rho_uv_G}) is employed and the LR ERIs can be evaluated in reciprocal space
\begin{equation}
	(\mu^{\mathbf{k}_\mu}\nu^{\mathbf{k}_\nu}|
	\kappa^{\mathbf{k}_\kappa}\lambda^{\mathbf{k}_\lambda})_{\text{LR}}
	= \frac{4\pi}{\Omega}\sum_{\mathbf{G}+\mathbf{k}_{\mu \nu} \neq 0}
	\frac{\exp\! (-\left|\mathbf{G}+\mathbf{k}_{\mu\nu}\right|^{2}\!/4\omega^{2})}
	{\left|\mathbf{G}+\mathbf{k}_{\mu\nu}\right|^{2}}
	\rho_{\mu\nu}^{\mathbf{k}_\mu \mathbf{k}_\nu}\!\left(\mathbf{G}\right)
	\rho_{\kappa\lambda}^{\mathbf{k}_\kappa \mathbf{k}_\lambda}\!\left(-\mathbf{G}\right)
\end{equation}
with the $\mathbf{G} = 0$ singularity is removed when $\mathbf{k}_{\mu \nu} = 0$. 

Using the Ewald summation method, the divergent $\mathbf{G} = 0$ contribution of the electron-electron Coulomb interaction is treated together with the electron-nuclear and nuclear-nuclear $\mathbf{G} = 0$ terms, yielding an additive constant to the total energy. However, the $\mathbf{G} = 0$ term arising from the exact-exchange-like functionals in RDMFT (such as Power and the LR part of $\omega$P22 functionals) requires explicit and dedicated treatment. The treatment of the exchange potential at $\mathbf{G} = 0$ follows Ref.~\cite{paier2005perdew}, which uses standard Ewald summation to calculate the energy difference between an isolated probe charge and a periodically repeated charge in a homogeneous compensating background to yield the $\mathbf{G} = 0$ correction potential. Specifically, the exchange-correlation term of $\omega$P22 comprises only the LR Coulomb operator $\text{erf}(\omega \mathbf{r})/\mathbf{r}$, whose reciprocal-space kernel, $4 \pi \text{exp}(-\mathbf{G}^2/4\omega^2)/\mathbf{G}^2$, decays rapidly. Consequently, the $\mathbf{G} = 0$ correction potential can be constructed from the LR component of the Ewald summation alone.

\clearpage

\section{Integral screening}
\addSIcontents{section}{Integral Screening}

The integral screening scheme adopted in this work is identical to that implemented in PySCF \cite{pyscf2018,pyscf2020recent,sun2023various}. The primitive Gaussian function centered at  $\mathbf{T}_{\mu}$ is defined as
\begin{equation}
	\phi_{\mu}(\mathbf{r}) = N_{\mu}(\mathbf{r} - \mathbf{T}_{\mu})^{l_\mu} e^{-\alpha_{\mu}(\mathbf{r} - \mathbf{T}_{\mu})^2}
\end{equation}
where $N_{\mu}$  is the normalization factor,  $\alpha_{\mu}$ is the Gaussian exponent, and $l_\mu$ is the angular momentum quantum number. Due to the exponential decay of Gaussian functions, the contribution from distant image cells in the periodic lattice summation becomes negligible. To ensure computational efficiency without sacrificing accuracy, the distance cutoff $R_{\text{cut}}$ is utilized to truncate the lattice summation. In this work, the two-center and three-center SR integrals are evaluated in real space using this summation technique.
The two-center SR integrals are expressed as
\begin{equation}
	g_{\mu, \nu}^{\mathbf{k}} = \sum_{\mathbf{N}} e^{i \mathbf{k}_{\nu} \cdot \mathbf{N}} \int d\mathbf{r}_1 \int  d\mathbf{r}_2 \frac{\chi_{\mu}^*(\mathbf{r}_1) \operatorname{erfc}\left(\omega |\mathbf{r}_1 - \mathbf{r}_2|\right) \chi_{\nu}\left(\mathbf{r}_2 - \mathbf{N}\right)}{|\mathbf{r}_1 - \mathbf{r}_2|} 
\end{equation}

To determine $R_{\text{cut}}$ for a given precision $\tau$, we consider the most diffuse primitive Gaussian basis functions. The truncation error $\varepsilon$ is bounded by
\begin{equation}
	\varepsilon \lesssim \frac{2 \pi^{3.5} N_{\mu} N_{\nu} e^{-\theta_{\mu \nu \omega} R_{\text{cut}}^2} (\theta_{\mu \nu \omega} R_{\text{cut}})^{l_{\mu \nu} - 1}}{\Omega \sqrt{\theta_{\mu \nu \omega}} \alpha_{\mu}^{l_{\mu} + 3/2} \alpha_{\nu}^{l_{\nu} + 3/2}} < \tau
\end{equation}
where 
\begin{equation}
	\theta_{\mu \nu \omega}=\left(\alpha_{\mu}^{-1}+\alpha_{\nu}^{-1}+\omega^{-2}\right)^{-1}
\end{equation}

Similarly, for the three-center SR integrals 
\begin{equation}
	g_{\mu\nu,\kappa}^{\mathbf{k}_{\mu}\mathbf{k}_{\nu}} = \sum_{\mathbf{MN}} e^{i\mathbf{k}_{\nu} \cdot \mathbf{N} - i\mathbf{k}_{\mu} \cdot \mathbf{M}} \int d\mathbf{r}_1 \int  d\mathbf{r}_2 \frac{\chi_{\kappa}(\mathbf{r}_1) \chi_{\mu}^*(\mathbf{r}_2 - \mathbf{M}) \chi_{\nu}(\mathbf{r}_2 - \mathbf{N}) }{|\mathbf{r}_1 - \mathbf{r}_2|} 
\end{equation}

The corresponding error bound is given by
\begin{equation}
	\varepsilon < \frac{2^{l_{\mu}+1} \pi^{3.5} N_{\mu} N_{\nu} N_{\kappa} e^{-\theta_{\nu\kappa\omega} R_{cut}^2} \theta_{\mu\nu\kappa\omega}^{3/2} (\theta_{\nu\kappa\omega} R_{cut})^{l_{\mu\nu\kappa}-2} R_{cut}}{\Omega \theta_{\nu\kappa\omega} \alpha_{\mu\nu}^{l_{\mu\nu}+3/2} \alpha_{\kappa}^{l_{\kappa}+3/2} \alpha_{\nu}^{l_{\nu}}} f_{l_{\mu\nu\kappa}} (\theta_{\mu\nu\kappa\omega}^{-1} \theta_{\nu\kappa\omega}^2 R_{cut}^2) < \tau
\end{equation}
where
\begin{equation}
	\theta_{\mu \nu \kappa \omega} = (\alpha_{\mu \nu}^{-1} + \alpha_{\kappa}^{-1} + \omega^{-2})^{-1}
\end{equation}
\begin{equation} 
	\quad \theta_{\nu \kappa \omega} = (\alpha_{\nu}^{-1} + \alpha_{\kappa}^{-1} + \omega^{-2})^{-1}
\end{equation}
\begin{equation} 
	f_l(x) = \sum_{k=0}^{l-1} \frac{(2l-1)!!}{(2l-2k-1)!!(2x)^k}
\end{equation}

In the large-$R_{\text{cut}}$ limit, $f_l$ typically ranges between 1 and 2. For simplicity and conservative estimation, we set $f_l = 2$ in our screening protocol.

For the LR part, the convergence is governed by the energy cutoff $E_{\text{cut}}$, which determines the number of plane waves in the basis. Since the $E_{\text{cut}}$ derived from two-center LR integrals often underestimates the error, we employ a more stringent criterion based on the four-center integral error to define the final $E_{\text{cut}}$  used fot the evaluation of the two-center LR integrals, which reads 
\begin{equation} 
	\varepsilon(E_{cut}) = \frac{1}{\Omega} \sum_{|\mathbf{G}|^2 > 2 E_{\text{cut}}} \frac{4\pi}{|\mathbf{G}|^2} \rho_{\mu\nu}(\mathbf{G}) \rho_{\kappa\lambda}(-\mathbf{G}) < 16\pi^2 \int_{\sqrt{2E_{cut}}}^{\infty} \rho_{\mu\nu}(G) \rho_{\kappa\lambda}(G) dG < \tau
\end{equation}

Here, $\rho_{\mu\nu}(\mathbf{G})$  denotes the Fourier transforms of the density pair $\rho_{\mu\nu}(\mathbf{r})$, which is approximated as
\begin{equation} 
	|\rho_{\mu\nu}(\mathbf{G})| =  \rho_{\mu\nu}(G) \approx N_{\mu}N_{\nu} e^{-\frac{G^2}{4\alpha_{\mu\nu}}} e^{-\theta_{\mu\nu}d_{\mu\nu}^2} \left( \frac{G}{2\alpha_{\mu\nu}} \right)^{l_{\mu\nu}} \left( \frac{\pi}{\alpha_{\mu\nu}} \right)^{3/2}
\end{equation}
where
\begin{equation} 
	\theta_{\mu\nu} = (\alpha_{\mu}^{-1} + \alpha_{\nu}^{-1})^{-1}
\end{equation}
\begin{equation} 
	d_{\mu\nu} = |\mathbf{T}_{\mu} - \mathbf{T}_{\nu}|
\end{equation}

To obtain a robust estimate of the plane-wave energy cutoff $E_{\text{cut}}$, we consider the integral associated with the most compact Gaussian basis functions, whose Fourier components exhibit the slowest decay in reciprocal space. The resulting upper bound for the truncation error can be written as
\begin{equation} 
	\varepsilon(E_{\text{cut}}) < 16\pi^2 \int_{\sqrt{2E_{\text{cut}}}}^{\infty} \rho_{\kappa\kappa}^2(G) dG.
\end{equation}

Substituting the density pair expression into the above inequality yields the final relation for $E_{\text{cut}}$ with respect to the target precision $\tau$

\begin{equation}
	8\pi^2 N_{\kappa}^4 \left( \frac{\pi}{2\alpha_{\kappa}} \right)^3 \left( \frac{E_{\text{cut}}}{8\alpha_{\kappa}^2} \right)^{2l_{\kappa} - 1/2} e^{-\frac{E_{\text{cut}}}{\theta_{\kappa \omega}}} < \tau
\end{equation}
where
\begin{equation}
	\theta_{\kappa \omega} = ((2\alpha_{\kappa})^{-1} + (2\omega^2)^{-1})^{-1}
\end{equation}

\clearpage

\section{Supplementary Figures and Tables}
\addSIcontents{section}{Supplementary Figures and Tables}

\begin{figure}[H]
	\centering
	\includegraphics[width=0.5\textwidth]{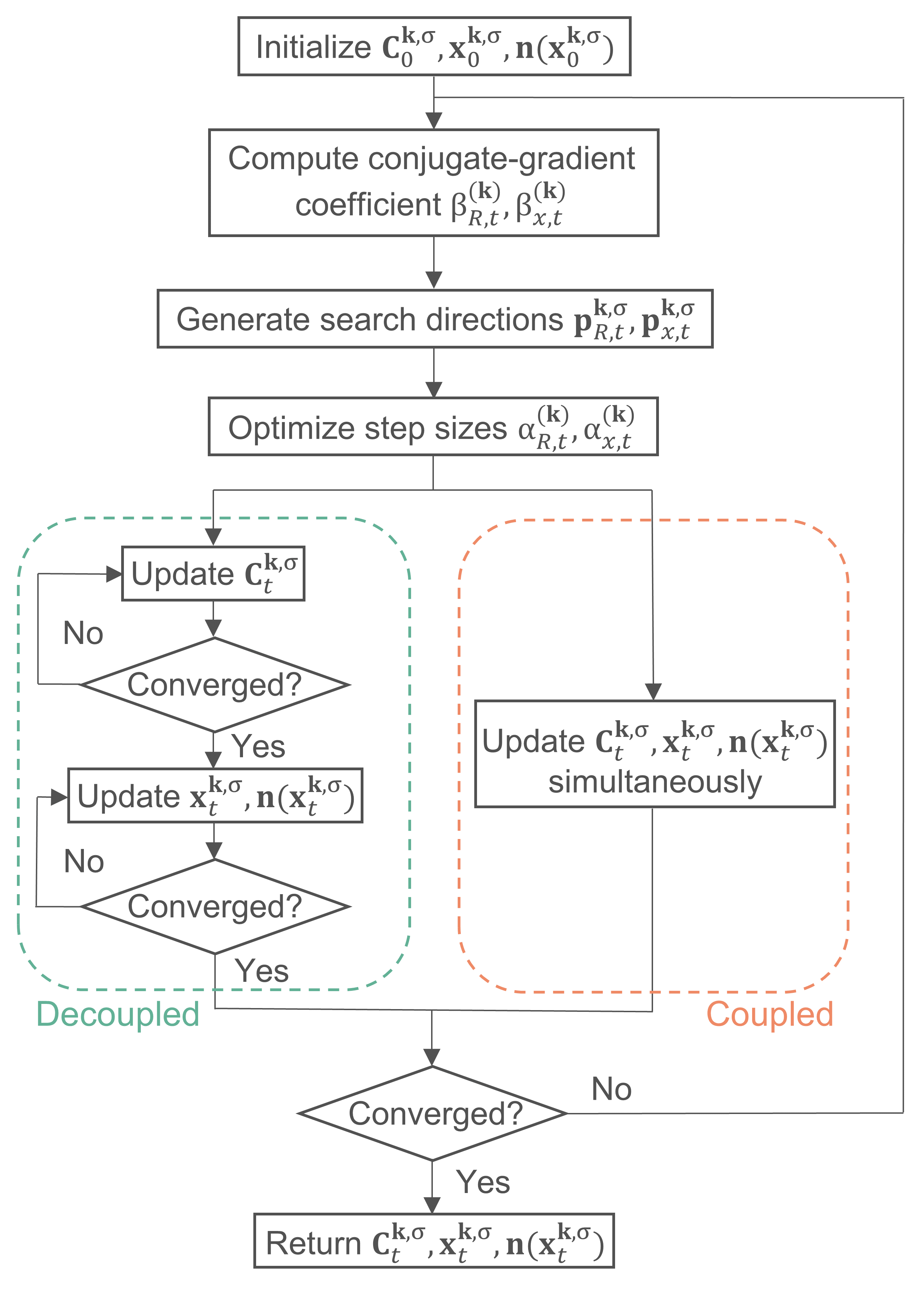}
	\caption{\label{fig:30_workflow} Computational workflow of the PCG algorithm integrated with coupled and decoupled optimization schemes.}  
\end{figure}

\clearpage

\begin{table}[H]
	\caption{\label{tab:table1_mol_iter} 
		Iteration counts for various molecules under molecular and periodic calculation schemes within the Power(0.6), Power(0.7), and $\omega$P22 functionals using coupled optimization (CO) and decoupled optimization (deCO) algorithms. All calculations are performed with the cc-pVTZ basis set. Mean iteration counts across all test systems are provided at the bottom.}
	\begin{ruledtabular}
		\renewcommand{\arraystretch}{1.3}
		\begin{tabular}{lcccccccccccc}
			& \multicolumn{6}{c}{Molecular} & \multicolumn{6}{c}{Periodic} \\
			\cline{2-7} \cline{8-13}
			Mol & \multicolumn{2}{c}{Power(0.6)} & \multicolumn{2}{c}{Power(0.7)} & \multicolumn{2}{c}{$\omega$P22} & \multicolumn{2}{c}{Power(0.6)} & \multicolumn{2}{c}{Power(0.7)} & \multicolumn{2}{c}{$\omega$P22} \\
			\cline{2-3} \cline{4-5} \cline{6-7} \cline{8-9} \cline{10-11} \cline{12-13}
			& CO & deCO & CO & deCO & CO & deCO & CO & deCO & CO & deCO & CO & deCO \\
			\hline
			$\text{H}_2$           &  41 &  81 &  54 &  96 &  24 &  74 &  35 &   80 &  29 & 101 &  24 &  75 \\
			$\text{H}_2\text{O}$   &  80 & 210 &  90 & 161 &  50 & 184 & 143 &  333 &  85 & 267 &  44 & 183 \\
			$\text{O}_2$           & 108 & 255 & 112 & 221 &  90 & 236 & 156 & 349 & 121 & 418 &  77 & 283 \\
			$\text{CO}$            &  86 & 187 &  71 & 218 &  86 & 170 & 182 &  479 & 123 & 383 &  52 & 175 \\
			$\text{CO}_2$          & 165 & 225 &  71 & 230 &  66 & 207 & 301 &  655 & 157 & 461 &  57 & 234 \\
			$\text{CH}_4$          &  60 & 175 & 101 & 206 &  31 & 145 & 239 &  495 &  91 & 201 &  41 & 125 \\
			$\text{CH}_2\text{O}$  & 142 & 303 & 127 & 275 &  82 & 233 & 253 &  593 & 444 & 695 &  79 & 237 \\
			$\text{C}_2\text{H}_4$ & 105 & 314 & 107 & 306 &  65 & 292 & 269 &  593 & 167 & 385 &  94 & 282 \\
			$\text{NH}_3$          &  73 & 190 &  90 & 214 &  38 & 191 & 157 &  395 &  93 & 311 &  74 & 161 \\
			$\text{N}_2$           & 122 & 182 &  59 & 199 &  57 & 200 & 215 &  370 & 145 & 313 &  42 & 149 \\
			\hline
			Mean                   & 98.2 & 212.2 & 88.2 & 212.6 & 58.9 & 193.2 & 195.0 & 434.2 & 145.5 & 353.5 & 58.4 & 190.4\\
		\end{tabular}
	\end{ruledtabular}
\end{table}

\begin{table}[h!]
	\caption{\label{tab:table2_iter_counts} 
		Iteration counts for 3D LiH with a $3 \times 3 \times 3$ $\mathbf{k}$-point mesh and diamond with a $2 \times 2 \times 2$ $\mathbf{k}$-point mesh using the Power(0.6), Power(0.7), and $\omega$P22 functionals within Uniform, Distinct, Partially Distinct, and Hybrid calculation schemes. The GTH-SZV-MOLOPT-SR basis set is used.}
	\centering
		
		\renewcommand{\arraystretch}{1.3} 
		
		\begin{tabular}{lccccccc}
			\hline\hline
			& \multicolumn{3}{c}{$\text{LiH}$} & & \multicolumn{3}{c}{Diamond} \\
			\cline{2-4} \cline{6-8}
			& Power(0.6) & Power(0.7) & $\omega$P22 & & Power(0.6) & Power(0.7) & $\omega$P22 \\
			\hline
			Uniform            & 80  & 86  & 31 & & 111 & 62 & 60 \\
			Distinct           & 173 & 49  & 57 & & 227 & 71 & 55 \\
			Partially Distinct & 152 & 104 & 90 & & 375 & 95 & 87 \\
			Hybrid             & 89  & 84  & 39 & & 72  & 63 & 80 \\ \hline\hline
		\end{tabular}
\end{table}

\begin{table}[h!]
	\caption{\label{tab:table3_e_error} 
		Energy errors (a.u.) for 3D LiH with a $3 \times 3 \times 3$ $\mathbf{k}$-point mesh and diamond with a $2 \times 2 \times 2$ $\mathbf{k}$-point mesh using Power(0.6), Power(0.7), and $\omega$P22 functionals under Uniform, Distinct, Partially Distinct, and Hybrid calculation schemes. The GTH-SZV-MOLOPT-SR basis set is used.}
	
	\centering
	\renewcommand{\arraystretch}{1.3} 
	\setlength{\tabcolsep}{10pt}      
	
	\begin{tabular}{lccccccc}
		\hline\hline
		& \multicolumn{3}{c}{$\text{LiH}$} & & \multicolumn{3}{c}{Diamond} \\
		\cline{2-4} \cline{6-8}
		& Power(0.6) & Power(0.7) & $\omega$P22 & & Power(0.6) & Power(0.7) & $\omega$P22 \\
		\hline
		Uniform            & 3.88e-7 & 6.22e-7 & 1.30e-6 & & 2.53e-6 & 0        & 4.53e-8 \\
		Distinct           & 1.46e-3 & 7.52e-6 & 6.28e-6 & & 1.21e-5 & 1.22e-5 & 6.29e-5 \\
		Partially Distinct & 6.32e-6 & 2.87e-6 & 4.40e-6 & & 2.34e-5 & 3.52e-5 & 3.41e-6 \\
		Hybrid             & 0        & 0        & 0        & & 0        & 1.08e-6 & 0        \\
		\hline\hline
	\end{tabular}
\end{table}

\begin{table}[h!]
	\caption{\label{tab:table4_co_deco} 
		Iteration counts for LiH, diamond, graphene and h-BN within Power(0.6), Power(0.7), and $\omega$P22 functionals using coupled optimization (CO) and decoupled optimization (deCO) algorithms. The $\mathbf{k}$-mesh for each system are chosed as follows: $3 \times 3 \times 3$ for $3$D LiH and Diamond, $6 \times 6 \times 1$ for $2$D Graphene and h-BN. Mean iteration counts across all test systems are provided at the bottom. The GTH-SZV-MOLOPT-SR basis set is used.}
	
	\centering
	\renewcommand{\arraystretch}{1.3} 
	\setlength{\tabcolsep}{8pt}       
	
	\begin{tabular}{lcccccccc}
		\hline\hline
		& \multicolumn{2}{c}{Power(0.6)} & & \multicolumn{2}{c}{Power(0.7)} & & \multicolumn{2}{c}{$\omega$P22} \\
		\cline{2-3} \cline{5-6} \cline{8-9}
		Crystal & CO & deCO & & CO & deCO & & CO & deCO \\
		\hline
		$\text{LiH}$     &  75 &   97 & &  86 &  400 & & 31 & 352 \\
		Diamond          & 159 & 1999 & & 150 & 1186 & & 97 & 589 \\
		Graphene         & 134 &  488 & & 116 &  472 & & 87 & 353 \\
		$\text{h-BN}$    &  74 &  282 & &  94 &  220 & & 52 & 371 \\
		\hline
		$\text{Mean}$    & 110.5 & 716.5 & & 111.5 & 569.5 & & 66.75 & 416.25 \\
		\hline\hline
	\end{tabular}
\end{table}

\begin{figure}
	\includegraphics[width=0.5\textwidth]{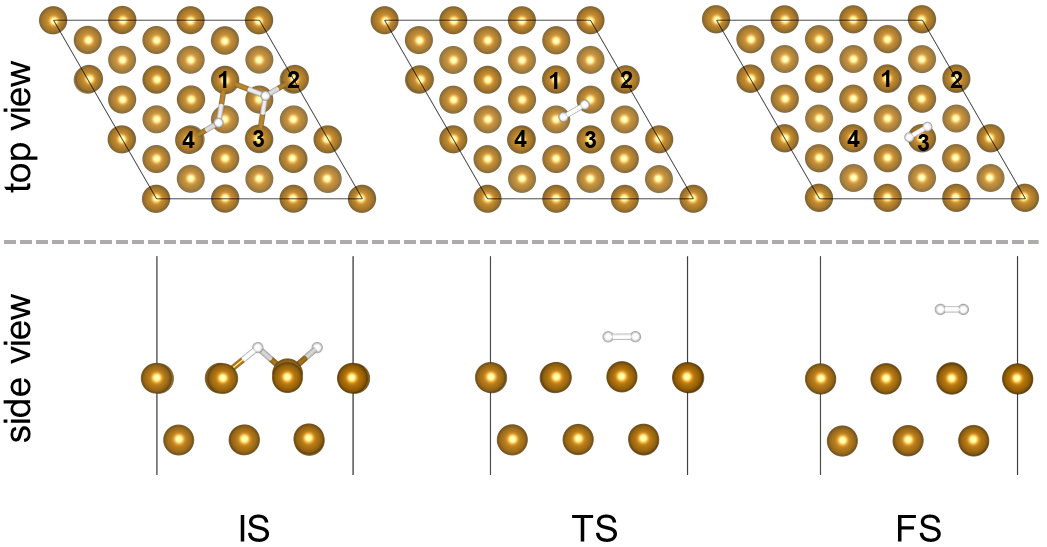}
	\caption{\label{fig:MEP} Selected structures (top and side views) of the initial state (IS), transition state (TS), and final state (FS) optimized by DFT-PBE-D3 along the minimum energy paths for $\text{H}_2$ associative desorption from Cu(111) on the five-layer $3\times3$ Cu(111) slab models. For visual clarity, the side view shows only the top two layers of the five-layer slab models. Atom colors: Cu in brown and H in light grey. In this work, the GTH-DZVP-MOLOPT-SR basis set was applied to the two H atoms and the four labeled Cu atoms, whereas the smaller GTH-SZV-MOLOPT-SR basis set was used for the remaining Cu atoms. Structures are taken with permission from Ref.~\cite{zhou2020nature}. Copyright 2020 Springer Nature.}  
\end{figure}

\begin{figure}
	\includegraphics[width=0.8\textwidth]{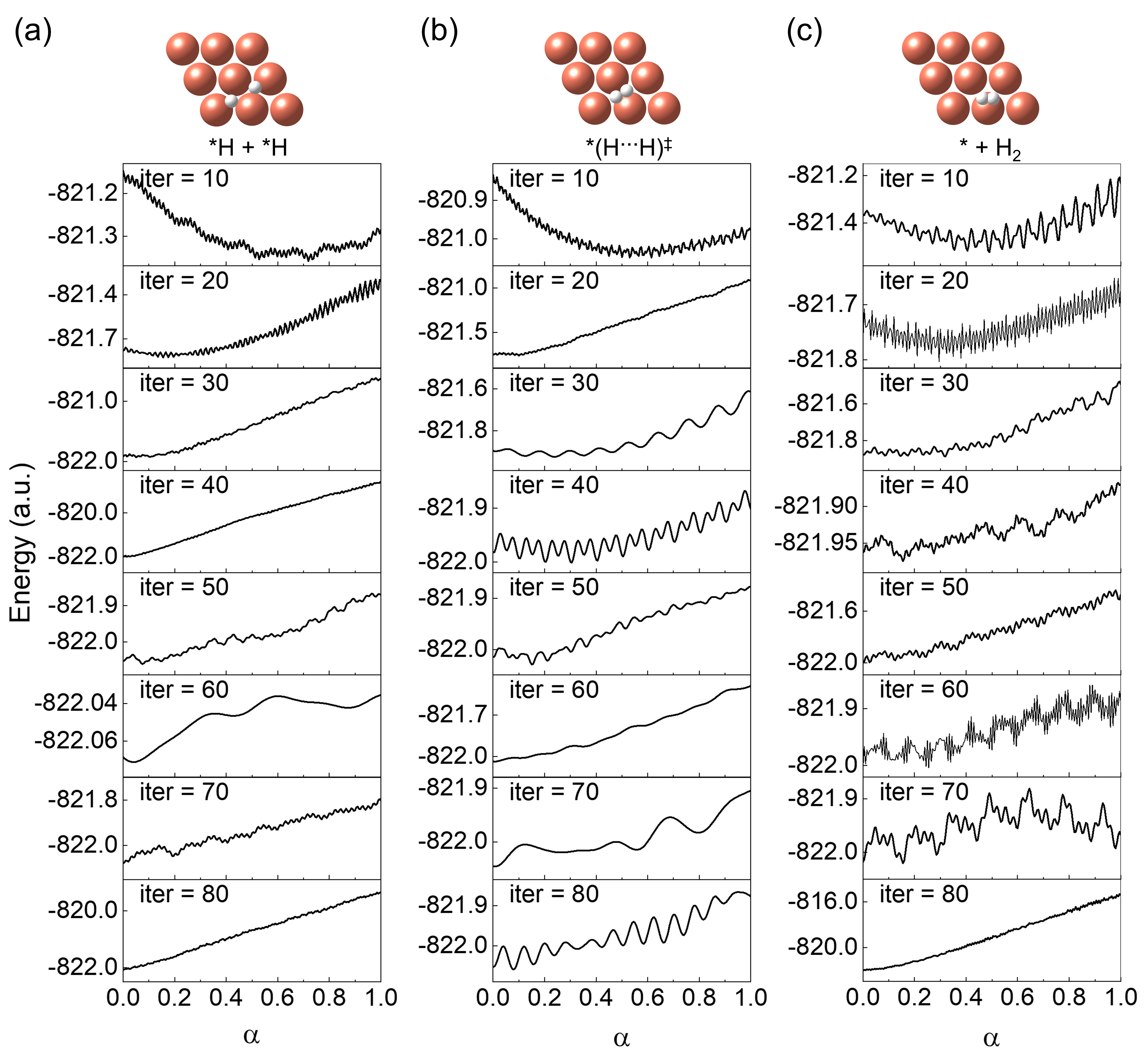}
	\caption{\label{fig:29_E_iter_merge} Energy landscapes computed using the Power($0.6$) functional as functions of the NOs update step size $\alpha_R$ and ONs update step size $\alpha_x$ at optimization iterations 10, 20, 30, 40, 50, 60, 70, and 80 for the three adsorption configurations along the $\text{H}_2$ desorption pathway on Cu(111) with a $2 \times 2 \times 1$ $\mathbf{k}$-point mesh. The step size range from 0 to 1 ($\alpha_R = \alpha_x = \alpha \in [0, 1]$), sampled at 200 evenly spaced points. }  
\end{figure}

\begin{figure}
	\includegraphics[width=0.7\textwidth]{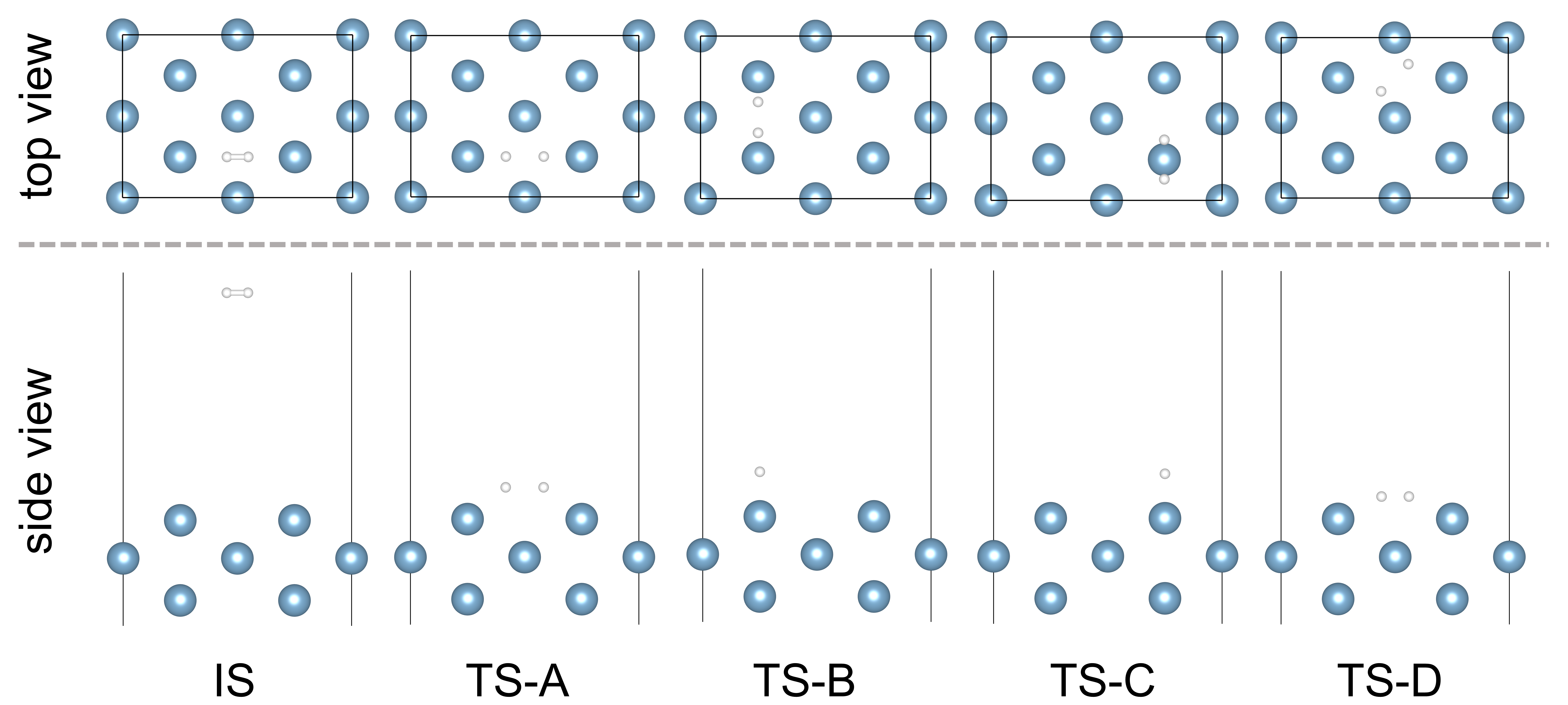}
	\caption{\label{fig:Al110_MEP} Structures (top and side views) of the initial state (IS) and the various transition states (TS-A--TS-D) considered in this work for $\text{H}_2$ dissociative adsorption on a three-layer $2\times2$ Al(110) slab model. Both TS-A and TS-B correspond to true transition states obtained using the PBE functional, except that TS-B was restricted to remain parallel to the surface. The two remaining configurations (TS-C and TS-D) were extracted from restricted 2D cuts through the PES. Atom colors: Al in blue and H in light grey. Structures are taken from Ref.~\cite{Al110DMC2020}.}  
\end{figure}

\begin{table}
	\caption{\label{tab:table1} 
		FCC Al bulk lattice constant $a_0$ and interlayer spacing used for the three layer Al(110) slab calculations. Values are taken from the experimental results obtained by low energy electron diffraction by G\"obel and van Blanckenhagen \cite{gobel1993Ala0} and are extrapolated to 0 K and corrected for zero-point anharmonic contributions \cite{staroverov2004Alinterlayer}.}
	
	\centering
	\renewcommand{\arraystretch}{1.3} 
	\setlength{\tabcolsep}{8pt}       
	
		\begin{tabular}{cc}
			\hline\hline
		& \textrm{Value (\AA)} \\ \hline
		$a_0$ & 4.0154 \\ 
		$d_{1,2}$ & 1.3367   \\ 
		$d_{2,3}$ & 1.4782   \\ 
		\hline\hline
	\end{tabular}
\end{table}

\begin{table}[htbp]
	\caption{\label{tab:table_geom_detailed} 
		Geometric parameters describing the initial state (IS) and the various transition states (TS-A--TS-D) considered in this work for $\text{H}_2$ dissociative adsorption on a three-layer $2\times2$ Al(110) slab model. Here, $\phi$ is the azimuthal angle relative to the long bridge direction, $r_{\text{H}-\text{H}}$ is the H--H bond distance, and $z$ denotes the molecule--surface distance. All values are taken from Ref.~\cite{Al110DMC2020}. 
	}
	
	\centering
	\renewcommand{\arraystretch}{1.3} 
	\setlength{\tabcolsep}{10pt}      
	
	\begin{tabular}{ccccc}
		\hline\hline
		\textrm{Geom} & Site & $\phi$ (deg) & $r_{\text{H}-\text{H}}$ (\AA) & $z$ (\AA) \\ 
		\hline
		\textrm{IS}  & Long bridge  & 0  & 0.741 & 8.000 \\ 
		\textrm{TS-A} & Long bridge  & 0  & 1.334 & 1.118 \\
		\textrm{TS-B} & Short bridge & 90 & 1.080 & 1.568 \\
		\textrm{TS-C} & Top          & 90 & 1.368 & 1.564 \\
		\textrm{TS-D} & Long bridge  & 45 & 1.361 & 0.786 \\
		\hline\hline
	\end{tabular}
\end{table}

\begin{table}
	\caption{\label{tab:k_convergence} 
		Convergence test of the reaction barrier of TS-A with respect to the $\mathbf{k}$-point mesh for the $\text{H}_2$ dissociative adsorption process on Al(110) using the PBE functional. A mixed basis set scheme is applied for the Al slab: the top layer is treated with GTH-DZVP-MOLOPT-SR, while the remaining two layers are described by GTH-SZV-MOLOPT-SR.}
	
	\centering
	\renewcommand{\arraystretch}{1.2} 
	\setlength{\tabcolsep}{18pt}      
	
	\begin{tabular}{cc}
		\hline\hline
		$\mathbf{k}$-points & $\text{E}_{\text{TS}}$ (eV) \\ 
		\hline
		$4 \times 4 \times 1$ & 0.648 \\  
		$5 \times 5 \times 1$ & 0.481 \\  
		$6 \times 6 \times 1$ & 0.507 \\  
		$7 \times 7 \times 1$ & 0.536 \\  
		$8 \times 8 \times 1$ & 0.563 \\  
		$9 \times 9 \times 1$ & 0.550 \\  
		\hline\hline
	\end{tabular}
\end{table}

\begin{table*}[htbp]
	\caption{\label{tab:table_barriers} 
		Calculated reaction barriers for the dissociative adsorption of $\text{H}_2$ (i.e. $\text{H}_2$ + * $\rightarrow$ *$\text{H}$ + *$\text{H}$) on the Cu(111) surface, its reverse associative desorption process, and the dissociative adsorption of $\text{H}_2$ on the Al(110) surface compared to experimental values. The results of PBE, PBE0, $\omega$P22, and DMC methods are presented. $E_{\text{TS}}$ and $E_{\text{TS}}^{\text{rev}}$ denote the activation barriers for the forward and reverse reactions, respectively. All the energies are given in eV.
	}
	
	\begin{ruledtabular}
		\begin{tabular}{ccccccccc}
			System & &  & $\mathbf{k}$-points & PBE & PBE0 & $\omega$P22 & DMC & Expt. \\
			\hline
			\multirow{3}{*}{$\text{H}_2+\text{Cu}(111)$} & & \multirow{3}{*}{$E_{\text{TS}}$} 
			& $2\times2\times1$ & 0.35 & 0.08 & 0.73 & &  \\
			& & & $3\times3\times1$ & 0.53 & 0.51 & 0.59 & 0.56\footnote{Data obtained in \cite{Cu111DMC2017}.} & 0.63\footnote{Data obtained in \cite{diaz2009science}.} \\
			& & & $4\times4\times1$ & 0.58 & 0.44 & 0.58 & &  \\
			\hline
			\multirow{3}{*}{$\text{H}_2+\text{Cu}(111)$} & &  \multirow{3}{*}{$E_{\text{TS}}^{\text{rev}}$} 
			& $2\times2\times1$ & 0.86 & 0.99 & 0.69 & &  \\
			& & & $3\times3\times1$ & 0.68 & 0.83 & 0.80 & - & 0.9\footnote{Data obtained in \cite{roberts1988,anger1989}.} \\
			& & & $4\times4\times1$ & 0.56 & 0.73 & 0.77 & &  \\
			\hline
			\multirow{4}{*}{$\text{H}_2+\text{Al}(110)$} & TS-A & \multirow{4}{*}{$E_{\text{TS}}$} & \multirow{4}{*}{$7\times7\times1$} & 0.54 & 0.63 & 0.80 & 1.09\footnote{Data obtained in \cite{Al110DMC2020}.} & 1.15\footnote{Experimentally motivated reference value obtained in \cite{Al110exp2024}.} \\
			& TS-B & & & 0.57 & 0.69 & 0.91 & 1.16$^\text{d}$ & - \\
			& TS-C & & & 1.05 & 1.23 & 1.44 & 1.59$^\text{d}$ & - \\
			& TS-D & & & 1.04 & 1.17 & 1.49 & 1.46$^\text{d}$ & - \\
		\end{tabular}
	\end{ruledtabular}
\end{table*}

\begin{figure}
	\includegraphics[width=0.5\textwidth]{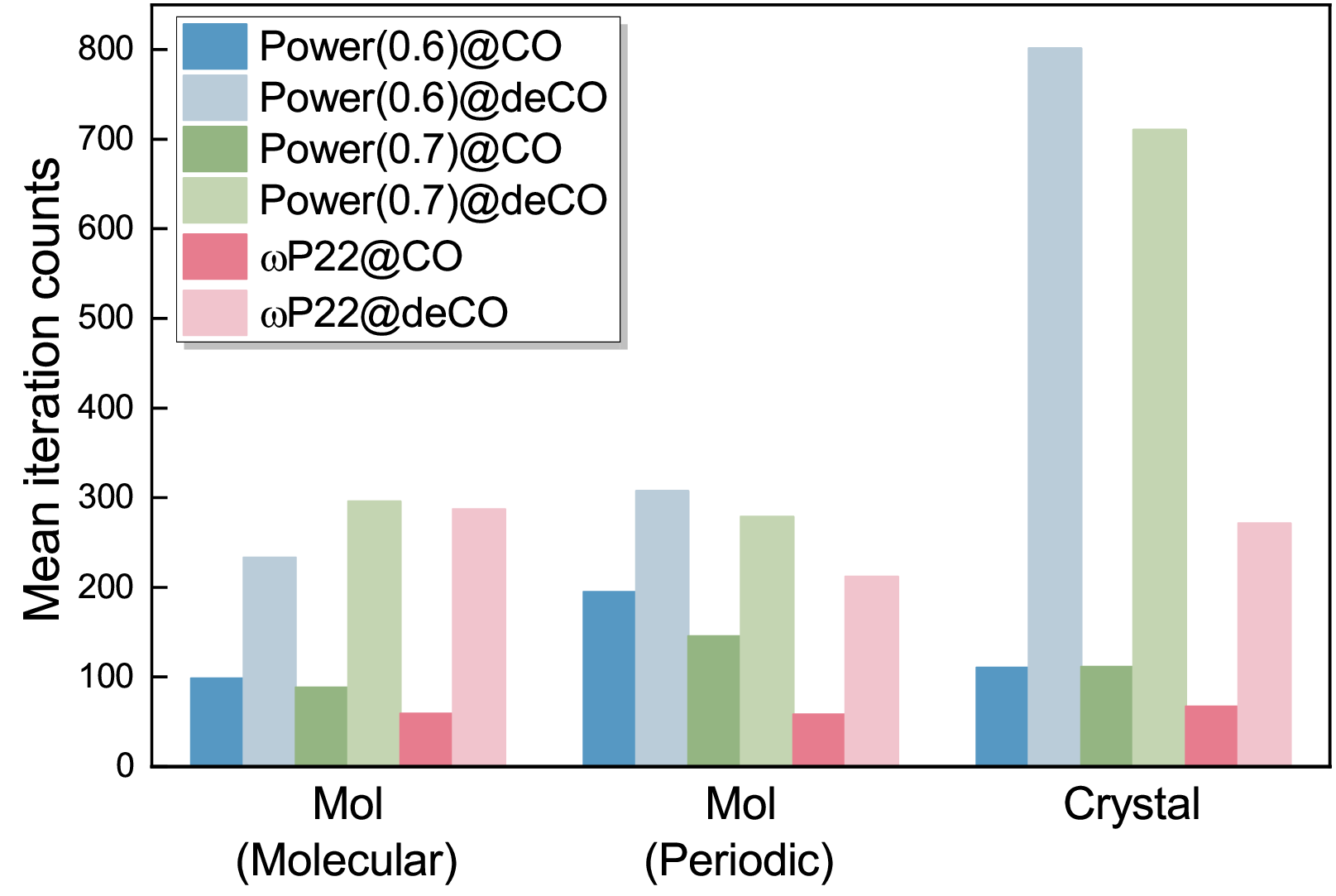}
	\caption{\label{fig:28_mean_iter} Mean iteration counts for ten molecules in both molecular and periodic calculations, as well as for four extended crystals, utilizing the Power($0.6$), Power($0.7$) and $\omega$P22 functionals under CO and deCO optimization methods. }  
\end{figure}

\clearpage

\begin{table*}[h]
	\caption{\label{tab:mol_barriers} 
		Calculated molecular reaction barriers under periodic boundary conditions for six displacement/addition reactions and their reverse reactions compared to the reference values. The results of the PBE, PBE0, Power(0.6), and $\omega$P22 functionals with the cc-pVTZ basis set are presented. The molecular geometries and reference barriers were adopted from the HTBH38 \cite{HTBH2009, HTBH2014} and NHTH38 \cite{HTBH2009, HTBH2014} benchmark sets. $E_{\text{TS}}$ and $E_{\text{TS}}^{\text{rev}}$ denote the activation barriers for the forward and reverse reactions, respectively. All values are in eV.
	}

	\centering
	\renewcommand{\arraystretch}{1.2} 
	\setlength{\tabcolsep}{18pt}      
	
	\begin{ruledtabular}
		\begin{tabular}{lcrrrrc}
			Reaction &  & PBE & PBE0 & Power & $\omega$P22 & Ref. \\
			\hline
			\multirow{2}{*}{$\text{H} + \text{HCl} \rightarrow \text{H}_2 + \text{Cl}$} 
			& $E_{\text{TS}}$ & 0.01 & 0.11 & $-0.46$ & 0.22 & 0.25 \\
			& $E_{\text{TS}}^{\text{rev}}$ & $-0.03$ & 0.12 & 0.73 & 0.37 & 0.34 \\
			
			\multirow{2}{*}{$\text{H} + \text{H}_2 \rightarrow \text{H}_2 + \text{H}$} 
			& $E_{\text{TS}}$ & 0.16 & 0.25 & 0.59 & 0.42 & 0.42 \\
			& $E_{\text{TS}}^{\text{rev}}$ & 0.16 & 0.25 & 0.59 & 0.42 & 0.42 \\
			
			\multirow{2}{*}{$\text{F} + \text{H}_2 \rightarrow \text{HF} + \text{H}$} 
			& $E_{\text{TS}}$ & $-0.49$ & $-0.16$ & 0.25 & $-0.12$ & 0.06 \\
			& $E_{\text{TS}}^{\text{rev}}$ & 1.02 & 1.16 & $-0.31$ & 1.36 & 1.45 \\
			
			\multirow{2}{*}{$\text{H} + \text{H}_2\text{S} \rightarrow \text{H}_2 + \text{HS}$} 
			& $E_{\text{TS}}$ & $-0.06$ & 0.04 & 0.01 & 0.20 & 0.15 \\
			& $E_{\text{TS}}^{\text{rev}}$ & 0.43 & 0.55 & 1.12 & 0.83 & 0.73 \\
			
			\multirow{2}{*}{$\text{H} + \text{F}_2 \rightarrow \text{HF} + \text{F}$} 
			& $E_{\text{TS}}$ & $-0.36$ & $-0.15$ & 0.02 & 0.00 & 0.10 \\
			& $E_{\text{TS}}^{\text{rev}}$ & 3.35 & 4.16 & 2.49 & 4.19 & 4.59 \\
			
			\multirow{2}{*}{$\text{H} + \text{N}_2 \rightarrow \text{HN}_2$} 
			& $E_{\text{TS}}$ & 0.23 & 0.38 & $-0.29$ & 0.59 & 0.62 \\
			& $E_{\text{TS}}^{\text{rev}}$ & 0.37 & 0.49 & 0.69 & 0.51 & 0.46 \\
		\end{tabular}
	\end{ruledtabular}
\end{table*}


\end{document}